\documentclass[a4paper,12pt,draft]{article}
\usepackage{amssymb,euscript,bbold}
\newcommand{\be}{\begin{equation}}
\newcommand{\ee}{\end{equation}}
\newcommand{\ba}{\begin{eqnarray}}
\newcommand{\ea}{\end{eqnarray}}

\def\appendix{{\newpage\section*{Appendix}}\let\appendix\section%
        {\setcounter{section}{0}
        \gdef\thesection{\Alph{section}}}\section}

\def\be{\begin{equation}}
\def\ee{\end{equation}}
\def\ba{\begin{eqnarray}}
\def\ea{\end{eqnarray}}

\def\Dslash{\,\,{\raise.15ex\hbox{/}\mkern-12mu D}}
\def\Dbarslash{\,\,{\raise.15ex\hbox{/}\mkern-12mu {\bar D}}}
\def\delslash{\,\,{\raise.15ex\hbox{/}\mkern-9mu \partial}}
\def\delbarslash{\,\,{\raise.15ex\hbox{/}\mkern-9mu {\bar\partial}}}
\def\pslash{\,\,{\raise.15ex\hbox{/}\mkern-9mu p}}
\def\calDslash{\,\,{\raise.15ex\hbox{/}\mkern-12mu {\cal D}}}

\def\CN{{\mathcal{N}}}
\def\a{\alpha}

\begin{document}
\input{epsf}

\begin{titlepage}

\begin{center}
September, 2004
\hfill hep-th/0409191\\
\hfill HUTP-03/A053\\
\hfill RUNHETC-2003-26\\

\vskip 1.5 cm
{\large \bf $M$ theory and Singularities of Exceptional Holonomy Manifolds}\\
\vskip 0.16cm

\vskip 1 cm
{Bobby S. Acharya$^1$ and Sergei Gukov$^2$}\\
\vskip 1cm
$^1${\sl Abdus Salam International Centre for Theoretical Physics, \\
Strada Costiera 11, 34100 Trieste, Italy.
\\ {\tt bacharya@ictp.trieste.it}\\}
\vskip 0.5cm
$^2${\sl Jefferson Physical Laboratory, Harvard University, \\
Cambridge, MA 02138, U.S.A. \\ {\tt gukov@schwinger.harvard.edu}\\}

\end{center}

\vskip 0.5 cm
\begin{abstract}

$M$ theory compactifications on $G_2$ holonomy manifolds,
whilst supersymmetric, require {\it singularities} in order to obtain
non-Abelian gauge groups, chiral fermions and other properties
necessary for a realistic model of particle physics.
We review recent progress in understanding the
physics of such singularities. Our main aim is to describe
the techniques which have been used to develop our understanding
of $M$ theory physics near these singularities. 
In parallel, we also describe similar
sorts of singularities in $Spin(7)$ holonomy manifolds which
correspond to the properties of three dimensional field theories.
As an application, we review how
various aspects of strongly coupled gauge theories, such as
confinement, mass gap and non-perturbative phase transitions
may be given a simple explanation in $M$ theory.

\end{abstract}

\end{titlepage}

\pagestyle{plain}
\setcounter{page}{1}
\newcounter{bean}
\baselineskip16pt
\tableofcontents

\newpage

\section{Introduction}
\label{intro}

$M$ theory is a promising candidate for a consistent quantum
theory of gravity. The theory unifies all of the five consistent
superstring theories. $M$ theory is locally
supersymmetric and at long distances describes physics
in spacetimes with eleven dimensions. The traditional
approach to obtaining large four dimensional universes
{}from theories with more than
four dimensions is to assume that the ``extra dimensions''
are small.
At energies below the compactification scale of the extra
dimensions, the physics is four dimensional and the
detailed properties of that physics is determined by
the properties of the metric of the extra dimensions.

Recently, $M$ theory compactifications on manifolds with exceptional
holonomy have attracted considerable attention.
The main motivation to study such models is that they
have all the ingredients required to embed phenomenologically
interesting quantum field theories with minimal supersymmetry
into a unified theory containing gravity.

Perhaps the most intriguing reason that makes supersymmetry
one of our best candidates
for physics beyond the Standard Model is the unification of
gauge couplings and a possible mechanism for understanding
the large hierarchy in scale between the masses
of particles at the electroweak scale and the much higher
unification scale
%
\be
M_{GUT} \approx 10^{16}~ GeV
\label{gutmass}
\ee
The main idea of Grand Unification is that three out of four
fundamental forces in nature (strong, weak, and electro-magnetic)
combine into a single force at high energies. At low energies
all of these forces are mediated by an exchange of gauge fields
and to very high accuracy can be described by the Standard Model
of fundamental interactions with the gauge group
\be
U_Y (1) \otimes SU_L (2) \otimes SU_c (3)
\label{smgroup}
\ee
Even though the coupling constants, $\a_i$, associated
with these interactions are dimensionless,
in quantum field theory they become functions of the energy scale $\mu$.
In particular, using the experimental data from
the Large Electron Positron accelerator (LEPEWWG) and from Tevatron
for the values of $\a_i$ at the electroweak scale \cite{smstatus},
one can predict the values of $\a_i (\mu)$ at arbitrary energy scale,
using quantum field theory.
Then, if one plots all $\a_i (\mu)$ as functions of $\mu$
on the same graph, one finds that near the Planck scale three curves come
close to each other, but do not meet at one point.
The latter observation means that the unification can only be
achieved if new physics enters between the electroweak and
the Planck scale, so that at some point all coupling constants
can be made equal,
\be
\a_1 (M_{GUT}) = \a_2 (M_{GUT}) = \a_3 (M_{GUT})
\label{guta}
\ee
When this happens, the three gauge interactions have
the same strength and can be ascribed a common origin.

An elegant and simple solution to the unification problem
is supersymmetry, which leads to a softening of the short distance
singularities and, therefore, modifies the evolution of coupling constants.
In fact, if we consider a minimal supersymmetric generalization
of the Standard Model (MSSM) where all the superpartners of
the known elementary particles have masses above
the effective supersymmetry scale
\be
M_{SUSY} \approx 1~ TeV,
\label{susymass}
\ee
then a perfect unification (\ref{guta}) can be obtained at
the GUT scale (\ref{gutmass}) \cite{superGUT}.

Moreover, supersymmetry might naturally explain the large difference between
the unification scale $M_{GUT}$ and the electroweak
scale ($m_{EW} \approx 100 GeV$) called the ``hierarchy problem''.
Even if some kind of fine tuning in a GUT theory can lead to
a very small number $m_{EW} / M_{GUT} \sim 10^{-14}$,
the problem is to preserve the hierarchy after properly accounting for
quantum corrections.
For example, the one-loop correction to the
Higgs mass $m_H$ in a non-supersymmetric theory
is quadratically divergent, hence
$\delta m_H^2 \sim \Lambda^2$. This is too large if the
cutoff scale $\Lambda$ is large.
Clearly, such quantum corrections destroy the hierarchy,
unless there is a mechanism to cancel these quadratic divergences.
Again, supersymmetry comes to the rescue.
In supersymmetric quantum field theory all quadratic corrections
automatically cancel to all orders in perturbation theory due to
opposite contributions from bosonic and fermionic fields.

The unification of gauge couplings and a possible solution of
the hierarchy problem demonstrate some of the remarkable properties
of supersymmetry.
Namely, the dynamics of supersymmetric field theories is
usually rather constrained (an example of this is the cancellation
of ultraviolet divergences that was mentioned above),
and yet rich enough to exhibit many interesting phenomena,
such as confinement, Seiberg duality,
non-perturbative phase transitions, {\it etc.}
It turns out that many of these phenomena can receive a relatively simple
and elegant explanation in the context of $M$ theory
and its string theory approximations.

In $M$ theory,
there are several natural looking ways to obtain four large
space-time dimensions with minimal ($\CN=1$) supersymmetry from
compactification on a manifold $X$ with special holonomy.
The most well studied possibility is the
heterotic string theory on a Calabi-Yau space $X$ \cite{chsw}.
A second way to obtain
vacua with $\CN=1$ supersymmetry
came into focus with the discovery of string dualities,
which allow definite statements to be made even
in the regions where perturbation theory can not be used \cite{MV}.
It consists of taking $X$ to be an elliptically fibered
Calabi-Yau four-fold as a background in F theory.
These examples are all limits of $M$ theory on $X$.
The third possibility, which will be one of the main
focal points of this review
is to take $M$ theory on a 7-manifold $X$ of $G_2$ holonomy.
A central point concerning such $G_2$ compactifications
is that, if $X$ is smooth, the four dimensional
physics contains at most an abelian gauge group and
no light charged particles. In fact, as we will see,
in order to obtain more interesting four dimensional physics,
$X$ should possess very particular kinds of singularity.
At these
singularities, extra light charged degrees of freedom are
to be found.

Many of these compactifications are related by
various string dualities that we will exploit below
in order to study the dynamics of $M$ theory
on singular manifolds of $G_2$ holonomy.
An example of such a duality --- which may be also of
interest to mathematicians, especially those with an interest
in mirror symmetry --- is a duality
between $M$ theory on $K3$-fibered $G_2$-manifolds and the heterotic string
theory on $T^3$-fibered Calabi-Yau threefolds. On the string theory side
the threefold is endowed with a Hermitian-Yang-Mills connection $A$ and
chiral fermions emerge from zero modes of the Dirac operator
twisted by $A$, whereas on the $M$ theory side this gets mapped
to a statement about the singularities of $X$.

Within the past few years there has been a tremendous amount of
progress in understanding  $M$ theory physics near singularities in
manifolds of exceptional holonomy.
In particular we now understand at
which kinds of singularities in $G_2$-manifolds
the basic requisites of the Standard Model --- non-Abelian gauge groups
and chiral fermions --- are to be found \cite{bsa1,bsa2,AMV,AW,wanom,bsaw}.
One purpose of this review is to explain how this picture
was developed in detail. We mainly aim to equip the reader with
techniques and refer the interested reader to
\cite{Wdeconstr,FriedmannW,Bobbymoduli,Bobbyyukawa}
for more detailed discussions
of phenomenological applications.
Among other things, we shall see how
important properties of strongly coupled gauge theories such as
confinement and the mass gap can receive a semi-classical
description in $M$ theory on $G_2$-manifolds.
Similarly, $Spin(7)$ manifolds expose additional aspects
of $M$ theory, related to the interesting dynamics of
minimally supersymmetric gauge theories in $2+1$ dimensions.

In order to make this review self-contained and pedagogical,
in the next section we start with an introduction to special
holonomy.
In section 3, we review the construction of manifolds with
exceptional holonomy.
Then, in section 4, we derive the basic properties of $M$ theory
on such a manifold, in the limit when the manifold is large and smooth.
Using string dualities, in section 5 we explain how one can
understand many aspects of the physics
when the compactification manifold
has various kinds of singularity.
These techniques are used in later sections to explain various phenomena
in $M$ theory on singular manifolds with exceptional holonomy.
In section 6 we describe in detail the singularities
of $G_2$ manifolds which give rise to chiral fermions.
In section 7, we review topology changing transitions in manifolds
with $G_2$ and $Spin(7)$ holonomy, and their relation to
the so-called {\it geometric transition} in string theory.
Finally, in section 8, we shall see how interesting aspects of
Yang-Mills theory, such as confinement and a mass gap,
receive a very simple explanation within the context
of $M$ theory on a $G_2$ manifold.

We emphasise that, whilst manifolds of special holonomy provide
elegant models of supersymmetric particle physics and gravity,
there is a very important gap in our understanding: how supersymmetry
is broken and why the cosmological constant is so small?

\newpage
\section{Riemannian Manifolds of Special Holonomy}
\label{basics}

\subsection{Holonomy Groups}

\begin{figure}
\begin{center}
\epsfxsize=2.6in\leavevmode\epsfbox{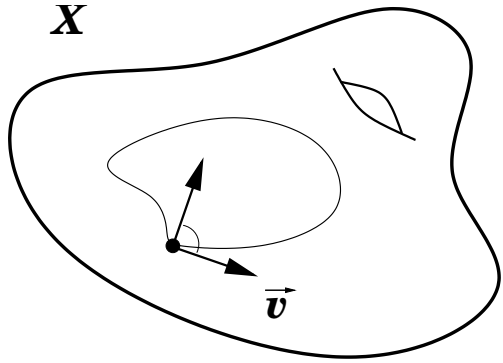}
\end{center}
\caption{Parallel transport of a vector $\vec v$ along
a closed path on the manifold $X$.}
\label{holonomy}
\end{figure}

Consider an oriented manifold $X$ of real dimension $n$
and a vector $\vec v$ at some point on this manifold.
One can explore the geometry of $X$ by
parallel transporting $\vec v$
along a closed contractible path in $X$, see Figure \ref{holonomy}.
Under such an operation the vector $\vec v$
may not come back to itself.
In fact, generically it will transform into  a different vector
that depends on the geometry of $X$, on the path,
and on the connection which was used to transport $\vec v$.
For a Riemannian manifold $X$ with metric $g(X)$, the natural connection
is the Levi-Cevita connection.
Furthermore, Riemannian geometry also tells us that
the length of the vector covariantly transported along
a closed path should be the same as the length of the original vector.
But the direction may be different,
and this is precisely what leads to the concept of holonomy.

The relative direction of
the vector after parallel transport relative to that of
the original vector $\vec v$ is described by {\it holonomy}.
This is simply an $n \times n$ matrix, which
on an $n$-dimensional, oriented manifold is
an element of the special orthogonal group, $SO(n)$.
It is not hard to see that the set of all holonomies themselves form
a group, called the {\it holonomy group}, where the group
structure is induced by the composition of paths and its inverse
corresponds to a path traversed in the opposite direction.
{}From the way we introduced the holonomy group, $Hol(g(X))$,
it seems to depend upon the choice of the base point.
However, for generic
choices of base points the holonomy group is in fact the same,
and therefore $Hol(g(X))$ becomes a true geometric characteristic
of the space $X$ with metric $g(X)$. By definition, we have
\begin{equation}
Hol(g(X)) \subseteq SO(n)
\end{equation}
where the equality holds for sufficiently generic metric on $X$.

In some special instances, however, one finds that $Hol(g(X))$
is a proper subgroup of $SO(n)$. In such cases, we say
that $(X, g(X))$ is a {\it special holonomy} manifold or a manifold
with restricted holonomy.
These manifolds are in some sense
distinguished, for they exhibit special geometric properties.
As we explain later in this section,
these properties are typically associated with
the existence of non-degenerate (in some suitable sense)
$p$-forms which are covariantly constant.
Such $p$-forms also serve as calibrations,
and are related to the subject of minimal varieties.

The possible choices for $Hol(g(X)) \subset SO(n)$ are limited,
and were classified by M.~Berger in 1955 \cite{Berger}.
Specifically, for $(X, g(X))$ simply-connected and neither
locally a product nor symmetric,
the only possibilities for $Hol(g(X))$,
other than the generic case of $SO(n)$,
are $U\left(\frac{n}{2}\right)$,
$SU\left(\frac{n}{2}\right)$,
$Sp\left(\frac{n}{4}\right) \cdot Sp(1)$,
$Sp\left(\frac{n}{4}\right)$,
$G_2$, $Spin(7)$ or $Spin(9)$, see Table 1.
The first four of these correspond, respectively,
to K\"ahler, Calabi-Yau, Quaternionic K\"ahler or
hyper-K\"ahler manifolds. The last three possibilities
are the so-called exceptional cases, which occur only
in dimensions $7$, $8$ and $16$, respectively.
The case of $16$-manifolds with $Spin(9)$ holonomy
is in some sense trivial since the Riemannian metric on
any such manifold is always symmetric \cite{spinnine}. Manifolds
with $G_2$ or $Spin(7)$ holonomy --- which will be our main
subject --- are called {\it exceptional holonomy}
manifolds, since they occur only in dimension seven or eight.
Let us say a few more words about these cases, in particular,
remind the definition and properties of $G_2$ and $Spin(7)$ groups.

\begin{table}\begin{center}
\begin{tabular}{|c|c|c|}
\hline
\rule{0pt}{5mm}
Metric & Holonomy & Dimension \\[3pt]
\hline
\hline
\rule{0pt}{5mm}
K\"ahler & $U \left( \frac{n}{2} \right)$ & $n=$ even \\[3pt]
\cline{1-3}
\rule{0pt}{5mm}
Calabi-Yau & $SU\left(\frac{n}{2}\right)$ & $n=$ even \\[3pt]
\cline{1-3}
\rule{0pt}{5mm}
HyperK\"ahler & $Sp\left(\frac{n}{4}\right)$ & $n=$ multiple of 4 \\[3pt]
\cline{1-3}
\rule{0pt}{5mm}
Quaternionic & $Sp\left(\frac{n}{4}\right)Sp(1)$ & $n=$ multiple of 4 \\[3pt]
\cline{1-3}
\rule{0pt}{5mm}
Exceptional & $G_2$ & 7 \\[3pt]
\cline{1-3}
\rule{0pt}{5mm}
Exceptional & $Spin(7)$ & 8 \\[3pt]
\cline{1-3}
\rule{0pt}{5mm}
Exceptional & $Spin(9)$ & 16 \\[3pt]
\hline
\end{tabular}\end{center}
\caption{Berger's list of holonomy groups.}
\end{table}

The 14-dimensional simple Lie group $G_2$
is precisely the automorphism group of the octonions, $\mathbb{O}$.
It may be defined as the set of elements of $GL(7,\mathbb{R})$,
which preserves the following 3-form on $\mathbb{R}^7$,
\begin{equation}
\label{assocformr}
\Phi = {1 \over 3!} \psi_{ijk}\ dx_i \wedge dx_j \wedge dx_k
\end{equation}
where $x_1, \ldots, x_7$ are coordinates on $\mathbb{R}^7$,
and $\psi_{ijk}$ are totally antisymmetric
structure constants of the imaginary octonions,
\begin{equation}
\label{octo}
\sigma_i \sigma_j
= -\delta_{ij} + \psi_{ijk}\ \sigma_k ~,~~ i,j,k=1, \ldots 7~
\end{equation}
In a particular choice of basis the non-zero structure constants are given by
\begin{equation}
\label{octoconst}
\psi_{ijk} = +1 ~,~~ (ijk) = \{(123),
(147), (165), (246), (257), (354), (367) \} ~.
\end{equation}

The 21-dimensional Lie group $Spin(7)$ is usually
defined as the double cover of $SO(7)$.
However, by analogy with the above definition of $G_2$ group,
it is convenient to define $Spin(7)$ as a subgroup of $GL(8,\mathbb{R})$,
which preserves the following 4-form on $\mathbb{R}^8$,
\begin{eqnarray}
\Omega & = & 
e^{1234} + e^{1256} + e^{1278} + e^{1357}
- e^{1368} - e^{1458} - e^{1467} - \nonumber \\
&&- e^{2358} - e^{2367} - e^{2457} + e^{2468}
+ e^{3456} + e^{3478} + e^{5678}
\label{cayleyform}
\end{eqnarray}
where $e^{ijkl} = dx_i \wedge dx_j \wedge dx_k \wedge dx_l$
and $x_1, \ldots, x_8$ are coordinates on $\mathbb{R}^8$.

Finally, we note that by allowing $\pi_1 (X)$ to be non-trivial
one can obtain proper subgroups of the above list of groups
as holonomy groups of $(X, g(X))$.

\subsection{Relation Between Holonomy and Supersymmetry}

Roughly speaking, one can think of the holonomy group as
a geometric characteristic of the manifold that restricts
the properties that $X$ has.
Namely, as the holonomy group becomes smaller
the more constrained the properties of $X$ become.
Conversely, for manifolds with larger holonomy groups
the geometry is less restricted.

This philosophy becomes especially helpful in the physical
context of superstring/$M$ theory compactifications on $X$.
There, the holonomy of $X$ becomes related to the degree
of supersymmetry preserved in compactification:
{\it manifolds with larger holonomy group
preserve a smaller fraction of the supersymmetry.}
This provides a nice link between the `geometric symmetry'
(holonomy) and the `physical symmetry' (supersymmetry).
In Table 2 we illustrate this general pattern with
a few important examples, which will be used later.

The first example in Table 2 is a torus, $T^n$,
which we view as a quotient of $n$-dimensional
real vector space, $\mathbb{R}^n$, by a lattice.
In this example, if we endow $\mathbb{R}^n$ with a flat metric
then $X=T^n$ has trivial holonomy group, since the Levi-Cevita
connection is zero.
Indeed, no matter which path we choose on $T^n$,
the parallel transport of a vector $\vec v$ along
this path always brings it back to itself.
Hence, this example is the most symmetric one,
in the sense of the previous paragraph, $Hol(X) = {\bf 1}$.
Correspondingly, in $M$ theory, toroidal compactifications
preserve {\it all} of the original supersymmetries.

Our next example is $Hol(X)=SU(3)$ which corresponds to
Calabi-Yau manifolds of complex dimension $3$ (real dimension $6$).
These manifolds exhibit a number of remarkable properties,
such as mirror symmetry, and are reasonably well studied
both in the mathematical and in the physical literature.
We just mention here that compactification on Calabi-Yau
manifolds preserves $1/4$ of the original supersymmetry.
In particular, compactification of heterotic string theory
on $X=CY_3$ yields an ${\mathcal N}=1$ effective field theory
in $3+1$ dimensions.

The last two examples in Table 2 are $G_2$ and $Spin(7)$
manifolds; that is, manifolds with holonomy group $G_2$
and $Spin(7)$, respectively.
They nicely fit into the general pattern, so that as we read
Table 2 from left to right the holonomy increases,
whereas the fraction of unbroken supersymmetry decreases.
Specifically, compactification of $M$ theory on a manifold with
$G_2$ holonomy leads to an ${\mathcal N}=1$ four-dimensional theory
and is therefore of phenomenological interest.
This is similar to the compactification of heterotic string theory
on Calabi-Yau three-folds.
Compactification on $Spin(7)$ manifolds breaks supersymmetry
even further.

\begin{table}\begin{center}
\begin{tabular}{|c|ccccccc|}
\hline
\rule{0pt}{5mm}
Manifold $X$ & $T^n$ && CY$_3$ && $X_{G_2}$ && $X_{Spin(7)}$ \\[3pt]
\hline
\hline
\rule{0pt}{5mm}
$\dim_{\mathbb{R}} (X)$ & $n$ && $6$ && $7$ && $8$ \\[3pt]
\cline{1-8}
\rule{0pt}{5mm}
$Hol(X)$ & {\bf 1} & $\subset$ & $SU(3)$ & $\subset$ & $G_2$ &
$\subset$ & $Spin(7)$ \\[3pt]
\cline{1-8}
\rule{0pt}{5mm}
SUSY & $1$ & $>$ & $1/4$ & $>$ & $1/8$ & $>$ & $1/16$ \\[3pt]
\hline
\end{tabular}\end{center}
\caption{Relation between holonomy and supersymmetry
for certain manifolds.}
\end{table}

Mathematically, the fact that all these manifolds preserve some
supersymmetry is related to the existence of covariantly
constant spinors:
\begin{equation}
\nabla \xi = 0
\label{covspinor}
\end{equation}

In fact, with all bosonic fields apart from the metric set
to zero, \ref{covspinor} is precisely the condition for unbroken
supersymmetry in string
or $M$ theory compactification.
This condition on a spinor field automatically implies
a holonomy reduction: since $\xi$ is invariant
under parallel transport,  $Hol(g(X))$ must be such that
the spinor representation contains the trivial representation. This is
impossible if $Hol(g(X))$ is $SO(n)$, since the spinor representation
is irreducible. Therefore, $Hol(g(X)) \subset SO(n)$.

For example, if $Hol(X)=G_2$ the covariantly constant spinor
is the singlet in the decomposition of the spinor of $SO(7)$
into representations of $G_2$:
$$
{\bf 8} \to {\bf 7} \oplus {\bf 1}
$$

Summarising, in Table 2 we listed some examples of special holonomies
that will be discussed below. All of these manifolds
preserve a certain fraction of supersymmetry, which depends on
the holonomy group. Moreover, all of these manifolds are Ricci-flat,
$$
R_{ij} = 0.
$$
This useful property guarantees that all backgrounds of the form
$$
\mathbb{R}^{11-n} \times X
$$
automatically solve the eleven-dimensional Einstein equations
with vanishing source terms for matter fields.

Of particular interest are $M$ theory compactifications
on manifolds with exceptional holonomy,
\begin{equation}
\matrix{
{\rm M \;\;theory~on} && {\rm M \;\;theory~on} \cr
G_2 {\rm ~manifold} && Spin(7) {\rm ~manifold} \cr
\Downarrow && \Downarrow \cr
&& \cr
{\rm D=3+1} {\rm ~~QFT} && {\rm D=2+1} {\rm ~~QFT}}
\end{equation}
since they lead to effective theories with minimal supersymmetry
in four and three dimensions, respectively.
As mentioned in the introduction,
in such theories one can find many interesting phenomena,
{\it e.g.} confinement, dualities, rich phase
structure, non-perturbative effects, {\it etc.}
This rich structure makes minimal supersymmetry very attractive
to study
and, in particular, motivates the study of $M$ theory
on manifolds with exceptional holonomy.
In this context, the spectrum of elementary particles
in the effective low-energy theory and their interactions
are encoded in the geometry of the space $X$.
Therefore, understanding the latter may help us to learn
more about dynamics of minimally supersymmetric field theories,
or even about $M$ theory itself!


\subsection{Invariant Forms and Minimal Submanifolds}

For a manifold $X$, we have introduced the notion of
special holonomy and related it to the existence of
covariantly constant spinors on $X$, {\it cf.} (\ref{covspinor}).
However, special holonomy manifolds can be also
characerised by the existence of certain
invariant forms and distinguished minimal submanifolds.

Indeed, one can sandwich antisymmetric combinations
of $\Gamma$-matrices with a covariantly constant
spinor $\xi$ on $X$ to obtain antisymmetric tensor
forms of various degree:
\begin{equation}
\omega^{(p)} = \xi^{\dagger} \Gamma_{i_1 \ldots i_p} \xi
\label{invform}
\end{equation}
By construction, the $p$-form $\omega^{(p)}$ is covariantly
constant and invariant under $Hol(g(X))$.
In order to find all possible invariant forms on a special
holonomy manifold $X$, we need to decompose
the space of differential forms on
$X$ into irreducible representations of $Hol(g(X))$
and identify all singlet components.
Since the Laplacian of $g(X)$ preserves this decomposition the harmonic
forms can also be decomposed this way. In a sense, for exceptional
holonomy manifolds, the decomposition of cohomology groups
into representations of $Hol(g(X))$ is analogous
to the Hodge decomposition in the realm of complex geometry.

For example, for a manifold with $G_2$ holonomy this
decomposition is given by \cite{joyce}:
\begin{eqnarray}
H^0 (X, \mathbb{R}) & = & \mathbb{R} \nonumber \\
H^1 (X, \mathbb{R}) & = & H^1_{{\bf 7}} (X, \mathbb{R}) \nonumber \\
H^2 (X, \mathbb{R}) & = & H^2_{{\bf 7}} (X, \mathbb{R})
\oplus H^2_{{\bf 14}} (X, \mathbb{R}) \nonumber \\
H^3 (X, \mathbb{R}) & = & H^3_{{\bf 1}} (X, \mathbb{R})
\oplus H^3_{{\bf 7}} (X, \mathbb{R})
\oplus H^3_{{\bf 27}} (X, \mathbb{R}) \nonumber \\
H^4 (X, \mathbb{R}) & = & H^4_{{\bf 1}} (X, \mathbb{R})
\oplus H^4_{{\bf 7}} (X, \mathbb{R})
\oplus H^4_{{\bf 27}} (X, \mathbb{R}) \label{hunderg2} \\
H^5 (X, \mathbb{R}) & = & H^5_{{\bf 7}} (X, \mathbb{R})
\oplus H^5_{{\bf 14}} (X, \mathbb{R}) \nonumber \\
H^6 (X, \mathbb{R}) & = & H^6_{{\bf 7}} (X, \mathbb{R}) \nonumber \\
H^7 (X, \mathbb{R}) & = & \mathbb{R} \nonumber
\end{eqnarray}
where $H^k_{{\bf n}} (X, \mathbb{R})$ is the subspace of
$H^k (X, \mathbb{R})$ with elements in
an $n$-dimensional irreducible representation of $G_2$.
The fact that the metric on $X$ has irreducible $G_2$-holonomy
implies global constraints on $X$ and this forces some of
the above groups to vanish when $X$ is compact. For example
a compact Ricci flat manifold with holonomy $Sp(k), SU(k), G_2$
or $Spin(7)$ has a finite fundamental group, $\pi_1 (X)$. This implies
that $H^1(X, \mathbb{R}) = {\bf 1}$, which in the $G_2$ case means
$$
H^k_{{\bf 7}} (X, \mathbb{R}) = 0, \quad k=1, \ldots, 6
$$

Let us now return to the construction (\ref{invform})
of the invariant forms on $X$.
{}From the above decomposition we see that on a $G_2$ manifold
such forms can appear only in degree $p=3$ and $p=4$.
They are called {\it associative} and {\it coassociative}
forms, respectively. In fact, a coassociative 4-form
is  the Hodge dual of the associative 3-form.
These forms, which we denote $\Phi$ and $* \Phi$,
enjoy a number of remarkable properties.

For example, the existence of $G_2$ holonomy metric on $X$
is equivalent to the closure and co-closure of the associative
form\footnote{Another, equivalent condition is to say that
the $G_2$-structure $(g,\Phi)$ is torsion-free: $\nabla \Phi =0$.},
\begin{eqnarray}
\label{harm}
d \Phi &=& 0 \\
d*\Phi &=& 0. \nonumber
\end{eqnarray}
This may look a little surprising,
especially since the number of metric components
on a 7-manifold is different from the number
of components of a generic 3-form.
However, given a $G_2$ holonomy metric,
\be
ds^2 = \sum_{i=1}^7 e^i \otimes e^i,
\ee
one can locally write the invariant 3-form $\Phi$
in terms of the vielbein $e^i$, {\it cf.} (\ref{assocformr}),
\begin{equation}
\label{assocform}
\Phi = {1 \over 3!} \psi_{ijk}\ e^i \wedge e^j \wedge e^k
\end{equation}
where $\psi_{ijk}$ are the structure constants
of the imaginary octonions (\ref{octo}).

It is, perhaps, less obvious that one can also
locally reconstruct a $G_2$ metric from the assoiative 3-form:
\begin{eqnarray}
g_{ij} & = & \det (B)^{-1/9} B_{ij} \label{gfromphi} \\
B_{jk} & = & - {1 \over 144} \Phi_{ji_1 i_2}
\Phi_{k i_3 i_4} \Phi_{i_5 i_6 i_7}
\epsilon^{i_1 \ldots i_7} \nonumber
\end{eqnarray}
This will be useful to us in the following sections.

Similarly, on a $Spin(7)$ manifold $X$ we find only one
invariant form (\ref{cayleyform})
in degree $p=4$, called the Cayley form, $\Omega$.
In this case, the decomposition of the cohomology groups of $X$
into $Spin(7)$ representations is \cite{joyce}:
\begin{eqnarray}
H^0 (X, \mathbb{R}) & = & \mathbb{R} \nonumber \\
H^1 (X, \mathbb{R}) & = & H^1_{{\bf 8}} (X, \mathbb{R}) \nonumber \\
H^2 (X, \mathbb{R}) & = & H^2_{{\bf 7}} (X, \mathbb{R})
\oplus H^2_{{\bf 21}} (X, \mathbb{R}) \nonumber \\
H^3 (X, \mathbb{R}) & = & H^3_{{\bf 8}} (X, \mathbb{R})
\oplus H^3_{{\bf 48}} (X, \mathbb{R}) \nonumber \\
H^4 (X, \mathbb{R}) & = & H^4_{{\bf 1}^+} (X, \mathbb{R}) \oplus
H^4_{{\bf 7}^+} (X, \mathbb{R}) \oplus
H^4_{{\bf 27}^+} (X, \mathbb{R}) \oplus
H^4_{{\bf 35}^-} (X, \mathbb{R}) \label{hunderspin} \\
H^5 (X, \mathbb{R}) & = & H^5_{{\bf 8}} (X, \mathbb{R})
\oplus H^5_{{\bf 48}} (X, \mathbb{R}) \nonumber \\
H^6 (X, \mathbb{R}) & = & H^6_{{\bf 7}} (X, \mathbb{R})
\oplus H^6_{{\bf 21}} (X, \mathbb{R}) \nonumber \\
H^7 (X, \mathbb{R}) & = & H^7_{{\bf 8}} (X, \mathbb{R}) \nonumber \\
H^8 (X, \mathbb{R}) & = & \mathbb{R} \nonumber
\end{eqnarray}
The additional label ``$\pm$'' 
denotes self-dual/anti-self-dual four-forms, respectively.
The cohomology class of the 4-form
$\Omega$ generates $H^4_{{\bf 1}^+} (X, \mathbb{R})$,
$$
H^4_{{\bf 1}^+} (X, \mathbb{R}) = \langle [\Omega] \rangle
$$
Again, on a compact manifold with exactly $Spin(7)$-holonomy
we have extra constraints,
\be
H^1_{{\bf 8}} =
H^3_{{\bf 8}} =
H^5_{{\bf 8}} =
H^7_{{\bf 8}} =0, \quad
H^2_{{\bf 7}} =
H^4_{{\bf 7}} =
H^6_{{\bf 7}} =0
\ee

\begin{table}\begin{center}
\begin{tabular}{|c|c|c|c|c|}
\hline
\rule{0pt}{5mm}
$Hol(X)$ & cycle $S$ & $p= \dim (S)$ & Deformations & $\dim ({\rm Def})$ 
\\[3pt]
\hline
\hline
\rule{0pt}{5mm}
$SU(3)$ & SLAG & $3$ & unobstructed & $b_1 (S)$ \\[3pt]
\cline{1-5}
\rule{0pt}{5mm}
$G_2$ & associative & $3$ & obstructed & --- \\[3pt]
\cline{2-5}
\rule{0pt}{5mm}
& coassociative & $4$ & unobstructed & $b_2^+ (S)$  \\[3pt]
\cline{1-5}
\rule{0pt}{5mm}
$Spin(7)$ & Cayley & $4$ & obstructed & --- \\[3pt]
\hline
\end{tabular}\end{center}
\caption{Deformations of calibrated submanifolds.}
\end{table}

Another remarkable propety of the invariant forms is that
they represent the volume forms of minimal
submanifolds in $X$. The forms with these properties are
called {\it calibrations}, and the corresponding submanifolds
are called {\it calibrated submanifolds} \cite{HL}.
More precisely,
we say that a closed $p$-form
$\Psi$ is a calibration if it is less than or equal to
the volume on each oriented $p$-dimensional submanifold $S \subset X$.
Namely, combining the orientation of $S$ with the restriction of
the Riemann metric on $X$ to the subspace $S$, we can define
a natural volume form ${\rm vol} (T_x S)$ on the tangent
space $T_x S$ for each point $x \in S$.
Then, $\Psi \vert_{T_x S} = \a \cdot {\rm vol} (T_x S)$
for some $\a \in \mathbb{R}$, and we write:
$$
\Psi \vert_{T_x S} \le {\rm vol} (T_x S)
$$
if $\a \le 1$. If equality holds for all points $x \in S$,
then $S$ is called a calibrated submanifold with respect
to the calibration $\Psi$. According to this definition, the volume
of a calibrated submanifold $S$ can be expressed in terms of $\Psi$ as:
\begin{equation}
\label{Svol}
{\rm Vol} (S) = \int_{x \in S} \Psi \vert_{T_ x S} = \int_S \Psi
\end{equation}
Since $d\Psi = 0$ the right-hand side depends only on the cohomology
class, so:
$$
{\rm Vol} (S) = \int_S \Psi = \int_{S'} \Psi
= \int_{x \in S'} \Psi \vert_{T_ x S'}
\le \int_{x \in S'} {\rm vol} (T_x S') = {\rm Vol} (S')
$$
for any other submanifold $S'$ in the same homology class.
Therefore, we see that a calibrated submanifold
has minimal volume in its homology class.
This important property of calibrated submanifolds
allows us to identify them with supersymmetric cycles, where the
bound in volume becomes equivalent to the BPS bound.
In particular, branes in string theory and $M$ theory
wrapped over calibrated submanifolds can give rise to
BPS states in the effective theory.

A familiar example of a calibrated submanifold is a special
Lagrangian (SLAG) cycle in a Calabi-Yau 3-fold $X$.
By definition, it is a 3-dimensional submanifold in $X$
calibrated with respect to the real part, ${\rm Re} (\Omega)$,
of the holomorphic 3-form $\Omega$ (more generally,
${\rm Re} (e^{i \theta} \Omega)$, where $\theta$ is an arbitrary phase).
Another class of calibrated submanifolds in Calabi-Yau spaces
consists of holomorphic subvarities, such as holomorphic curves,
surfaces, {\it etc.}
Similarly, if $X$ is a $G_2$ holonomy manifold,
there are associative 3-manifolds and coassociative 4-manifolds,
which correspond, respectively, to the associative 3-form $\Phi$
and to the coassociative 4-form $* \Phi$.
In a certain sense, the role of these two types of calibrated
submanifolds is somewhat similar to the holomorphic and
special Lagrangian submanifolds in a Calabi-Yau space.
In the case of $Spin(7)$ holonomy manifolds, there is only one kind
of calibrated submanifolds --- called Cayley 4-manifolds --- which
correspond to the Cayley 4-form (\ref{cayleyform}).

Deformations of calibrated submanifolds have been studied
by Mclean \cite{Mclean}, and are briefly summarised in Table 3.
In particular, deformations of special Lagrangian and
coassociative submanifolds are unobstructed in the sense that
the local moduli space Def$(S)$ has no singularities.
In both cases, the dimension of the moduli space
is determined by the topology of the calibrated submanifold $S$,
{\it viz.} by $b_1 (S)$ when $S$ is special Lagrangian,
and by $b_2^+ (S)$ when $S$ is coassociative.
These two types of calibrated submanifolds will play a special
role in what follows.


\subsection{Why Exceptional Holonomy is Hard}

Once we have introduced manifolds with special
holonomy, let us try to explain why, until recently,
so little was known about the exceptional cases, $G_2$ and $Spin(7)$.
Indeed, on the physics side, these manifolds are very natural
candidates for constructing minimally supersymmetric field
theories from string/$M$ theory compactifications. Therefore,
one might expect exceptional holonomy manifolds to be at least
as popular and attractive as, say, Calabi-Yau manifolds.
However, there are several reasons why exceptional holonomy
appeared to be a difficult subject;
here we will stress two of them:

{{\bf\large
\begin{itemize}
\item
Existence

\item
Singularities

\end{itemize}
}}

Let us now explain each of these problems in turn.
The first problem refers to the existence of an exceptional
holonomy metric on a given manifold $X$.
Namely, it would be useful to have a general theorem which,
under some favorable conditions, would guarantee
the existence of such a metric.
Indeed, Berger's classification, described earlier
in this section, only tells us which holonomy groups can occur,
but says nothing about examples of such manifolds
or conditions under which they exist.
To illustrate this further, let us recall
that when we deal with Calabi-Yau manifolds we
use such a  theorem all the time --- it is a theorem due
to Yau, proving a conjecture due to Calabi, which guarantees
the existence of a Ricci-flat metric on a compact, complex,
K\"ahler manifold $X$ with $c_1 (X)=0$ \cite{Yau}.
Unfortunately, no analogue of this theorem is known in
the case of $G_2$ and $Spin(7)$ holonomy
(the local existence of such manifolds
was first established in 1985 by Bryant \cite{Bryant}).
Therefore, until such a general theorem is found we are
limited to a case-by-case analysis of the specific examples.
We will return to this problem in the next section. We also note that,
to date, not a single example of a Ricci flat metric of special
holonomy is known {\it explicitly} for a compact, simply connected manifold!

The second reason
is associated with the singularities of these manifolds.
As will be explained in the sequel,
interesting physics occurs at the singularities.
Moreover, the most interesting physics is associated with the types
of singularities of maximal codimension, which exploit
the geometry of the special holonomy manifold to the fullest.
Until recently, little was known about these types of
degenerations of manifolds with $G_2$ and $Spin(7)$ holonomy.
Moreover, even for known examples of isolated singularities,
the dynamics of $M$ theory in these backgrounds was unclear.
Finally, it is important to stress that the mathematical understanding
of exceptional holonomy manifolds would be incomplete
without a proper understanding of singular limits.

\newpage
\section{Construction of Manifolds With Exceptional Holonomy}

In this section we review various methods of constructing
compact and non-compact manifolds with $G_2$ and $Spin(7)$ holonomy.
In the absence of general existence theorems,
akin to Yau's theorem \cite{Yau},
these methods become especially valuable.
It is hard to give full justice to all the existing techniques
in one section. So, we will try to explain only a few basic methods,
focusing mainly on those which played an important
role in recent developments in string theory.
We also illustrate these general techniques with several
concrete examples that will appear in the later sections.

\subsection{Compact Manifolds}
\label{compact}

The first examples of compact manifolds with $G_2$ and $Spin(7)$
holonomy were constructed by Joyce \cite{joyce}.
The basic idea is to start with toroidal orbifolds of the form
\begin{equation}
T^7 / \Gamma \quad \quad {\rm or} \quad \quad T^8 / \Gamma
\label{joyceorbifolds}
\end{equation}
where $\Gamma$ is a finite group,
{\it e.g.} a product of $\mathbb{Z}_2$ cyclic groups.
Notice that $T^7$ and $T^8$ themselves can be regarded as special
cases of
$G_2$ and $Spin(7)$ manifolds, respectively. This is because
their trivial holonomy group is a subgroup of $G_2$ or $Spin(7)$.
In fact,
they possess continuous families of $G_2$ and $Spin(7)$ structures.
Therefore, if $\Gamma$ preserves one of these structures
the quotient space automatically will be a space with
exceptional holonomy. However, since the holonomy of the
torus is trivial, the holonomy of the quotient is inherited from $\Gamma$
and is thus a {\it discrete}
subgroup of $G_2$ or $Spin(7)$. Joyce's idea was to take
$\Gamma$ to act with fixed points, so that ${T^n}/\Gamma$ is
a singular space, and then to repair the singularities to give
a smooth manifold with continuous holonomy $G_2$ or $Spin(7)$.

\bigbreak\hrule\medskip\nobreak\noindent

{\bf Example \cite{joyce}:}
Consider a torus $T^7$, parametrized by periodic variables
$x_i \sim x_i + 1$, $i=1, \ldots, 7$. As we pointed out,
it admits many $G_2$ structures. Let us choose one of them:
$$
\Phi =
e^1 \wedge e^2 \wedge e^3
+ e^1 \wedge e^4 \wedge e^5
+ e^1 \wedge e^6 \wedge e^7
+ e^2 \wedge e^4 \wedge e^6
- e^2 \wedge e^5 \wedge e^7
- e^3 \wedge e^4 \wedge e^7
- e^3 \wedge e^5 \wedge e^6
$$
where $e^j = dx_j$. Furthermore, let us take
\be
\Gamma = \mathbb{Z}_2 \times  \mathbb{Z}_2 \times  \mathbb{Z}_2
\label{joyceg}
\ee
generated by three involutions
\begin{eqnarray}
\begin{array}{l}
\alpha \quad : \quad (x_1, \ldots, x_7) \mapsto
(x_1, x_2, x_3, -x_4, -x_5, -x_6, -x_7) \nonumber \\
\beta \quad : \quad (x_1, \ldots, x_7) \mapsto
(x_1, - x_2, - x_3, x_4, x_5, {1 \over 2} -x_6, -x_7) \nonumber \\
\gamma \quad : \quad (x_1, \ldots, x_7) \mapsto
(- x_1, x_2, - x_3, x_4, {1 \over 2} -x_5, x_6, {1 \over 2} - x_7)
\end{array}
\end{eqnarray}
It is easy to check that these generators indeed satisfy
$\alpha^2 = \beta^2 = \gamma^2 = 1$
and that the group $\Gamma = \langle \alpha, \beta, \gamma \rangle$
preserves the associative three-form $\Phi$ given above.
It follows that the quotient space $X = T^7 / \Gamma$
is a manifold with $G_2$ holonomy. More precisely,
it is an orbifold since the group $\Gamma$ has fixed points
of the form $T^3 \times \mathbb{C}^2 / \mathbb{Z}_2$.
The existence of orbifold fixed points
is a general feature of the Joyce construction.

\par\nobreak\medskip\nobreak\hrule\bigbreak

\begin{figure}
\begin{center}
\epsfxsize=4in\leavevmode\epsfbox{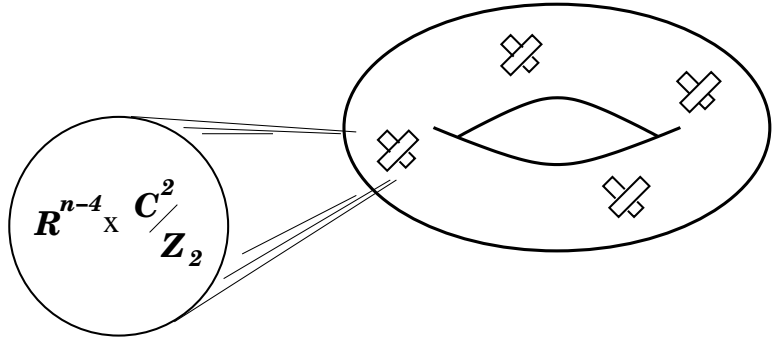}
\end{center}
\caption{A cartoon representing Joyce orbifold $T^n/\Gamma$ with
$\mathbb{C}^2 / \mathbb{Z}_2$ orbifold points.}
\label{joycefig}
\end{figure}

In order to find a nice manifold $X$ with $G_2$ or $Spin(7)$
holonomy one has to repair these singularities.
In practice, this means removing the local neighbourhood of
each singular point and replacing it with a smooth geometry,
in a way which enhances the holonomy group from a discrete group to
an exceptional holonomy group.
This may be difficult (or even impossible) for generic orbifold
singularities. However, if we have orbifold singularities
that can also appear as degenerations of Calabi-Yau manifolds,
then things simplify dramatically.

Suppose we have a $\mathbb{Z}_2$ orbifold, as in the previous example:
\be
\mathbb{R}^{n-4} \times \mathbb{C}^2 / \mathbb{Z}_2
\label{ztwoorb}
\ee
where $\mathbb{Z}_2$ acts only on the $\mathbb{C}^2$ factor
(by reflecting all the coordinates).
This type of orbifold singularity can be obtained as
a singular limit of the $A_1$ A(symptotically) L(ocally) E(uclidean) space,
as we will see in detail in section 5:
$$
\mathbb{R}^{n-4} \times {\rm ALE}_{A_1} ~~~\to~~~
\mathbb{R}^{n-4} \times \mathbb{C}^2 / \mathbb{Z}_2
$$
The ALE space has holonomy $SU(2)$, whereas its singular orbifold
limit, being locally flat, has holonomy $\mathbb{Z}_2 \subset SU(2)$.
So we see
that repairing the orbifold singularity enhances the holonomy from
$\mathbb{Z}_2$ to $SU(2)$. An important point is that the ALE space
is a non-compact Calabi-Yau 2-fold and we can use the
tools of algebraic geometry to study its deformations.
This is an important point;
we used it implicitly to resolve the orbifold singularity.
Moreover, Joyce proved that under certain conditions,
resolving orbifold singularities in this way can be used to produce
many manifolds of exceptional holonomy (\ref{joyceorbifolds}).
Therefore, by the end of the day, when all singularities
are removed, we can obtain a smooth, compact manifold $X$
with $G_2$ or $Spin(7)$ holonomy.

\bigbreak\hrule\medskip\nobreak\noindent

{\bf Example:}
In the previous example, one finds a smooth manifold $X$
with $G_2$ holonomy and Betti numbers \cite{joyce}:
\be
b^2 (X) = 12, \quad b^3 (X) = 43
\label{joycebetti}
\ee
These come from the $\Gamma$-invariant forms on $T^7$ and
also from the resolution of the fixed points on $T^7 / \Gamma$.
First, let us consider the invariant forms.
It is easy to check that the there are no 1-forms and
2-forms on $T^7$ invariant under (\ref{joyceg}),
and the only $\Gamma$-invariant 3-forms are
the ones that appear in the associative 3-form $\Phi$,
\be
\Lambda^3_{\Gamma} (T^7) =
\langle  e^{123}, e^{145}, e^{167},
e^{246}, e^{257}, e^{347}, e^{356} \rangle
\label{ginvforms}
\ee
where $e^{ijk} = e^i \wedge e^j \wedge e^k$.
Therefore, we find $b^1_{\Gamma}(T^7) = b^2_{\Gamma}(T^7) =0$
and $b^3_{\Gamma}(T^7) = 7$.

Now, let us consider the contribution of the fixed points to $b^i (X)$.
We leave it to the reader to verify that the fixed point set
of $\Gamma$ consists of 12 disjoint 3-tori,
so that the singularity near each $T^3$
is of the orbifold type (\ref{ztwoorb}).
Each singularity, modelled on
$T^3 \times \mathbb{C}^2 / \mathbb{Z}_2$,
can be resolved into a smooth space $T^3 \times {\rm ALE}_{A_1}$.
As will be explained in more detail in section 5,
via the resolution the second Betti number of the orbifold
space $\mathbb{C}^2 / \mathbb{Z}_2$ is increased by 1.
Therefore, by Kunneth formula, we find that
$b^4 (T^3 \times \mathbb{C}^2 / \mathbb{Z}_2)$
is increased by $1 \cdot b^2 (T^3) =3$,
while $b^5 (T^3 \times \mathbb{C}^2 / \mathbb{Z}_2)$
jumps by $1 \cdot b^3 (T^3) =1$.
Using Poincar\'e duality and
adding the contribution of all the fixed points
and the $\Gamma$-invariant forms (\ref{ginvforms}) together,
we obtain the final resuls (\ref{joycebetti})
\begin{eqnarray}
\begin{array}{l}
b^2 (X) = b^2_{\Gamma}(T^7) + 12 \cdot 1 = 12 \nonumber \\
b^3 (X) = b^3_{\Gamma}(T^7) + 12 \cdot 3 = 43
\end{array}
\end{eqnarray}

\par\nobreak\medskip\nobreak\hrule\bigbreak

There are many other examples of the above construction,
which are modelled not only on singularities of Calabi-Yau two-folds,
but also on orbifold singularities
of Calabi-Yau three-folds \cite{joyce}.
More examples can be found by replacing the tori
in (\ref{joyceorbifolds}) by products of the $K3$ 4-manifold\footnote{
$K3$ is the only compact 4-manifold admitting metrics with $SU(2)$ holonomy.}
or Calabi-Yau three-folds with lower-dimensional tori.
In such models, finite groups typically act as involutions
on $K3$ or Calabi-Yau manifolds, to produce fixed points
of a familiar kind. Again, repairing the singularities using algebraic
geometry techniques
one can obtain compact, smooth manifolds with exceptional holonomy.

It may look a little disturbing that in Joyce's construction
one always finds a compact manifold $X$ with exceptional holonomy
near a singular (orbifold) limit.
However, from the physics point of view,
this is not a problem at all since interesting
phenomena usually occur when $X$ develops a singularity.
Indeed, as will be explained in more detail in section 4,
compactification on a smooth manifold $X$
whose dimensions are very large (compared to the Planck scale)
leads to a very simple effective field theory;
it is abelian gauge theory with some number
of scalar fields coupled to gravity.
To find more interesting physics, such as non-abelian gauge
symmetry or chiral matter, one needs singularities.

Moreover, there is a close relationship between various types
of singularities and the effective physics they produce.
A simple, but very important aspect of this relation is
that a codimension $d$ singularity of $X$ can typically be associated
with the physics of a $D \ge 11-d$ dimensional field theory.
For example, there is no way one can obtain four-dimensional
chiral matter or parity symmetry breaking in $D=3+1$ dimensions from
a codimension four $\mathbb{C}^2 / \mathbb{Z}_2$ singularity in $X$.
As we will see, such singularities are associated with $D=7$ field theories.

Therefore, in order to reproduce properties specific to
field theories in dimension four or three from compactification
on $X$ one has to use the geometry of $X$ `to the fullest'
and consider singularities of maximal codimension.
This motivates us to study {\it isolated singular points}
in $G_2$ and $Spin(7)$ manifolds.

Unfortunately, even though Joyce manifolds naturally admit
orbifold singularities, none of them contains isolated
$G_2$ or $Spin(7)$ singularities close to the orbifold point
in the space of metrics.
Indeed, as we explained earlier, it is crucial that orbifold
singularities are modelled on Calabi-Yau singularities,
so that we can treat them using the familiar methods.
Therefore, at best, such singularities can give us the same
physics as one finds in the corresponding Calabi-Yau manifolds.

Apart from a large class of Joyce manifolds, very few explicit
constructions of compact manifolds with exceptional holonomy are known.
One nice approach was recently provided by A.~Kovalev \cite{Kovalev},
where a smooth, compact 7-manifold $X$ with $G_2$ holonomy
is obtained by gluing `back-to-back'
two asymptotically cylindrical Calabi-Yau
manifolds $W_1$ and $W_2$,
$$
X \cong (W_1 \times S^1) \cup (W_2 \times S^1)
$$
This construction is very elegant, but like Joyce's construction
produces smooth $G_2$-manifolds.
In particular, it would be interesting to study
deformations of these spaces and to see if they can
develop the kinds of isolated singularities of interest to physics.
This leaves us with the following
\begin{center}
{\large \bf Open Problem:} Construct compact $G_2$ and $Spin(7)$ manifolds\\
with various types of isolated singularities
\end{center}


\subsection{Non-compact Manifolds}
\label{noncompact}

\begin{figure}
\begin{center}
\epsfxsize=3in\leavevmode\epsfbox{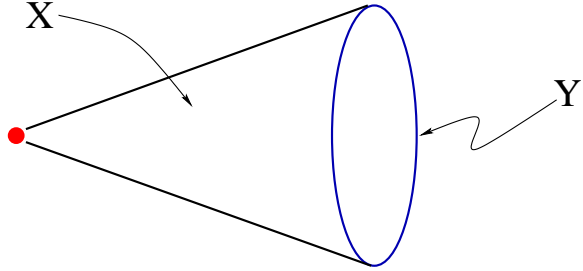}
\end{center}
\caption{A cone over a compact space $Y$.}
\label{cone}
\end{figure}

As we will demonstrate in detail, interesting physics occurs
at the singular points of the special holonomy manifold $X$.
Depending on the singularity, one may find, for example,
extra gauge symmetry or charged massless
states localized at the singularity.
Even though the physics depends strongly upon the
details of the singularity itself,
this physics typically
depends only on properties of $X$ in the {\it neighbourhood} of
the singularity.
Therefore, in order to study the physics associated with
a given singularity, one can imagine isolating the local
neighbourhood of the singular point and studying it separately.
In practice this means replacing the compact $X$ with a non-compact
manifold with singularity and
gives us the so-called `local model' of the singular point.
This procedure is somewhat analagous to considering one factor
in the standard model gauge group, rather than studying
the whole theory at once.
In this sense, non-compact manifolds provide us with the basic
building blocks for the low-energy energy physics
that may appear in vacua constructed from compact manifolds.

Here we discuss a particular class of isolated
singularities, namely {\it conical singularies}.
They correspond to degenerations of the metric
on the space $X$ of the form:
\begin{equation}
ds^2 (X) = dt^2 + t^2 ds^2 (Y),
\label{conicalmet}
\end{equation}
where a compact space $Y$ is the base of the cone;
the dimension of $Y$ is one less than the dimension of $X$.
$X$ has an isolated singular point
at the tip of the cone $t=0$, except for the special case
when $Y$ is a {\it round} sphere, ${\bf S}^{n-1}$, in which
case (\ref{conicalmet}) is just Euclidean space.

The conical singularities of the form (\ref{conicalmet})
are among the simplest isolated singularities
one could study, see Figure \ref{cone}.
In fact, the first examples of non-compact manifolds
with $G_2$ and $Spin(7)$ holonomy,
obtained by Bryant and Salamon \cite{BS} and rederived
by Gibbons, Page, and Pope \cite{GPP},
exhibit precisely this type of degeneration.
Specifically, the complete metrics constructed
in \cite{BS,GPP} are smooth everywhere,
and asymptotically look like (\ref{conicalmet}),
for various base manifolds $Y$. Therefore, they can be
considered as {\it smoothings} of conical singularities.
In Table 3 we list the currently known asymptotically conical (AC)
complete metrics with $G_2$ and $Spin(7)$ holonomy that
were originally found in \cite{BS,GPP}
and more recently \cite{garycohom, GTS}.

\begin{table}\begin{center}
\begin{tabular}{|c|c|c|}
\hline
\rule{0pt}{5mm}
Holonomy & Topology of $X$ & Base $Y$ \\[3pt]
\hline
\hline
\rule{0pt}{5mm}
& ${\bf S}^4 \times \mathbb{R}^3$
& $\mathbb{C}{\bf P}^3$ \\[3pt]
\cline{2-3}
\rule{0pt}{5mm}
$G_2$
& $\mathbb{C}{\bf P}^2 \times \mathbb{R}^3$
& $SU(3)/U(1)^2$ \\[3pt]
\cline{2-3}
\rule{0pt}{5mm}
& ${\bf S}^3 \times \mathbb{R}^4$
& $SU(2) \times SU(2)$ \\[3pt]
\cline{2-2}
\rule{0pt}{5mm}
& $T^{1,1} \times \mathbb{R}^2$
&  \\[3pt]
\hline
\hline
\rule{0pt}{5mm}
& ${\bf S}^4 \times \mathbb{R}^4$
& $SO(5)/SO(3)$ \\[3pt]
\cline{2-3}
\rule{0pt}{5mm}
$Spin(7)$
& $\mathbb{C}{\bf P}^2 \times \mathbb{R}^4$
& $SU(3)/U(1)$ \\[3pt]
\cline{2-2}
\rule{0pt}{5mm}
& ${\bf S}^5 \times \mathbb{R}^3$
&  \\[3pt]
\hline
\end{tabular}\end{center}
\caption{Asymptotically conical manifolds with $G_2$
and $Spin(7)$ holonomy.}
\end{table}

The method of constructing $G_2$ and $Spin(7)$ metrics
originally used in \cite{BS,GPP} was essentially based
on the direct analysis of the conditions for special holonomy $(2.10)$ or
the
Ricci-flatness equations,
\begin{equation}
\label{Ricciflat}
R_{ij} =0,
\end{equation}
for a particular metric ansatz. We will not go into details
of this approach here since it relies on finding the right
form of the ansatz and, therefore, is not practical for
generalizations.
Instead, following \cite{GYZ,gary6},
we will describe a very powerful approach,
recently developed by Hitchin \cite{Hitchin},
which allows one to construct all the $G_2$ and $Spin(7)$
manifolds listed in Table 3 (and many more !)
in a systematic manner.
Another advantage of this method is that it leads to
first-order differential equations, which are much easier
than the second-order Einstein equations (\ref{Ricciflat}).

\begin{figure}
\begin{center}
\epsfxsize=4in\leavevmode\epsfbox{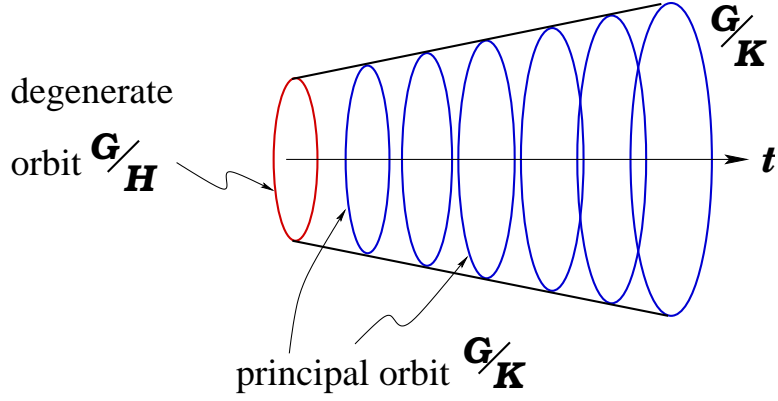}
\end{center}
\caption{A non-compact space $X$ can be viewed as a foliation
by principal orbits $Y = G/K$. The non-trivial cycle in $X$
correspond to the degenerate orbit $G/H$,
where $G \supset H \supset K.$}
\label{orbits}
\end{figure}

Before we explain the basic idea of Hitchin's construction,
notice that for all of the AC manifolds in Table 3
the base manifold $Y$ is a homogeneous quotient space
\begin{equation}
Y = G / K,
\label{gmodk}
\end{equation}
where $G$ is some group and $K \subset G$ is a subgroup.
Therefore, we can think of $X$ as being foliated by
{\it principal orbits} $G/K$ over a positive real line,
$\mathbb{R}_+$, as shown on Figure \ref{orbits}.
A real variable $t \in \mathbb{R}_+$ in this picture plays
the role of the radial coordinate;
the best way to see this is from the singular limit,
in which the metric on $X$ becomes exactly conical,
{\it cf.} eq. (\ref{conicalmet}).

As we move along $\mathbb{R}_+$, the size and the shape
of the principal orbit changes, but only in a way
consistent with the symmetries of the coset space $G/K$.
In particular, at some point the principal orbit $G/K$
may collapse into a degenerate orbit,
\begin{equation}
B = G/H
\label{boltgh}
\end{equation}
where symmetry requires
\begin{equation}
G \supset H \supset K
\label{ghkrel}
\end{equation}
At this point (which we denote $t=t_0$) the ``radial
evolution'' stops, resulting in a non-compact space $X$ with
a topologically non-trivial cycle $B$, sometimes called a {\it bolt}.
In other words, the space $X$ is contractible to a compact set $B$,
and from the relation (\ref{ghkrel}) we can easily deduce
that the normal space of $B$ inside $X$ is itself a cone on $H/K$.
Therefore, in general, the space $X$ obtained in this way
is a singular space, with a conical singularity along
the degenerate orbit $B=G/H$.
However, if $H/K$ is a round sphere, then the space $X$ is smooth,
$$
H/K = {\bf S}^k \quad \Longrightarrow \quad X ~~{\rm smooth}
$$
This simply follows from the fact that the normal space of
$B$ inside $X$ in such a case is non-singular,
$\mathbb{R}^{k+1}$ (= a cone over $H/K$).
It is a good exercise to check that for
all manifolds listed in Table 3, one indeed has $H/K = {\bf S}^k$,
for some value of $k$. To show this, one should first write down
the groups $G$, $H$, and $K$, and then find $H/K$.

The representation of a non-compact space $X$ in terms of
principal orbits, which are homogeneous coset spaces is
very useful. In fact, as we just explained, topology
of $X$ simply follows from the group data (\ref{ghkrel}).
For example, if $H/K = {\bf S}^k$ so that $X$ is smooth,
we have
\begin{equation}
X \cong (G/H) \times \mathbb{R}^{k+1}
\label{xtopology}
\end{equation}
However, this structure can be also used to find a $G$-nvariant
metric on $X$. In order to do this, all
we need to know are the groups $G$ and $K$.

First, let us sketch the basic idea of Hitchin's
construction \cite{Hitchin},
and then explain the details in some specific examples.
For more details and further applications
we refer the reader to \cite{GYZ,gary6}.
We start with a principal orbit $Y=G/K$ which can be,
for instance, the base of the conical manifold that
we want to construct. Let ${\cal P}$ be the space of
(stable\footnote{Stable forms are defined as follows \cite{Hit}.
Let $X$ be a manifold of real dimension $n$, and $V = TX$.
Then, the form $\rho \in \Lambda^p V^*$ is stable if it lies
in an open orbit of the (natural) $GL(V)$ action on $\Lambda^p V^*$.
In other words, this means that all forms in the neighborhood
of $\rho$ are $GL(V)$-equivalent to $\rho$.
This definition is useful because it allows
one to define a volume.
For example, a symplectic form $\omega$ is stable if and only if
$\omega^{n/2} \ne 0$.})
$G$-invariant differential forms on $Y$.
This space is finite dimensional and, moreover,
it turns out that there exists a symplectic structure on ${\cal P}$.
This important result allows us to think of the space ${\cal P}$
as the phase space of some dynamical system:
\begin{eqnarray}
{\cal P} = ~{\rm Phase~Space} \nonumber \\
\omega = \sum dx_i \wedge dp_i
\end{eqnarray}
where we parametrized ${\cal P}$ by
some coordinate variables $x_i$
and the conjugate momentum variables $p_i$.

Given a principal orbit $G/K$ and a space of $G$-invariant
forms on it, there is a canonical construction of
a Hamiltonian $H (x_i, p_i)$ for our dynamical system,
such that the Hamiltonian flow equations are equivalent
to the special holonomy condition \cite{Hitchin}:
\begin{equation}
\left\{
\begin{array}{rcl}
{dx_i \over dt} & = & {\partial H \over \partial p_i} \\
{dp_i \over dt} & = & - {\partial H \over \partial x_i} \\
\end{array}
\right.
\iff
\begin{array}{c}
{\rm ~Special~Holonomy~Metric} \\
{\rm ~on~~} (t_1, t_2) \times (G/K) \\
\end{array}
\label{Hamiltonflow}
\end{equation}
where the `time' in the Hamiltonian system is identified
with the radial variable $t$.
Thus, solving the Hamiltonian flow equations from $t=t_1$
to $t=t_2$ with a particular boundary condition leads to
the special holonomy metric on $(t_1, t_2) \times (G/K)$.
Typically, one can extend the boundaries of the interval
$(t_1, t_2)$ where the solution is defined to infinity on one side,
and to a point $t=t_0$, where the principal orbit degenerates,
on the other side. Then, this gives a complete metric with special
holonomy on a non-compact manifold $X$ of the form (\ref{xtopology}).
Let us now illustrate these general ideas in more detail
in a concrete example.

\bigbreak\hrule\medskip\nobreak\noindent

{\bf Example:}
Let us take $G=SU(2)^3$ and $K=SU(2)$, the diagonal subgroup of $G$.
We can form the following natural sequence of subgroups:
\begin{equation}
\begin{array}{ccccc}
G && H && K \\
\| && \| && \| \\
SU(2)^3 & \supset & SU(2)^2 & \supset & SU(2)
\end{array}
\label{ghkexamp}
\end{equation}
{}From the general formula (\ref{gmodk}) it follows that
in this example we deal with a space $X$, whose principal
orbits are
\be
Y  = SU(2) \times SU(2) \cong {\bf S}^3 \times {\bf S}^3
\ee
Furthermore, $G/H \cong H/K \cong {\bf S}^3$ implies that
$X$ is a smooth manifold with topology, {\it cf.} (\ref{xtopology}),
$$
X \cong {\bf S}^3 \times \mathbb{R}^4
$$
In fact, $X$ is one of the asymptotically conical manifolds
listed in Table 3.

In order to find a $G_2$ metric on this manifold, we need
to construct the ``phase space'', ${\cal P}$, that is
the space of $SU(2)^3$--invariant 3-forms and 4-forms on $Y = G/K$:
$$
{\cal P} = \Omega_G^3 (G/K) \times \Omega_G^4 (G/K)
$$
In this example, it turns out that each of the factors
is one-dimensional, generated by a 3-form $\rho$ and
by a 4-form $\sigma$, respectively,
\be
\label{formrho}
\rho= \sigma_1 \sigma_2 \sigma_3 - \Sigma_1 \Sigma_2 \Sigma_3
+ x \Big( d(\sigma_1 \Sigma_1)
+ d(\sigma_2 \Sigma_2)
+ d(\sigma_3 \Sigma_3) \Big),
\ee
\be
\label{formsig}
\sigma = p^{2/5} \Big( \sigma_2 \Sigma_2 \sigma_3 \Sigma_3
+ \sigma_3 \Sigma_3 \sigma_1 \Sigma_1
+ \sigma_1 \Sigma_1 \sigma_2 \Sigma_2 \Big).
\ee
where we introduced
two sets of left invariant 1-forms $(\sigma_a, \Sigma_a)$ on $Y$
\ba
& \sigma_1 = \cos \psi d \theta + \sin \psi \sin \theta d \phi,\quad~~
& \Sigma_1 = \cos \tilde \psi d \tilde \theta
+ \sin \tilde \psi \sin \tilde \theta d \tilde \phi, \nonumber\\
& \sigma_2 = - \sin \psi d \theta + \cos \psi \sin \theta d \phi, \quad
& \Sigma_2 = - \sin \tilde \psi d \tilde \theta
+ \cos \tilde \psi \sin \tilde \theta d \tilde \phi, \nonumber\\
& \sigma_3 = d \psi + \cos \theta d \phi, \quad\quad\quad\quad\quad\quad
& \Sigma_3 = d \tilde \psi + \cos \tilde \theta d \tilde \phi
\label{lsigmas}
\ea
written explicitly in terms of the Euler angles
$\psi,\tilde \psi, \theta, \tilde \theta \in [0,\pi]$
and $\phi, \tilde \phi \in [0,2\pi]$.
The invariant 1-forms $\sigma_a$ and $\Sigma_a$
satisfy the usual $SU(2)$ algebra
\be
\label{sualg}
d \sigma_a=-{1 \over 2} \epsilon_{abc} \sigma_a \wedge \sigma_b,
~~~~~
d \Sigma_a=-{1 \over 2} \epsilon_{abc} \Sigma_a \wedge \Sigma_b.
\ee

Therefore, we have only one ``coordinate'' $x$
and its conjugate ``momentum'' $p$,
parametrizing the ``phase space'' ${\cal P}$ of our model.
In order to see that there is a natural symplectic structure
on ${\cal P}$, note that $x$ and $p$ multiply exact forms.
For $x$ this obviously follows from (\ref{formrho}) and for $p$
this can be easily checked using (\ref{formsig}) and (\ref{sualg}).
This observation can be used to define a non-degenerate symplectic
structure on ${\cal P} =  \Omega^3_{exact} (Y) \times \Omega^4_{exact} (Y)$.
Explicitly, it can be written as
$$
\omega \left( (\rho_1, \sigma_1) , (\rho_2, \sigma_2) \right)
= \langle \rho_1, \sigma_2 \rangle - \langle \rho_2, \sigma_1 \rangle,
$$
where, in general, for
$\rho = d \beta \in \Omega^k_{exact} (Y)$
and $\sigma = d \gamma \in \Omega^{n-k}_{exact} (Y)$
one has a nondegenerate pairing
\begin{equation}
\langle \rho, \sigma \rangle = \int_Y d \beta \wedge \gamma
= (-1)^k \int_Y \beta \wedge d \gamma.
\end{equation}

Once we have the phase space ${\cal P}$, it remains to
write down the Hamiltonian flow equations (\ref{Hamiltonflow}).
In Hitchin's construction, the Hamiltonian $H(x,p)$ is
defined as an invariant functional on the space of
differential forms, ${\cal P}$.
Specifically, in the context of $G_2$ manifolds it is given by
\be
\label{hamilt}
H=2 V(\sigma)-V(\rho),
\ee
where $V(\rho)$ and $V(\sigma)$ are
suitably defined volume functionals \cite{Hitchin}.
Even though the explicit form of the volume functionals
$V(\sigma)$ and $V(\rho)$ is somewhat technical\footnote{
The volume $V(\sigma)$ for a 4-form
$\sigma \in \Lambda^4 TY^* \cong \Lambda^2 TY \otimes \Lambda^6 TY^*$
is very easy to define. Indeed, we have
$\sigma^3 \in \Lambda^6 TY \otimes (\Lambda^6 TY^*)^3
\cong (\Lambda^6 TY^*)^2$, and therefore we can take
\begin{equation}
V(\sigma)=\int_Y |\sigma^3|^{\frac{1}{2}}.
\label{defvos}
\end{equation}
to be the volume for $\sigma$.
In order to define the volume $V(\rho)$ for a 3-form
$\rho \in \Lambda^3 TY^*$, one first defines a map
$K_{\rho} \colon TY \to TY \otimes \Lambda^6 TY^*,$
such that for a vector $v \in TY$ it gives
$K (v) = \imath (v) \rho \wedge \rho \in \Lambda^5 TY^*
\cong TY \otimes \Lambda^6 TY^*.$
Hence, one can define $tr (K^2) \in (\Lambda^6 TY^*)^2.$
Since stable forms with stabilizer $SL(3,\mathbb{C})$
are characterised by $tr (K)^2 < 0$, following \cite{Hit},
we define
\begin{equation}
V(\rho)=\int_Y |\sqrt{- tr K^2}|.
\label{defvor}
\end{equation}
}, they can be systematically computed
for any given choice of the forms $\sigma$ and $\rho$.
Thus, evaluating (\ref{hamilt}) for the $G$-invariant
forms (\ref{formrho}) and (\ref{formsig})
we obtain the Hamiltonian flow equations:
$$
\left\{
\begin{array}{rcl}
\dot p & = & x(x-1)^2 \\
\dot x & = & p^2 \\
\end{array}
\right.
$$
These first-order equations can be easily solved,
and the solution for $x(t)$ and $p(t)$ determines
the evolution of the forms $\rho$ and $\sigma$, respectively.
On the other hand, these forms define the associative three-form
on the 7-manifold $Y \times (t_1 , t_2)$,
\begin{equation}
\Phi=dt \wedge \omega+\rho,
\label{formgtw}
\end{equation}
where $\omega$ is a 2-form on $Y$,
such that $\sigma = \omega^2 /2$.
For $x(t)$ and $p(t)$ satisfying the Hamiltonian flow equations
the associative form $\Phi$ is automatically closed and co-closed,
{\it cf.} (\ref{harm}).
Therefore, as we explained in section \ref{basics},
it defines a $G_2$ holonomy metric.
Specifically, one can use (\ref{gfromphi}) to find the explicit
form of the metric, which after a simple change of variables
becomes the $G_2$ metric on the spin bundle over ${\bf S}^3$,
originally found in \cite{BS,GPP}:
\begin{equation}
ds^2 = {dr^2 \over 1 - r_0^2/r^2}
+ {r^2 \over 12} \sum_{a=1}^3 (\sigma_a - \Sigma_a)^2
+ {r^2 \over 36} \left( 1 - {r_0^2 \over r^2} \right)
\sum_{a=1}^3 (\sigma_a + \Sigma_a)^2
\label{bsgppmet}
\end{equation}
Here, $r > r_0$ and $r_0$ determines the size
of the ${\bf S}^3$ generated by $(\sigma_a - \Sigma_a)$.
It is easy to check that this ${\bf S}^3$
is an associative submanifold
(with respect to the 3-form (\ref{formgtw})),
and the total space is topologically ${\bf S}^3 \times \mathbb{R}^4$.

\par\nobreak\medskip\nobreak\hrule\bigbreak

The above example can be easily generalised in a number of directions.
For example, if instead of (\ref{ghkexamp}) we take $G=SU(2)^2$
and $K$ to be its trivial subgroup, we end up with the same topology
for $X$ and $Y$, but with a larger space of $G_2$ metrics on $X$.
Indeed, for $G=SU(2)^2$ the space of $G$-invariant forms on $Y$
is much larger. Therefore, the corresponding dynamical system
is more complicated and has a richer structure.
Some specific solutions of this more general system have been
recently constructed in \cite{BGGG, G_2conifold, Brandhuber, Cveticunif},
but the complete solution is still not known.

There is a similar systematic method, also developed
by Hitchin \cite{Hitchin}, of constructing complete
non-compact manifolds with $Spin(7)$ holonomy.
Again, this method can be used to obtain the asymptotically
conical metrics listed in Table 4, as well as other $Spin(7)$ metrics
recently found in \cite{gary7,GS,garycohom,kanno}.
Another approach to constructing special holonomy metrics
is based on finding D6-brane solutions using gauged supergravity,
following \cite{MNunez}, see {\it e.g.} \cite{JoseN,JosePR,HSfetsos,Behrndt}.
Once can also use the technique of toric geometry
to construct (singular) metrics on certain noncompact
$G_2$ manifolds \cite{Lilia}.

As a final remark we note, that this construction leads
to non-compact exceptional holonomy manifolds with
continuous isometries determined by the symmetry groups (\ref{ghkrel}).
In the effective low-energy field theory,
these isometries play the role of the global symmetries.
On the other hand, when such a manifold is realized
as a part of a compact space $X$ with $G_2$ or $Spin(7)$ holonomy,
these isometries are broken by the global geometry of $X$.
In particular, in either situation we can not use the geometric
symmetries of $X$ to get non-abelian gauge fields
in the effective field theory.

\newpage
\section{$M$ theory on Smooth Special Holonomy Manifolds}
\label{kkreduction}

At low energies,  $M$ theory is well approximated by eleven dimensional
supergravity
when spacetime
is smooth and large compared to the eleven dimensional Planck length.
So, when $X$ is smooth and large enough, we can obtain an effective
low energy description by considering a Kaluza-Klein analysis of
the fields on $X$. This analysis was first carried out in \cite{paptown}.
In this section, we review the Kaluza-Klein reduction of
the eleven-dimensional supergravity on a manifolds with $G_2$ holonomy.
The arguments for a reduction on a manifold with $Spin(7)$ holonomy
are very similar.

The supergravity theory has two bosonic fields, the metric $g$ and a
3-form field $C$ with field strength $G=dC$. The equation of motion
for $C$ is $d*G = {1 \over 2} G\wedge G$.

In compactification of eleven dimensional supergravity, massless
scalars in four dimensions can originate from either the metric or the
$C$-field. If $g(X)$ contains $k$ parameters i.e. there is a $k$-dimensional
family of $G_2$-holonomy metrics on $X$, then there will be correspondingly
$k$ massless scalars in four dimensions.

The scalars in four dimensions which originate from $C$ arise via
the Kaluza-Klein ansatz,
\be
C = {\sum} {\omega}^I (x) {\phi}_I (y) + ...
\ee
where ${\omega}^I$ form a basis for the harmonic 3-forms on $X$. These are 
zero modes
of the Laplacian on $X$ (and are closed and co-closed). There are $b_3 (X)$ 
linearly
independent such forms.   The dots refer to further terms in the Kaluza-Klein 
ansatz which
will be prescribed later.
The ${\phi}_I (y)$ are scalar fields in four dimensional Minkowski space with 
coordinates $y$.
With this ansatz, these scalars are classically massless in four dimensions. 
To see this, note
that,
\be
G = \sum {\omega}^I {\wedge} d{\phi}_I
\ee
and $d*G$ is just
\be
d*G =  \sum *_7 {\omega}^I \wedge d *_4 d {\phi}_I
\ee
Since $G{\wedge}G$ vanishes to first order in the $C$-field modes,
the $C$-field equation of motion actually aserts that the
scalar fields ${\phi}_I$ are all massless in four dimensions. Thus, the 
$C$-field gives
rise to $b_3 (X)$ real massless scalars in four dimensions.

In fact it now follows from ${\cal N}$ $=$ $1$ supersymmetry in four 
dimensions that
the Kaluza-Klein analysis of $g$ will yield an additional $b_3 (X)$ scalars in 
four dimensions.
This is because the superpartners of $C$ should come from $g$ as these fields
are superpartners in eleven dimensions. We should also add that
(up to duality tranformations) all representations of the ${\cal N}$ $=$ $1$
supersymmetry algebra which contain one massless
real scalar actually contain two scalars in total which
combine into complex scalars.
We will now describe how these scalars arise explicitly.

We began with a
$G_2$-holonomy metric $g(X)$ on $X$. $g(X)$ obeys the vacuum Einstein
equations,
\be
R_{ij} (g(X)) = 0
\ee
To obtain the spectrum of modes originating from $g$ we look for fluctuations 
in $g(X)$
which also satisfy the vacuum Einstein equations. We take the fluctuations in
$g(X)$ to depend on the four dimensional coordinates $y$ in Minkowski space. 
Writing the
fluctuating metric as
\be
g_{ij} (x) + \delta g_{ij} (x,y)
\ee
and expanding to first order in the fluctuation yields the Lichnerowicz 
equation
\be
{\Delta}_L \delta g_{ij} \equiv -{\nabla}^2_{M} \delta g_{ij} -2R_{imjn} 
\delta g^{mn} +
2R_{(i}^k \delta g_{j)k} = 0
\ee
Next we make a Kaluza-Klein ansatz for the fluctuations as
\be
\delta g_{ij} = h_{ij}(x) \rho (y)
\ee
Note that the term ${\nabla}^2$ is the square of the full d=11 covariant 
derivative. If we separate
this term into two
\be
{\nabla}^2_{M} = {\nabla}^2_{\mu} + {\nabla}^2_{i}
\ee
we see that  the fluctuations are scalar fields in four dimensions with 
squared masses
given by the eigenvalues of the Lichnerowicz operator acting on the $h_{ij}$:
\be
h_{ij} {\nabla}^2_{\mu} \rho (y) = -( {\Delta}_L h_{ij} )\rho (y) = -\lambda 
h_{ij} \rho (y)
\ee
Thus, zero modes of the Lichnerowicz operator give rise to massless scalar 
fields in four
dimensions. We will now show that we have precisely $b_3 (X)$ such zero modes.

On a 7-manifold of $SO(7)$ holonomy, the $h_{ij}$ --- being symmetric 2-index 
tensors ---
transform in the ${\bf 27}$ dimensional representation. Under $G_2$ this 
representation
remains irreducible. On the other hand, the 3-forms on a $G_2$-manifold, which 
are
usually in the ${\bf 35}$ of $SO(7)$ decompose under $G_2$ as
\be
{\bf 35} \longrightarrow {\bf 1} + {\bf 7}  + {\bf 27}
\ee

Thus, the $h_{ij}$ can also be regarded as 3-forms on $X$. Explicitly,
\be
{\Phi}_{n[pq} h^n_{r]} = {\omega}_{pqr}
\ee

The $\omega$'s are 3-forms in the same representation as $h_{ij}$ since 
${\Phi}$
is in the trivial representation. The condition that $h$ is a zero mode of 
${\Delta}_L$ is
equivalent to $\omega$ being a zero mode of the Laplacian:
\be
{\Delta_L}h = 0 \leftrightarrow \Delta \omega = 0
\ee
This shows that there are precisely $b_3 (X)$ additional massless scalar 
fields coming from the
fluctuations of the $G_2$-holonomy metric on $X$.

As we mentioned above, these scalars combine with the $\phi$'s to give $b_3 (X)
$ massless
{\it complex} scalars, ${\Phi}^I (y)$,
which are the lowest components of massless chiral superfields in four
dimensions.          There is a very natural formula for the complex scalars 
${\Phi}^I (y)$.
Introduce a basis ${\alpha}_I$ for the third homology group of $X$, $H_3 (X, 
{\bf R})$.
This is a basis for the
incontractible holes in $X$ of dimension three. We can choose the ${\alpha}_I$ 
so that
\be
\int_{\alpha_I} {\omega}^J = \delta_I^J
\ee
Since the fluctuating $G_2$-structure is
\be
\Phi ' = \Phi + \delta{\Phi} = \Phi + \sum {\rho}^I (y) {\omega}^I (x)
\ee
we learn that
\be
\Phi^I (y) = \int_{\alpha_I}  (\Phi ' + i C)
\ee

The fluctuations of the four dimensional Minkowski metric give
us the usual fluctuations of four dimensional gravity, which due to 
supersymmetry implies that
the four dimensional theory is locally supersymmetric.

In addition to the massless chiral multiplets, we also get massless vector 
multiplets. The bosonic
component of such a mulitplet is a massless abelian gauge field which arises 
from the $C$-field
through the Kaluza-Klein ansatz,
\be
C   = \sum \;\;\beta^{\alpha}(x) \wedge A_{\alpha} (y)
\ee
where the $\beta$'s are a basis for the harmonic 2-forms and the $A$'s are 
one-forms in
Minkowski space i.e. Abelian gauge fields. Again, the equations of motion for 
$C$ imply that
the $A$'s are massless in four dimensions. This gives $b_2 (X)$ such gauge 
fields. As with the
chiral multiplets above, the fermionic superpartners of the gauge fields arise 
from the gravitino
field.  Note that we could have also included an ansatz giving 2-forms in four 
dimensions by
summing over harmonic 1-forms on $X$. However, since $b_1 (X) = 0$, this does 
not produce
any new massless fields in four dimensions.

We are now in a position to summarise the basic effective theory for the 
massless fields.
The low energy effective theory is an ${\cal N}$ $=1$ supergravity theory 
coupled to
$b_2 (X)$ abelian vector multiplets and $b_3 (X)$ massless, {\it neutral} 
chiral multiplets.
This theory is relatively {\it uninteresting} physically. In particular, the 
gauge group is abelian and
there are no light charged particles.
We will thus have to work harder to obtain the basic requisites
of the standard model --- non-Abelian gauge fields and chiral
fermions --- from $G_2$-compactifications.
The basic point of the following two sections is to emphasise that these
features emerge naturally from singularities in $G_2$-manifolds.

Using similar arguments one can derive the effective low energy
description of $M$ theory on a manifold $X$ with $Spin(7)$ holonomy,
in the regime where the size of $X$ is large (compared to the Planck scale).
In this case, one also finds abelian $\mathcal{N}=1$ gauge theory
in three dimensions with some number of (neutral) matter fields,
coupled to supergravity.
Specifically, from the Kaluza-Klein reduction
one finds $b_2$ abelian vector fields, $\mathcal{A}^i$,
which come from the modes of the three-form field $C$,
and $(b_3 + b_4^- + 1)$ scalars, $\phi^a$.
Some of these scalar fields, namely $b_3$ of them,
come from the $C$-field, whereas the others correspond
to deformations of the ${Spin}(7)$ structure.

A novel feature of compactification on manifolds with
$Spin(7)$ holonomy is that, due to the membrane
anomaly \cite{vw1,DLM,witten1}, one often has to study
compactifications with non-trivial background $G$-flux.
Recall that the membrane path-integral is well-defined only
if the $G$-field satisfies the
shifted quantization condition \cite{witten1}
\be
\left[\frac{G}{2\pi}\right]-\frac{\lambda}{2}\in
H^4(X;\mathbb{Z})
\label{flux}\ee
where $\lambda(X)=p_1(X)/2 \in H^4(X;\mathbb{Z})$ is an integral class
for a spin manifold $X$. If $\lambda$ is even, one may consistently
set $G=0$. However, if  $\lambda$ is not divisible by two as an
element of $H^4(X;\mathbb{Z})$, one must turn on a half-integral
$G$-flux in order to have a consistent vacuum.

The background $G$-flux generates
Chern-Simons terms and a superpotential ${\cal W} (\phi)$
in the effective field theory \cite{GVW,sergei2,GS,adelag,BCkk,BCGates}:
\be
k_{ij} = {1 \over 2\pi} \int_X \beta_i \wedge \beta_j \wedge G
\label{fluxcs}
\ee
\be
{\cal W} = {1 \over 2\pi} \int_X \Omega \wedge G
\label{fluxw}
\ee
Taking these terms into account,
we may write the complete supersymmetric action at a generic
point in the moduli space of $X$:
\begin{eqnarray}
S_{3d} &= \int d^3 x \Big[
\sum_{k} {1 \over 4 g_k^2} \big( \mathcal{F}_{\mu \nu}^k \mathcal{F}^{k~ \mu 
\nu}
+ \bar \psi^k i \Gamma \cdot D \psi^k \big)
+ {1 \over 2} \sum_{a,b} g_{ab} \big(
\partial_{\mu} \phi^a \partial^{\mu} \phi^b
+ \bar \chi^a i \Gamma \cdot D \chi^b \big) - \nonumber \\
& - {1 \over 2} \sum_{a,b} g^{ab} \big(
{\partial {\cal W} (\phi) \over \partial \phi^a}
{\partial {\cal W} (\phi) \over \partial \phi^b}
- {1 \over 2}{\partial^2 {\cal W} (\phi) \over \partial \phi^a \partial 
\phi^b}
\bar \chi^a \chi^b \big) \Big]
- \sum_{i,j} {i k_{ij} \over 4 \pi} \int \big( {\cal A}^i \wedge d {\cal A}^j
+  \bar \psi^i \psi^j \big)~~~~~~~~~~~
\label{effaction}
\end{eqnarray}
Here, $\psi^i$ are the gaugino fields, $\chi^a$ represent the fermionic
superpartners of the scalar fields $\phi^a$, $g^i$ are the gauge couplings,
and $g^{ab}$ denotes the scalar field metric.
In this Lagrangian we suppressed the terms corresponding
to interactions with supergravity.

Before we conclude this section, let us remark that
similar techniques can be applied to non-compact
manifolds $X$ with $G_2$ or $Spin(7)$ holonomy.
In such cases, instead of the Betti numbers $b_k$
one should use the dimension of the space
of $L^2$-normalisable $k$-forms on $X$.

\newpage
\section{$M$ theory Dynamics on Singular Special Holonomy Manifolds}
\label{singular}

Since compactification on smooth manifolds does not produce
interesting physics --- in particular, does not lead to realistic
quantum field theories --- one has to study dynamics of string
theory and $M$ theory on {\it singular} $G_2$ manifolds.
This is a very interesting problem which can provide us with
many insights about the infra-red behaviour of minimally supersymmetric
gauge theories and even about $M$ theory itself.
The new physics one might find
at the singularities of $G_2$ and $Spin(7)$ manifolds,
could be,

\begin{itemize}

\item New light degrees of freedom

\item Extra gauge symmetry

\item Restoration of continuous/discrete symmetry

\item Topology changing transitions

$\vdots$

\end{itemize}

Before one talks about the physics associated with $G_2$ and $Spin(7)$
singularities, it would be nice to have a classification
of all such degenerations. Unfortunately,
this problem is not completely solved even for Calabi-Yau
manifolds (apart from complex dimension two ), and seems
even less promising for real manifolds with exceptional holonomy.
Therefore, one starts with some simple examples.

One simple kind of singularity --- which we already encountered
in section \ref{compact}
in the Joyce construction of compact manifolds with
exceptional holonomy --- is an orbifold
singularity\footnote{The classification of local $G_2$-orbifold
singularities is reviewed in \cite{YHHe}.}.
Locally, an orbifold singularity can be represented
as a quotient of $\mathbb{R}^n$
by some discrete group $\Gamma$,
\begin{equation}
\mathbb{R}^n / \Gamma
\end{equation}

In perturbative string theory, the physics associated with such
singularities can be systematically extracted from
the orbifold conformal field theory \cite{DHVW}.
(See \cite{Shatashvili,bsa1,Eguchi1,Sugiyama,BraunG_2,Eguchi2,Roiban}
for previous work on conformal field theories associated
with $G_2$ manifolds, and \cite{Shatashvili,bsa1,BraunSpin7}
for CFT's associated with $Spin(7)$ manifolds.)
Typically, one finds new massless degrees of freedom localized
at the orbifold singularity and other phenomena listed above.
However, the CFT technique is not applicable for studying
$M$ theory on singular $G_2$ and $Spin(7)$ manifolds.
Moreover, as we explained earlier, many interesting
phenomena occur at singularities which are not of the orbifold
type, and to study the physics of those we need some new methods.
In the rest of this section, we describe two particularly
useful methods of analyzing $M$ theory dynamics on singularities
of special holonomy manifolds, which are based on the duality
with the heterotic and type IIA string, respectively.


\subsection{Low Energy Dynamics via Duality with the Heterotic String}
\label{heterotic}

We start with a duality between $M$ theory and the heterotic string theory,
which, among other things, will help us to understand the origin of
non-Abelian degrees of freedom arising from certain orbifold singularities.
We have known for some time now that non-Abelian gauge groups
emerge from $M$ theory when space has a so-called ${\sf ADE}$-singularity.
We learned this in the context of the duality between M theory on $K3$
and the heterotic string on a flat three torus, ${T^3}$ \cite{WittenM}.
So, our strategy for obtaining non-Abelian gauge symmetry from
$G_2$ or $Spin(7)$ compactifications
will be to embed ${\sf ADE}$-singularities into
special holonomy manifolds.
After reviewing the basic features of the duality between $M$ theory on $K3$ 
and
heterotic string theory on ${T^3}$, we describe ${\sf ADE}$-singularities 
explicitly.
We then develop a picture of a $G_2$-manifold near an embedded ${\sf 
ADE}$-singularity.
Based on this picture we analyse what kinds of {\it four} dimensional gauge 
theories
these singularities give rise to.
We then go on to describe local models for such singular
$G_2$-manifolds as finite quotients of smooth ones.

\subsection*{$M$ theory - Heterotic Duality in Seven Dimensions}

$M$ theory compactified on a $K3$ manifold is strongly believed to be 
equivalent to
the heterotic string theory compactified on a 3-torus ${T^3}$. As with 
$G_2$ compactification,
both of these are compactifications to flat Minkowski space.
Up to diffeomorphisms,
$K3$ is the only simply connected, compact 4-manifold admitting metrics of $SU(
2)$-holonomy.
$SU(2)$ is the analog in four dimensions of $G_2$ in seven dimensions. 
Interestingly enough
in this case $K3$ is the only simply connected example, whereas there are many 
$G_2$-manifolds.

There is a 58-dimensional moduli space of $SU(2)$-holonomy metrics on 
$K3$.This space ${\cal M}(K3)$ is locally a coset space:
\be
{\cal M}(K3)  = {\bf R^+} {\times} {SO(3,19) \over SO(3) {\times} SO(19)}
\ee

An $SU(2)$ holonomy metric admits two parallel spinors, which when tensored 
with
the ${\bf 8}$ constant spinors of 7-dimensional Minkowski space give 16 global 
supercharges.
This corresponds to minimal supersymmetry in seven dimensions (in the same way 
that
$G_2$-holonomy corresponds to minimal supersymmetry in four dimensions).
If we work at a smooth point in ${\cal M}$ we can use Kaluza-Klein analysis 
and we learn immediately
that the effective d=7 supergravity has 58 massless scalar fields which 
parametrise ${\cal M}$.
These are the fluctuations of the metric on $K3$. Additionally, since $H^2 (K3,
 {\bf R} )$ $\cong$
${\bf R^{22}}$ there are twenty-two linearly independent classes of harmonic 
2-forms. These
may be used a la equation  $(33)$ to give a $U(1)^{22}$ gauge group in seven 
dimensions.
We now go on to describe how this spectrum is the same as that of the 
heterotic string theory
on ${T^3}$, at generic points in ${\cal M}$.

The heterotic string in ten dimensions has a low energy description in terms 
of a supergravity
theory whose massless bosonic fields are a metric, a 2-form $B$, a dilaton 
$\phi$ and
non-Abelian gauge fields of structure group $SO(32)$ or ${E_8} {\times} 
{E_8}$. There are
sixteen global supersymetries. Compactification on a flat ${T^3}$ 
preserves all supersymmetries
which are all products of constant spinors on both ${T^3}$ and Minkowski 
space.
A flat metric on ${T^3}$ involves six parameters so the metric gives rise 
to six massless scalars.
and since there are three independent harmonic
two forms we obtain from $B$ three more. The condition for the gauge fields to 
be supersymmetric
on ${T^3}$ is that their field strengths vanish: these are so called flat 
connections. They are
parametrised by Wilson lines around the three independent circles in ${ 
T^3}$. These are
representations of the fundamental group of ${T^3}$ in the gauge group. 
Since the fundamental
group has three commuting generators we are looking for commuting triples of 
elements in
$G$.

Most of the flat connections actually arise from Wilson loops which are 
actually in the
maximal torus of the gauge group, which in this case is $U(1)^{16}$. Clearly, 
this gives
a 48 dimensional moduli space giving 58 scalars altogether. Narain showed by
direct computation that this moduli
space is actually also locally the same form as ${\cal M}$ \cite{narain}.

{}From the point of view of the heterotic string on ${T^3}$, the effective 
gauge group in
7 dimensions (for generic metric and $B$-field) is the subgroup of $SO(32)$ or 
${E_8}{\times}{E_8}$
which commutes with the flat connection on ${T^3}$.
At generic points in the
moduli space of flat connections, this gauge group will be $U(1)^{16}$.
This is because the
generic flat connection defines three generic elements in $U(1)^{16}$ 
$\subset$ $G$. We can think of
these as diagonal 16 by 16 matrices with all elements on the diagonal 
non-zero. Clearly,
only the diagonal elements of $G$ will commute with these. So, at a generic 
point in moduli
space the gauge group is abelian.

Six more
$U(1)$ gauge fields arise as follows from the metric and $B$-field. ${
T^3}$ has three harmonic
one forms, so Kaluza-Klein reduction of $B$ gives three gauge fields. 
Additionally, since
${T^3}$ has a $U(1)^3$ group of isometries, the metric gives three more.
In fact, the local action for supergravity theories in seven dimensions are
actually determined by the number of massless vectors.
So, in summary, we have
shown that at generic points in ${\cal M}$ the low energy supergravity 
theories arising from
$M$ theory on $K3$ or heterotic string on ${T^3}$ are the same.

At special points, some of the eigenvalues of the flat connections will 
vanish. At these points the
unbroken gauge group can get enhanced to a non-Abelian group. This is none 
other than the
Higgs mechanism: the Higgs fields are just the Wilson lines. Additionally, 
because seven dimensional
gauge theories are infrared trivial (the gauge coupling has dimension a 
positive power of length),
the low energy quantum theory actually has a non-Abelian gauge 
symmetry\footnote{Note
that because of the dimensionful coupling constant, 7d Yang-Mills is ill 
defined in the UV.
Here it is embedded into a consistent 11d theory which therefore provides a UV 
completion
of this gauge theory. We will mainly be interested in low energy properties in 
this article.}

If $M$ theory on $K3$
is actually equivalent to the heterotic string in seven dimensions, it too 
should therefore
exihibit non-Abelian symmetry enhancement at special points in the moduli 
space.
These points are precisely the points in moduli space where the $K3$ develops 
orbifold singularities.
We will not provide a detailed proof of this statement, but will instead look 
at the $K3$
moduli space in a neighbourhood of this singularity, where all the interesting 
behaviour of the
theory is occuring. So, the first question is what do these orbifold 
singularities look like?

\subsubsection*{ADE-singularities}

An orbifold singularity in a Riemannian 4-manifold can locally be described as
$\mathbb{R^4}/\mathbb{\Gamma}$, where $\mathbb{\Gamma}$ is a finite subgroup 
of $SO(4)$.
For generic enough $\mathbb{\Gamma}$, the only singular point of this orbifold 
is the
origin. These are the points in $\mathbb{R^4}$ left invariant under 
$\mathbb{\Gamma}$.
A very crucial point is that on the heterotic side, supersymmetry is 
completely unbroken
all over the moduli space, so our orbifold singularities in $K3$
should also preserve supersymmetry.
This means that $\mathbb{\Gamma}$ is a finite subgroup of $SU(2)$ $\subset$ 
$SO(4)$.
This $SU(2)$ is the holonomy group of the global $K3$ manifold.
The particular $SU(2)$ can easily be identified as follows. Choose some set
of complex coordinates so that $\mathbb{C^2}$ $\equiv$ $\mathbb{R^4}$. Then, a 
point
in $\mathbb{C^2}$ is labelled by a 2-component vector. The $SU(2)$ in question 
acts on
this vector in the standard way:
\begin{eqnarray}
\left(\begin{array}{c} {u} \\ {v} \end{array} \right) \longrightarrow
\left( \begin{array}{c} a \;\;\;b \\ c \;\;\;d \end{array} \right)
\left(\begin{array}{c} {u} \\ {v} \end{array} \right)
\end{eqnarray}

The finite subgroups of $SU(2)$ have a classification which may be described 
in terms
of the simply laced semi-simple Lie algebras: ${\sf A_n}$,
${\sf D_k}$, ${\sf E_6}$, ${\sf E_7}$ and ${\sf E_8}$.
There are two infinite series corresponding to $SU(n+1)$ $=$ ${\sf A_n}$ and 
$SO(2k)$ $=$
${\sf D_k}$
and three exceptional subgroups corresponding to the three exceptional Lie 
groups of $E$-type.
The subgroups, which we will denote by $\mathbb{\Gamma_{A_n}}$,
$\mathbb{\Gamma_{D_k}}$, $\mathbb{\Gamma_{E_i}}$ can be described explicitly.

$\mathbb{\Gamma_{A_{n-1}}}$ is isomorphic to $\mathbb{Z_n}$ -
the cyclic group of order $n$ - and is generated by
\ba
\left( \begin{array}{c} e^{{2\pi i \over n}} \;\;\;0 \\ 0
\;\;\;e^{{-2\pi i \over n}} \end{array} \right)
\ea

$\mathbb{\Gamma_{D_k}}$ is isomorphic to
$\mathbb{D_{k-2}}$ - the binary dihedral group
of order $4k-8$ - and has two generators $\alpha$ and $\beta$ given by
\ba
\alpha =
\left( \begin{array}{c} e^{{\pi i \over k-2}} \;\;\;0 \\ 0 \;\;\;e^{{-\pi i 
\over k-2}} \end{array} \right)
\;\;\;\beta =  \left( \begin{array}{c} 0 \;\;\;i \\ i \;\;\;0 \end{array} 
\right)
\ea

$\mathbb{\Gamma_{E_6}}$ is isomorphic to $\mathbb{T}$ -
the binary tetrahedral group
of order $24$ - and has two generators  given by
\ba
\left( \begin{array}{c} e^{{\pi i \over 2}} \;\;\;0 \\ 0 \;\;\;e^{{-\pi i 
\over 2}} \end{array} \right)
\;\;\; and \;\;\;{1 \over \sqrt{2}}  \left( \begin{array}{c} e^{{2\pi i 7 
\over 8}} \;\;\; e^{{2\pi i 7 \over 8}} \\
e^{{2\pi i 5 \over 8}} \;\;\;e^{{2\pi i  \over 8}} \end{array} \right)
\ea

$\mathbb{\Gamma_{E_7}}$ is isomorphic to $\mathbb{O}$ -
the binary octohedral group
of order $48$ - and has three generators. Two of these are the generators of 
$\mathbb{T}$ and the
third is
\ba
\left( \begin{array}{c} e^{{2\pi i \over 8}} \;\;\;0 \\ 0 \;\;\;e^{{2\pi i 7 
\over 8}} \end{array} \right)
\ea

Finally, $\mathbb{\Gamma_{E_8}}$ is isomorphic to $\mathbb{I}$ -
the binary icosahedral group
of order $120$ - and has two generators  given by

\ba
- \left( \begin{array}{c} e^{{2\pi i 3\over 5}} \;\;\;0 \\ 0 \;\;\;e^{{2\pi i2 
\over 5}} \end{array} \right)
\;\;\; and \;\;\;{1 \over e^{{2\pi i 2 \over 5}} - e^{{2\pi i 3 \over 5}}  }
\left( \begin{array}{c} e^{{2\pi i  \over 5}} + e^{{-2\pi i  \over 5}} \;\;\; 
1 \\
1 \;\;\;- e^{{2\pi i  \over 5}} - e^{{-2\pi i  \over 5}}\end{array} \right)
\ea

Since all the physics of interest is happening near the orbifold singularities 
of $K3$, we can
replace the $K3$ by $\mathbb{C^2/{\Gamma_{ADE}}}$ and study the physics of $M$ 
theory on
$\mathbb{C^2/{\Gamma_{ADE}}}{\times}\mathbb{R^{6,1}}$ near its singular set 
which is just
${\bf 0}{\times}\mathbb{R^{6,1}}$. Since the $K3$ went from smooth to singular 
as we varied its moduli
we expect that the singular orbifolds $\mathbb{C^2/{\Gamma_{ADE}}}$ are 
singular limits of
{\it non-compact} smooth 4-manifolds
$X^{ADE}$. Because of supersymmetry,
these should have $SU(2)$-holonomy. This is indeed the case. The metrics of 
$SU(2)$-holonomy
on the $X^{ADE}$ are known as ALE-spaces, since they asymptote to the
locally Euclidean metric on $\mathbb{C^2/{\Gamma_{ADE}}}$. Their existence
was proven by Kronheimer \cite{kron} - who constructed a gauge theory whose 
Higgs branch
is precisely the $X^{ADE}$ with its $SU(2)$-holonomy
(or hyper-Kahler) metric.

A physical description of this gauge theory arises in string theory.
Consider Type IIA or IIB string theory on $\mathbb{C^2 /{\Gamma_{ADE}}}
{\times}{\mathbb{R^{5,1}}}$. Take a flat D$p$-brane (with $p\leq5$)
whose world-volume directions span $\mathbb{R^{p,1}}$ $\subset$
$\mathbb{R^{5,1}}$ i.e. the D-brane is sitting at a point on the
orbifold. Then the world-volume gauge theory, which was first derived in 
\cite{quiver}.,
is given by the Kronheimer gauge theory. This theory has eight
supersymmetries which implies that its Higgs branch is a hyper-Kahler
manifold.
For one D-brane this
theory has a gauge group which is a product of unitary groups of ranks
given by the
Dynkin indices (or dual Kac labels)
of the affine Dynkin diagram of the corresponding ${\sf ADE}$-group. So,
for the ${\sf A_n}$-case the gauge group is $U(1)^{n+1}$. The matter content
is also given by the affine Dynkin diagram - each link between a pair
of nodes represents a hyper-multiplet transforming in the bi-fundamental
representation of the two unitary groups. This is an example of a
quiver gauge theory - a gauge theory determined by a quiver diagram.

We will make this explicit in the simplest case of
$\mathbb{\Gamma_{A_1}}$. $\mathbb{\Gamma_{A_1}}$ is
isomorphic to $\mathbb{Z_2}$ and is in fact the center of $SU(2)$. Its 
generator acts on $\mathbb{C^2}$ as
\begin{eqnarray}
\left(\begin{array}{c} {u} \\ {v} \end{array} \right) \longrightarrow
\left(\begin{array}{c} {-u} \\ {-v} \end{array} \right)
\end{eqnarray}

In this case, the Kronheimer gauge theory has a gauge group which is $G = U(1)
^2$
and has two fields $\Phi_1$ and $\Phi_2$ transforming as $(+,-)$ and $(-,+)$.
These are hypermultiplets in the
string theory realisation on a D-brane. Clearly, the diagonal $U(1)$ in $G$ 
acts trivially and
so can be factored out to give a gauge group $G/U(1) = U(1)$ under which (
after rescaling the
generator) $\Phi_1$ and $\Phi_2$ transform with charge $+1$ and $-1$.

The hypermultiplets each contain
two complex scalars $(a_i , b_i )$. The $a$'s transform with charge $+1$
under $U(1)$, whilst the $b$'s transform with charge $-1$.

The potential energy of these scalar fields on the D-brane is
\be
V = |\vec{D}|^2 = |\vec{\mu} |
\ee

where the three $D$-fields $\vec{D}$ (which are also known as the
hyper-Kahler moment maps $\vec{\mu}$ associated with $U(1)$ action
on the ${\mathbb C^4}$ parameterised by the fields) are given by
\be
D_1 = |a_1|^2 + |b_1|^2 - |a_2|^2 - |b_2|^2
\ee
and
\be
D_2 + i D_3 = a_1 b_1 - a_2 b_2
\ee

The space of zero energy minima of $V$ is the space of supersymmetric
ground states $S$ of the theory on the brane up to gauge transformations:

\be
S = \{ \vec{D} = 0 \}/U(1)
\ee

In supersymmetric field theories, instead of solving these equations directly.
it is equivalent to simply construct the space of gauge invariant
holomorphic polynomials
of the fields and impose only the holomorphic equation above (this is the
$F$-term in the language of four dimensional supersymmetry).
A solution to the $D_1$ equation is then guaranteed to exist because of
invariance under the complexification of the gauge group.

In the case at hand the gauge invvariant polynomials are simply
\be
X = a_1 b_1 \;\;Y = a_2 b_2 \;\; Z = a_1 b_2 \;\; W = a_2 b_1
\ee

These obviously parameterise $\mathbb{C^4}$ but are subject to the
relation
\be
XY = WZ
\ee

However, the complex D-term equation asserts that
\be
X=Y
\ee
hence
\be
X^2 = WZ
\ee

The space of solutions is precisely a copy of $\mathbb{C^2 / \Gamma_{A_1}}$.
To see this,
we can parametrise $\mathbb{C^2/{\Gamma_{A_1}}}$
algebraically in terms of the $\mathbb{\Gamma_{A_1}}$
invariant coordinates on $\mathbb{C^2}$. These are $u^2$, $v^2$ and $uv$.
If we denote these three coordinates as $w,z,x$, then obviously
\be
x^2 = wz
\ee

We prefer to re-write this equation by changing coordinates again.
Defining
$x$  $=$ $u^2 - v^2$,
$y$ $=$ $iu^2 + iv^2$ and $z$ $=$ $2uv$ gives a map from 
$\mathbb{{C^2}/\Gamma_{A_1}}$
to $\mathbb{C^3}$. Clearly
however,
\be
x^2 + y^2 + z^2 = 0
\ee
which means that $\mathbb{C^2/\Gamma_{A_1}}$ is the hypersurface in 
$\mathbb{C^3}$ defined by
this equation.

The orbifold can be deformed by adding a small constant to the right hand side,

\be
x^2 + y^2 + z^2= r^2
\ee
If we take $x$, $y$ and $z$ to all be real and $r$ to be real then it is clear 
that the
deformed 4-manifold contains a 2-sphere of radius $r$.
This 2-sphere contracts to zero size as $r$ goes to zero. The total space of 
the
deformed 4-manifold is in fact the co-tangent bundle of the 2-sphere, ${\bf 
T^* {S^2}}$.
To see this write the real parts of the $x$, $y$ and $z$ as $x_i$ and their 
imaginary
parts as $p_i$. Then, since $r$ is real, the $x_i$ are coordinates on the 
sphere which
obey the relation
\be
\Sigma_i x_i p_i = 0
\ee
This means that the $p_i$'s parametrise tangential directions.  The radius $r$ 
sphere in the
center is then the zero section of the tangent bundle.
Since the manifold
is actually complex it is natural to think of this as the co-tangent bundle of 
the Riemann
sphere, ${\bf T^* {\mathbb{C}P^1}}$.
In the context of Euclidean quantum gravity, Eguchi and Hanson constructed a 
metric of
$SU(2)$-holonomy on this space,
asymptotic to the locally flat metric on $\mathbb{C^2/\Gamma_{A_1}}$.

In the Kronheimer gauge theory on the D-brane, deforming the singularity
corresponds to setting the D-terms or moment maps, not to zero but to 
constants.
On the D-brane these cosntants represent the coupling of the background
closed string fields to the brane. These fields parameterise precisely
the metric moduli of the Eguchi-Hanson metric.

\subsubsection*{$M$ theory Physics at The Singularity}

This metric, whose precise form we will not require actually has three 
parameters (since their are three $D$-terms in this case) which control the
size and shape of the two-sphere which desingularises the orbifold.
{}From a distance it
looks as though there is an orbifold singularity, but as one looks more 
closely one
sees that the singularity has been smoothened out by a two-sphere. The 
2-sphere is
dual to a compactly supported harmonic 2-form, $\alpha$.
Thus, Kaluza-Klein reducing the $C$-field using $\alpha$ gives a $U(1)$ gauge 
field in seven
dimensions. A vector multiplet in seven dimensions contains precisely one 
gauge field
and three scalars and the latter are the parameters of the ${\bf S^2}$. So, 
when ${\bf T^* {\mathbb{C}P^1}}$
is smooth the massless spectrum is an abelian vector multiplet.

{}From the duality with the heterotic string we expect to see an enhancement 
in the
gauge symmetry when we vary the scalars to zero i.e. when the sphere shrinks to
zero size. In order for this to occur, $W^{\pm}$-bosons must become massless 
at
the singularity. These are electrically
charged under the $U(1)$ gauge field which originated from
$C$. From the eleven dimensional point of view the object which is charged 
under $C$
is the $M2$-brane. The reason that this is natural is the equation of motion 
for $C$
is an eight-form in eleven dimensions. A source for the $C$-field is thus 
generated by an
8-form closed current. In eleven dimensions such currents are naturally 
supported along
three dimensional manifolds. These can be identified with $M2$-brane 
world-volumes.

If the $M2$-brane wraps around the two-sphere, it appears as
a particle from the seven dimensional point of view. This particle is 
electrically
charged under the $U(1)$ and has a mass which is classically given by the 
volume of
the sphere. Since, the $M2$-brane has tension its dynamics will push it to 
wrap the
smallest volume two-sphere in the space. This least mass configuration is in 
fact
invariant under half of the supersymmetries
\footnote{This is because the least volume two-sphere is an example of a 
calibrated
or supersymmetric cycle.} --- a fact which means that it lives in a
short representation of the supersymmetry algebra. This in turn means that its 
classical
mass is in fact uncorrected quantum mechanically. The $M2$-brane wrapped
around this cycle with the opposing orientation has the opposing $U(1)$ charge 
to
the previous one.

Thus, when the two-sphere shrinks to zero size we find two oppositely charged 
BPS
multiplets become massless. These have precisely the right quantum numbers to
enhance the gauge symmetry from $U(1)$ to ${\sf A_1}$ $=$ $SU(2)$. Super 
Yang-Mills
theory in seven dimensions depends only on its gauge group, in the sense
that its low energy lagrangian is uniquely determined by supersymmetry and
the gauge group.
In this case we are asserting that,
in the absence of gravity, the low energy physics of $M$ theory on
$\mathbb{C^2 /{\Gamma_{A_1}}}{\times}\mathbb{R^{6,1}}$ is described by
super Yang-Mills theory on ${\bf 0}{\times}\mathbb{R^{6,1}}$ with gauge group
${\sf A_1}$.

The obvious generalisation also applies:
in the absence of gravity, the low energy physics of $M$ theory on
$\mathbb{C^2 /{\Gamma_{ADE}}}{\times}\mathbb{R^{6,1}}$ is described by
super Yang-Mills theory on ${\bf 0}{\times}\mathbb{R^{6,1}}$ with {\sf ADE} 
gauge group.
To see this, note that the smoothing out of the
orbifold singularity in $\mathbb{C^2 /{\Gamma_{ADE}}}$
contains ${\sf rank(ADE)}$ two-spheres
which intersect according to the Cartan matrix of the ${\sf ADE}$ group. At 
smooth points in the
moduli space the gauge group is thus $U(1)^{\sf rank(ADE)}$.
The corresponding wrapped membranes give rise to massive BPS multiplets with 
precisely
the masses and quantum numbers required to enhance the gauge symmetry to the
full ${\sf ADE}$-group at the origin of the moduli space, where the orbifold
singularity appears.

\subsection*{ADE-singularities in $G_2$-manifolds.}

We have thus far restricted our attention to the ${\sf ADE}$ singularities in 
$K3 {\times}
\mathbb{R^{6,1}}$. However, the ${\sf ADE}$ singularity is a much more local 
concept.
We can consider more complicated spacetimes $X^{10,1}$ with ${\sf ADE}$ 
singularities
along more general seven-dimensional spacetimes, $Y^{6,1}$. Then, if $X$ has a 
modulus
which allows us to scale up the volume of $Y$, the large volume limit is a 
semi-classical limit
in which $X$ approaches the previous maximally symmetric situation discussed 
above.
Thus, for large enough volumes we can assert that the description of the 
classical physics
of $M$ theory near $Y$ is in terms of seven dimensional super Yang-Mills 
theory on
$Y$ - again with gauge group determined by which {\sf ADE} singularity lives 
along $Y$.

In the context of $G_2$-compactification on $X {\times}\mathbb{R^{3,1}}$, we 
want $Y$ to be
of the form $W{\times}\mathbb{R^{3,1}}$, with $W$ the locus of {\sf ADE} 
singularities inside $X$.
Near $W{\times}\mathbb{R^{3,1}}$, $X{\times}\mathbb{R^{3,1}}$ looks like
$\mathbb{C^2 /\Gamma_{ADE}}{\times}W{\times}\mathbb{R^{3,1}}$.  In order to 
study the gauge theory
dynamics without gravity, we can again focus on the physics near the 
singularity itself.
So, we want to focus on seven-dimensional super Yang-Mills theory on 
$W{\times}\mathbb{R^{3,1}}$.


In flat space the super Yang-Mills theory has a global symmetry group which is
$SO(3) {\times} {SO(6,1)}$. The second factor is the Lorentz group, the first 
is the R-symmetry.
The theory has gauge fields transforming as ${\bf (1,7)}$, scalars in the 
${\bf (3,1)}$ and fermions
in the ${\bf (2,8)}$ of the universal cover. All fields transform in the 
adjoint representation of the
gauge group. Moreover the sixteen supersymmetries also transform as ${\bf (2,8)
}$.

On $W{\times}\mathbb{R^{3,1}}$ --- with an arbitrary $W$ ---
the symmetry group gets broken to
$$SO(3){\times}{SO(3)'} {\times}{SO(3,1)}$$
Since $SO(3)'$ is the structure group of the tangent bundle on $W$,
covariance requires that the theory is coupled to a background $SO(3)'$ gauge 
field - the
spin connection on $W$. Similarly, though perhaps less intuitively, $SO(3)$ 
acts on the
normal bundle to $W$ inside $X$, hence there is a background $SO(3)$ gauge 
field also.

The supersymmetries transform as
${\bf (2,2,2) + (2,2,{\bar 2})}$. For large enough $W$ and at energy scales 
below the inverse
size of $W$, we can describe the physics in terms of a {\it four} dimensional 
gauge theory.
But this theory as we have described it is not supersymmetric as this requires 
that we have
covariantly constant spinors on $W$. Because $W$ is curved, there are none. 
However, we
actually want to consider the case in which $W$ is embedded inside a 
$G_2$-manifold $X$.
In other words we require that our local model - $\mathbb{C^2 
/\Gamma_{ADE}}{\times}W$ -
admits a $G_2$-holonomy metric. When $W$ is curved this metric cannot be the
product of the locally flat metric on $\mathbb{C^2}/{\Gamma_{ADE}}$ and a 
metric on $W$.
Instead the metric is warped and is more like the metric on a fiber bundle in 
which the
metric on $\mathbb{C^2}$ varies as we move around in $W$. Since the space has 
$G_2$-holonomy
we should expect the four dimensional gauge theory to be supersymmetric. We 
will now
demonstrate that this is indeed the case by examining the $G_2$-structure more 
closely.
In order to do this however, we need to examine the $SU(2)$ structure on
$\mathbb{C^2 /\Gamma}$
as well.

4-dimensional spaces of $SU(2)$-holonomy are actually examples of hyperKahler 
manifolds.
They admit   three parallel 2-forms ${\omega}_i$. These are analogous to the 
parallel
forms on $G_2$-manifolds.
These three forms transform locally under
$SO(3)$ which locally rotates the complex structures.
On $\mathbb{C^2}$ these forms can be given explicitly as
\ba
\omega_1 + i \omega_2 &=& du\wedge dv \\
\omega_3  &=& {i \over 2} du\wedge d{\bar u} + dv\wedge d{\bar v}
\ea

$\mathbb{\Gamma_{ADE}}$
is defined so that it preserves all three of these forms. The $SO(3)$ which
rotates these three forms is identified with the $SO(3)$ factor in our seven 
dimensional gauge theory
picture. This is because the moduli space of
$SU(2)$-holonomy metrics is the moduli space of the gauge theory and this
has an action of $SO(3)$.

In a locally flat frame we can write down a formula for the $G_2$-structure on
$\mathbb{C^2/\Gamma_{ADE}}{\times}W$,
\be
\Phi = {1 \over 6} \omega_i \wedge e_j \delta^{ij} + e_1 \wedge e_2 \wedge e_3
\ee
where $e_i$ are a flat frame on $W$. Note that this formula is manifestly 
invariant
under the $SO(3)$ which rotates the $w_i$ {\it provided} that it also acts
on the $e_i$ in the same way.

The key point is that when the $SO(3)$ of the gauge theory acts, in order for 
the $G_2$-structure
to be well defined , the $e_i$'s must transform in precisely the same way as 
the $\omega_i$.
But $SO(3)'$ acts on the $e_i$, because it is the structure group of the 
tangent bundle to $W$.
Therefore, if  $\mathbb{C^2 /{\Gamma_{ADE}}}{\times}W$,
admits a $G_2$-holonomy metric, we must identify $SO(3)$ with $SO(3)'$. In 
other words,
the connection on the tangent bundle is identified with the connection on the 
normal $SO(3)'$
bundle. This breaks the symmetries to the diagonal subgroup of the two $SO(3)
$'s and
implies
that the effective four dimensional field theory is classically 
supersymmetric.
Identifying the two groups breaks the symmetry group down to $SO(3)
''{\times}SO(3,1)$
under which the supercharges
transform as ${\bf (1,2) + (3,2) + cc}$. We now have supersymmetries
since the ${\bf (1,2)}$ and its conjugate can be taken to be constants on $W$.

An important point which we will not actually prove here, but will require in
the sequel is that the locus of ${\sf ADE}$-singularities - namely the
copy of $W$ at the center of $\mathbb{C^2 /\Gamma}$ is actually a 
supersymmetric
cycle {\it i.e.} calibrated by $\Phi$. This follows essentially from
the fact that $\mathbb{\Gamma_{ADE}}$ fixes $W$ and therefore the $\Phi$
restricts to be the volume form on $W$. This is the condition for $W$
to be supersymmetric.

Supposing we could find a $G_2$-manifold of this type, what exactly is the
four dimensional supersymmetric gauge theory it corresponds to? This we can
answer also by Kaluza-Klein analysis \cite{bsa1,bsa2},
since $W$ will be assumed to be smooth and
`large'. Under $SO(3)''{\times}SO(3,1)$, the seven dimensional gauge fields 
transform as
${\bf (3,1) + (1,4)}$, the three scalars give ${\bf (3,1)}$ and the fermions 
give
${\bf (1,2) + (3,2) + cc}$. Thus the fields which are scalars under the four 
dimensional
Lorentz group are two copies of the ${\bf 3}$ of $SO(3)''$. These may be 
interpreted as
two one forms on $W$. These will be massless if they are zero modes of the 
Laplacian on
$W$ (wrt its induced metric from the $G_2$-manifold). There will be precisely
$b_1 (W)$ of these i.e. one for every harmonic one form.
Their superpartners are clearly the ${\bf (3,2)+cc}$ fermions, which will be 
massless by
supersymmetry. This is precisely the field content of $b_1 (W)$ chiral 
supermultiplets
of the supersymmetry algebra in four dimensions.

The ${\bf (1,4)}$ field is massless if it is constant on $W$ and this gives 
one
gauge field in four dimensions. The requisite superpartners are the remaining
fermions which transform as ${\bf (1,2) + cc}$.

All of these fields transform in the adjoint representation of the seven 
dimensional gauge group.
Thus the final answer for the massless fields is that they are described by 
${\cal N}$ $=1$
super Yang-Mills theory with $b_1 (W)$ massless adjoint chiral 
supermultiplets. The
case with pure ``superglue'' i.e. $b_1 (W)$ = $0$ is a particularly interesting 
gauge theory at
the quantum level: in the infrared the theory is believed to
confine colour, undergo chiral symmetry breaking and have a mass gap. We will 
actually
exhibit some of these very interesting properties semi-classically in $M$ 
theory ! Section 8 will be devoted to explaining this.
But before we can do that we must first
describe concrete examples of $G_2$-manifolds with the properties we desire.

One idea is to simply look for {\it smooth} $G_2$-manifolds which are 
topologically
$\mathbb{C^2}{\times}W$ but admit an action of $SU(2)$ which leaves $W$ 
invariant but
acts on $\mathbb{C^2}$ in the natural way. Then we simply pick a
$\mathbb{\Gamma_{ADE}}$ $\subset$
$SU(2)$ and form the quotient space     $\mathbb{C^2/\Gamma_{ADE}}{\times}W$.

Luckily, such non-compact $G_2$-manifolds were
constructed some time ago in \cite{BS,GPP},
see section \ref{noncompact}.
Moreover, in these examples, $W$ $={\bf S^3}$,
the simplest possible compact 3-manifold with $b_1 (W)$ = $0$.


\subsection{Low Energy Dynamics via IIA Duals}
\label{typeiia}

Another useful method of analyzing $M$ theory
on singular manifolds with special holonomy follows from
the duality between type IIA string theory and $M$ theory
compactified on a circle \cite{Townsend,WittenM}:

\smallskip
\begin{center}
\begin{tabular}{ccc}
\framebox[3.3cm][c]{\parbox{2.3cm}{IIA Theory}}
& \quad $\iff$ \quad &
\framebox[3.8cm][c]{\parbox{2.8cm}{$M$ theory on ${\bf S}^1$}} \\
\end{tabular}
\end{center}
\smallskip

Among other things, this duality implies that any state in
IIA theory can be identified with the corresponding state
in $M$ theory. In this identification, some of the states
acquire a geometric origin when lifted to eleven dimensions.
In order to see this explicitly, let us write
the eleven-dimensional metric in the form
\begin{equation}
ds^2 = e^{ - {2 \over 3} \phi} g_{ij} dx^{i} dx^{j}
+ e^{{4 \over 3} \phi} (d x_{11} + A_{i} dx^{i})^2
\label{11dmetric}
\end{equation}
Upon reduction to ten-dimensionsional space-time
(locally parametrized by $x_{i}$),
the field $\phi$ is identified with the dilaton,
$A_{i}$ with Ramond-Ramond 1-form,
and $g_{ij}$ with the ten-dimensional metric.
Therefore, any IIA background that involves excitations
of these fields
uplifts to a purely geometric background in eleven dimensions.
Moreover, from the explicit form of the metric (\ref{11dmetric})
it follows that the eleven-dimensional geometry is a circle fibration
over the ten-dimensional space-time, such that the topology of this
fibration is determined by the Ramond-Ramond 1-form field.
This important observation will play
a central role in this section.

To be specific, let us consider a D6-brane in type IIA string theory.
Since a D6-brane is a source for the dilaton, for Ramond-Ramond 1-form,
and for the metric, it is precisely the kind of state
that uplifts to pure geometry. For example, if both
D6-brane world-volume and the ambient space-time are flat,
the dual background in $M$ theory is given by the Taub-NUT space:

\smallskip
\begin{center}
\begin{tabular}{ccc}
\framebox[4cm][c]{\parbox{3cm}{
$M$ theory on \\ Taub-Nut space}}
& \quad $\iff$ \quad &
\framebox[5.5cm][c]{\parbox{4.5cm}{
IIA theory in flat space-time with a D6-brane}} \\
\end{tabular}
\end{center}
\smallskip

\begin{figure}
\begin{center}
\epsfxsize=3in\leavevmode\epsfbox{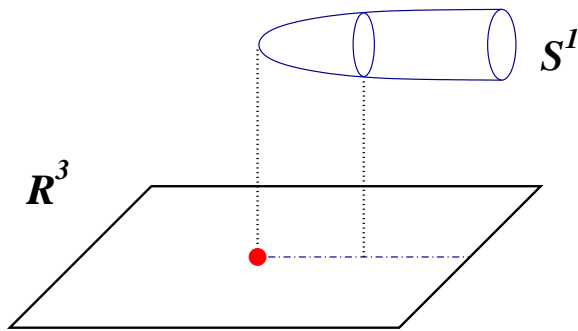}
\end{center}
\caption{A cartoon representing Taub-NUT space as a circle
fibration over a 3-plane.}
\label{taubnut}
\end{figure}

The Taub-NUT space is a non-compact four-manifold
with $SU(2)$ holonomy. It can be viewed as a circle fibration
over a 3-plane, see Figure \ref{taubnut}.
The ${\bf S}^1$ fiber degenerates at a single point ---
the origin of the $\mathbb{R}^3$ --- which is identified
with the location of the D6-brane in type IIA string theory.
On the other hand, at large distances the size of
the `$M$ theory circle' stabilizes at some constant value
(related to the value of the string coupling constant in IIA theory).
Explicitly, the metric on the Taub-NUT space is given by \cite{EH}:
\begin{eqnarray}
ds^2_{TN} & = & H d {\vec x}^{~2} + H^{-1} (d x_{11} + A_{i} dx^{i})^2,
\label{TaubNUT} \\
\vec{\nabla} \times \vec A & =&  - \vec \nabla H, \nonumber \\
H & = & 1 + {1 \over 2 |\vec x|} \nonumber
\end{eqnarray}
This form of the metric makes especially clear the structure
of the circle fibration. Indeed, if we fix a constant-$r$
sphere inside the $\mathbb{R}^3$, then it is easy to see that
${\bf S}^1$ fiber has `winding number one' over this sphere.
This indicates that there is a topological defect --- namely,
a D6-brane --- located at $\vec x = 0$, where ${\bf S}^1$ fiber degenerates.

The relation between D6-branes and geometry can be extended
to more general manifolds that admit a smooth $U(1)$ action.
Indeed, if $X$ is a space (not necessarily smooth and/or compact)
with $U(1)$ isometry, such that $X/U(1)$ is smooth,
then the fixed point set, $L$, of the $U(1)$ action
must be of codimension 4 inside $X$ \cite{MZ,Levine}.
This is just the right codimension to identify
$L$ with the D6-brane
locus\footnote{The part of the D-brane world-volume that is transverse
to $X$ is flat and does not play an important r$\mathrm{\hat{o}}$le
in our discussion here.}
in type IIA theory on $X/U(1)$:

\smallskip
\begin{center}
\begin{tabular}{ccc}
\framebox[5cm][c]{\parbox{4cm}{
~~~~$M$ theory on  $X$}}
& \quad $\iff$ \quad &
\framebox[5cm][c]{\parbox{4cm}{
IIA theory on $X/U(1)$ \rule{0pt}{3mm} with D6-branes on $L$}} \\
\end{tabular}
\end{center}
\smallskip

\noindent
For example, if $X$ is the Taub-NUT space,
then $U(1)$ action is generated by the shift of
the periodic variable $x_{11}$ in (\ref{TaubNUT}).
Dividing by this action one finds $X/U(1) \cong \mathbb{R}^3$
and that $L = \{ {\rm pt} \} \in X$ indeed has codimension 4.

It may happen that a space $X$ admits more than one $U(1)$ action.
In that case, $M$ theory on $X$ will have several IIA duals,
which may look very different but, of course, should exhibit
the same physics. This idea was used by Atiyah, Maldacena,
and Vafa to realise a geometric duality between certain IIA
backgrounds as topology changing transitions in $M$ theory \cite{AMV}.
We will come back to this in section \ref{transitions}.

Now let us describe a particularly
useful version of the duality between $M$ theory on
non-compact space $X$ and IIA theory with D6-branes on $X/U(1)$
which occurs when $X$ admits a $U(1)$ action, such that
the quotient space is isomorphic to a flat Euclidean space.
Suppose, for example, that $X$ is a non-compact $G_2$ manifold,
such that
\begin{equation}
X/U(1) \cong \mathbb{R}^6
\end{equation}
Then, the duality statement reads:

\smallskip
\begin{center}
\begin{tabular}{ccc}
\framebox[6cm][c]{\parbox{5cm}{
$M$ theory on non-compact $G_2$ manifold $X$}}
& \quad $\iff$ \quad &
\framebox[6cm][c]{\parbox{5cm}{
IIA theory in flat space-time with
D6-branes on $\mathbb{R}^{3,1} \times L$}} \\
\end{tabular}
\end{center}
\smallskip

\noindent
On the left-hand side of this equivalence the space-time in $M$ theory
is $\mathbb{R}^{3,1} \times X$.
On the other hand, the geometry of space-time in IIA theory
is trivial (at least topologically) and all the interesting
information about $X$ is mapped into the geometry of the D6-brane
locus $L$. For example, the Betti numbers of $L$ are determined
by the corresponding Betti numbers of the space $X$ \cite{GS}:
\begin{eqnarray}
b_k (L) & = & b_{k+2} (X), \quad k>0 \label{bettirels} \\
b_0 (L) & = & b_2 (X) + 1 \nonumber
\end{eqnarray}
Notice the shift in degree by 2, and also that the number
of the D6-branes (= the number of connected components of $L$)
is determined by the second Betti number of $X$.
We will not present the derivation of this formula here.
However, it is a useful exercise to check
({\it e.g.} using the Lefschetz fixed point theorem)
that the Euler numbers of $X$ and $L$
must be equal, in agreement with (\ref{bettirels}).

The duality between $M$ theory on a non-compact manifold $X$
and a configuration of D6-branes in a (topologically) flat space
can be used to study singular limits of $X$.
Indeed, when $X$ develops a singularity, so does $L$.
Moreover, $L$ must be a supersymmetric (special Lagrangian)
submanifold in $\mathbb{R}^6 \cong \mathbb{C}^3$
in order to preserve the same amount
of supersymmetry\footnote{
In general, in a reduction from $M$ theory down
to Type IIA one does not obtain the standard flat
metric on $X/U(1) \cong \mathbb{R}^{6}$ due
to non-constant dilaton and other fields in the background.
However, one would expect that near the singularities of
the D-brane locus $L$ these fields exhibit a regular behavior,
and the metric on $X/U(1)$ is approximately flat, {\it cf.} \cite{AW}.
In this case the condition for the Type IIA background
to be supersymmetric can be expressed as a simple geometric
criterion: it says that the D-brane locus $L$ should be
a calibrated submanifold in $X/U(1)$.} as the $G_2$-holonomy space $X$.
Therefore, the problem of studying the dynamics of $M$ theory on
$G_2$ singularities can be recast as a problem
of studying D6-brane configurations on singular
special Lagrangian submanifolds in flat space \cite{AW}.

Following Atiyah and Witten \cite{AW}, let us see
how this duality can help us to analyze one of the conical
$G_2$ singularities listed in Table 4.

\bigbreak\hrule\medskip\nobreak\noindent

{\bf Example:}
Consider an asymptotically conical $G_2$ manifold $X$
with $SU(3)/U(1)^2$ principal orbits and topology
\begin{equation}
X \cong \mathbb{C}{\bf P}^2 \times \mathbb{R}^3
\label{xcp2top}
\end{equation}

Assuming that $M$ theory on this space $X$ admits a circle
reduction to IIA theory with D6-branes in flat space,
we can apply the general formula (\ref{bettirels})
to find the topology of the D6-brane locus $L$.
For the manifold (\ref{xcp2top}) we find
the following non-zero Betti numbers:

\begin{center}
\begin{tabular}{ccc}
Betti numbers of $X$ &  & Betti numbers of $L$ \\
\cline{1-1}\cline{3-3}
\rule{0pt}{5mm}
$b_4=1$ & $\longrightarrow$ & $b_2 = 1$ \\
$b_2=1$ & $\longrightarrow$ & $b_0 = 2$ \\
\end{tabular}
\end{center}

\begin{figure}
\begin{center}
\epsfxsize=5in\leavevmode\epsfbox{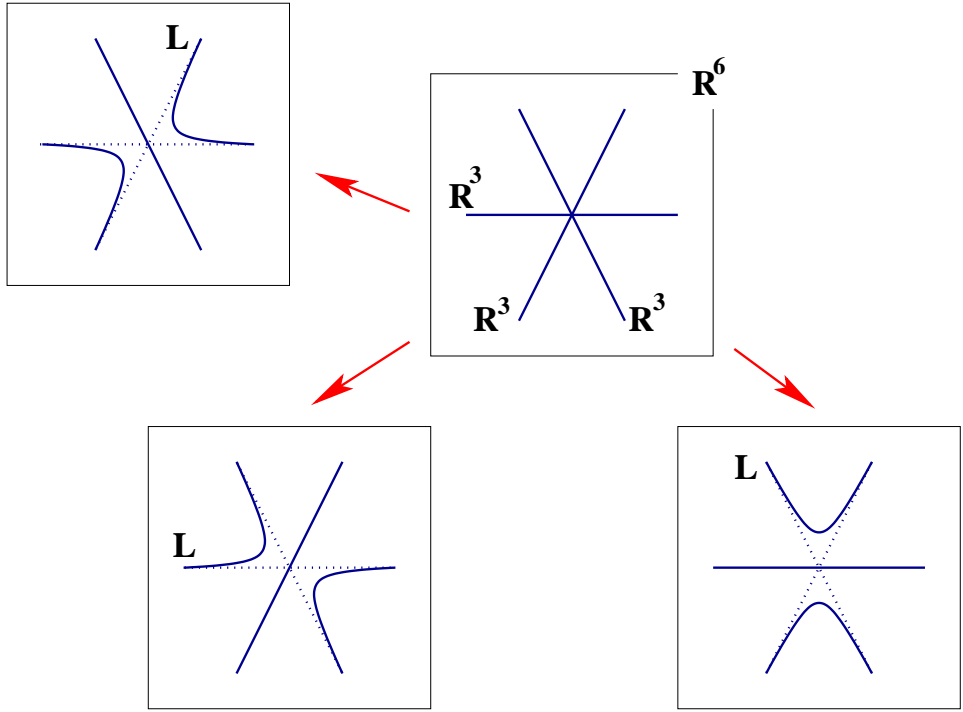}
\end{center}
\caption{Intersection of special Lagrangian D6-branes
dual to a $G_2$ cone over the flag manifold $SU(3)/U(1)^2$,
and its three non-singular deformations.}
\label{3d6s}
\end{figure}

\noindent
Therefore, we conclude that $L$ should be a non-compact
3-manifold with two connected components (since $b_0=2$) and with
a single topologically non-trivial 2-cycle (since $b_2=1$).
A simple guess for a manifold that has these properties is
\begin{equation}
L \cong \left( {\bf S}^2 \times \mathbb{R} \right) \cup \mathbb{R}^3
\label{slagcp2top}
\end{equation}
It turns out that there indeed exists a special Lagrangian
submanifold in $\mathbb{C}^3$ with the right symmetries and
topology (\ref{slagcp2top}), see \cite{AW} for an explicit construction
of the circle action on $X$ that has $L$ as a fixed point set.
If we choose to parametrize ${\rm Re} (\mathbb{C}^3) = \mathbb{R}^3$
and ${\rm Im} (\mathbb{C}^3) = \mathbb{R}^3$ by 3-vectors
$\vec{\phi}_1$ and $\vec{\phi}_2$, respectively,
then one can explicitly describe $L$ as the zero locus
of the polynomial relations \cite{joycesymm}:
\begin{equation}
L = \left\{ \vec{\phi}_1\cdot\vec{\phi}_2=-|\vec{\phi}_1||\vec{\phi}_2|
\ ; \quad|\vec \phi_1|(3 |\vec \phi_2|^2 - |\vec \phi_1|^2) = \rho
\right\}
\cup \left\{ |\vec{\phi}_1-\sqrt{3}\vec{\phi}_2|=0 \right\}
\label{slagcp2}
\end{equation}
It is easy to see that the first component of this manifold looks
like a hyperboloid in $\mathbb{C}^3 = \mathbb{R}^3 \times \mathbb{R}^3$.
It has a `hole' in the middle,
resulting in the ${\bf S}^2 \times \mathbb{R}$ topology,
required by (\ref{slagcp2top}).
The second component in (\ref{slagcp2}) is clearly a 3-plane,
which goes through this hole, as shown in Figure \ref{3d6s}.

When $X$ develops a conical singularity
$L$ degenerates into a collection of three planes,
\begin{equation}
L_{{\rm sing}} \cong \mathbb{R}^3 \cup \mathbb{R}^3 \cup \mathbb{R}^3
\end{equation}
intersecting at a single point, see Figure \ref{3d6s}.
This can be seen explicitly by taking $\rho \to 0$ limit
in the geometry (\ref{slagcp2}),
\begin{eqnarray}
D6_1 :&&\ \ \vec{\phi}_1=0 \nonumber \\
D6_2 :&&\ \ {1 \over 2} \vec{\phi}_1
+{\sqrt{3} \over 2} \vec{\phi}_2=0 \nonumber \\
D6_3 :&&\ - {1 \over 2} \vec{\phi}_1 + {\sqrt{3} \over 2}\vec{\phi}_2=0
\nonumber
\end{eqnarray}
which corresponds to the limit of collapsing ${\bf S}^2$ cycle.

Therefore, in this example the physics of $M$ theory on conical
$G_2$ singularity can be reduced to the physics of three intersecting
D6-branes in type IIA string theory. In particular, since D6-branes
appear symmetrically in this dual picture, one can conclude that
there must be three ways of resolving this singularity,
depending on which pair of D6-branes is deformed into
a smooth special Lagrangian submanifold (\ref{slagcp2}).
This is precisely what Atiyah and Witten found in a more careful
analysis \cite{AW}. We will come back to this example later,
in section \ref{transitions}.

\par\nobreak\medskip\nobreak\hrule\bigbreak

There is a similar duality that relates $M$ theory on
$Spin(7)$ manifolds to D6-brane configurations
on coassociative cycles \cite{Gomis}.
In particular, if $X$ is a non-compact $Spin(7)$
manifold with a $U(1)$ action, such that
\begin{equation}
X/U(1) \cong \mathbb{R}^7
\end{equation}
then one finds the following useful duality:

\smallskip
\begin{center}
\begin{tabular}{ccc}
\framebox[6cm][c]{\parbox{5cm}{
$M$ theory on non-compact $Spin(7)$ manifold $X$}}
& \quad $\iff$ \quad &
\framebox[6cm][c]{\parbox{5cm}{
IIA theory in flat space-time with
D6-branes on $\mathbb{R}^3 \times L$}} \\
\end{tabular}
\end{center}
\smallskip

\noindent
where $L$ is a coassociative submanifolds in $\mathbb{R}^7$.
Again, on the left-hand side the geometry of space-time is
$\mathbb{R}^3 \times X$, whereas on the right-hand side
space-time is (topologically) flat and all the interesting
information is encoded in the geometry of the D6-brane locus $L$.
Topology of the latter can be determined from the general
formula (\ref{bettirels}).
Therefore, as in the $G_2$ case, $M$ theory dynamics on
singular $Spin(7)$ manifolds can be obtained
by investigating D6-brane configurations on singular
coassociative submanifolds in flat space \cite{GS}.

\newpage
\section{Chiral Fermions from Conical Singularities in $G_2$-manifolds}
\label{chiralmatter}

In the previous section we introduced various techniques
for studying $M$ theory dynamics on singular manifolds,
and discussed their application in a few simple examples.
In particular, using duality with the heterotic string
theory we explained the origin of the non-abelian gauge
symmetries on ${\sf ADE}$ orbifold singularities,
and using duality with the IIA string we derived the physics
of $M$ theory at some simple conical singularities
in $G_2$ and $Spin(7)$ manifolds.
Locally, a conical singularity can be
described by the metric of the form (\ref{conicalmet}),
where $Y$ is the base of the cone\footnote{Known examples of
$G_2$ and $Spin(7)$ metrics that admit this kind of
degeneration have been discussed in section \ref{noncompact}.}.

In this section, we continue the study of conical singularities
using these methods. In particular, we will show that,
for many different choices of $Y$, the dynamics of $M$ theory on
a $G_2$ manifold $X$ with a conical singularity leads to
chiral fermions, which are necessary for construction of
realistic models of particle physics in four dimensions.
We remind the reader that charged chiral fermions are important in nature
since they are massless as long as the gauge symmetries they
are charged under are unbroken. This enables us to understand, say,
the lightness of the electron in terms of the Higgs vacuum expectation value
and the Higgs-electron Yukawa coupling.

\subsection{Hints from Anomaly-Inflow.}

The basic strategy of this subsection will be to assume there is a 
$G_2$-manifold with
a conical singularity of the above type and consider the variation of bulk 
terms in the
$M$ theory effective action under various gauge symmetries. These will be
shown to be non-zero if $Y$ obeys certain conditions. If the theory is to be 
consistent,
these anomalous variations must be cancelled and this suggests the presence of
chiral fermions in the spectrum. This is based upon \cite{wanom} which also
demonstrates that when $X$ is compact all these variations add up to zero!

The gauge symmetries we will consider
are the ones we have focussed on in this paper: the $U(1)$ gauge symmetries 
from
Kaluza-Klein reducing the $C$-field and the ${\sf ADE}$ symmetries from the
${\sf ADE}$-singularities.

We begin with the former case. We take $M$ theory on $X{\times}{\mathbb{R^{3,
1}}}$
with $X$ a cone on $Y$ so that $X=\mathbb{R}{\times}Y$. The Kaluza-Klein 
ansatz for
$C$ which gives gauge fields in four dimensions is
 \be
C   = \sum \;\;\beta^{\alpha}(x) \wedge A_{\alpha} (y)
\ee
where the $\beta$'s are harmonic 2-forms on $X$. With this ansatz, consider 
the
eleven dimensional Chern-Simons interaction
\be
S = \int_{X{\times}\mathbb{R^{3,1}}} {1 \over 6} C\wedge G\wedge G
\ee
Under a gauge transformation of $C$ under which
\be
C \longrightarrow C + d\epsilon
\ee
$S$ changes by something of
the form\footnote{We will not be too careful about factors in this section.}
\be
\delta S \sim \int_{X{\times}\mathbb{R^{3,1}}} d(\epsilon\wedge\ G\wedge G)
\ee
We can regard $X$ as a manifold with boundary $\partial X = Y$ and hence
\be
\delta S \sim \int_{Y{\times}\mathbb{R^{3,1}}} \epsilon\wedge\ G\wedge G
\ee

If we now make the Kaluza-Klein ansatz for the 2-form $\epsilon$
\be
\epsilon = \sum \epsilon^{\alpha} \beta^{\alpha}
\ee
and use our ansatz for $C$, we find
\be
\delta S \sim \int_{Y} \beta^{\rho} \wedge \beta^{\sigma} \wedge 
\beta^{\delta}
\int_{\mathbb{R^{3,1}}} \epsilon^{\rho} dA^{\sigma} \wedge dA^{\delta}
\ee
Thus if the integrals over $Y$ (which are topological) are non-zero we obtain 
a non-zero
four dimensional interaction characteristic of an anomaly in an abelian gauge 
theory.
Thus, if the theory
is to be consistent, it is natural to expect a spectrum of chiral fermions at 
the conical singularity
which exactly cancels $\delta S$.

We now turn to non-Abelian gauge anomalies. We have seen that ${\sf ADE}$ 
gauge symmetries
in $M$ theory on a $G_2$-manifold $X$ are supported along a three-manifold $W$ 
in $X$.
If additional conical singularities of $X$ are to support chiral fermions 
charges under
the ${\sf ADE}$-gauge group, then these singularities should surely also be
points $P_i$ on $W$. So let us assume that near such a point, the metric on 
$X$
assumes the conical form. In four dimensional ${\sf ADE}$ gauge theories the 
triangle
anomaly is only non-trivial for ${\sf A_n}$-gauge groups. So, we restrict 
ourselves to this case.
In this situation, there is a seven dimensional interaction of the form
\be
S = \int_{W{\times}{\mathbb{R^{3,1}}}} K \wedge {\Omega_5}(A)
\ee
where $A$ is the $SU(n)$ gauge field and
\be
d{\Omega_5}(A) = trF\wedge F\wedge F
\ee
$K$ is a two-form which is the field strength of a $U(1)$ gauge field which is 
part of the
normal bundle to $W$. $K$ measures how the ${\sf A_n}$-singularity twists 
around $W$.
The $U(1)$ gauge group is the subgroup of $SU(2)$ which commutes with ${\sf 
\Gamma_{A_{n}}}$.

Under a gauge transformation,
\be
A \longrightarrow A + D_A \lambda
\ee
and
\be
\delta S \sim  \int_{W{\times}{\mathbb{R^{3,1}}}} (K\wedge dtr\lambda F\wedge 
F)
\ee
so if $K$ is closed, $\delta S$ $=0$. This will be the case if the ${\sf 
A_n}$-singularity
is no worse at the conical singularity $P$ than at any other point on $W$. If 
however, the
${\sf A_n}$-singularity actually increases rank at $P$, then
\be
dK = 2\pi q \delta_P
\ee
and we have locally a Dirac monopole of charge $q$ at $P$. The charge is an 
integer because
of obvious quantisation conditions. In this situation we have that
\be
\delta S \sim  \int_{W{\times}{\mathbb{R^{3,1}}}}d (K\wedge tr\lambda F\wedge 
F)
=  -q\int_{\mathbb{R^{3,1}}} tr \lambda F \wedge F
\ee
which is precisely the triangle anomaly in an $SU(n)$ gauge theory. Thus, if 
we have
this sort of situation in which the ${\sf ADE}$-singularity along
$W$ degenerates further at $P$ we also expect chiral fermions to be present.

We now go on to utilise the $M$ theory heterotic duality of section 
\ref{heterotic} to
construct explicitly conically singular manifolds at which we know the 
existence
of chiral fermions.

\subsection{Chiral Fermions via Duality With The Heterotic String}

In section \ref{heterotic} we utilised duality with the heterotic string
on ${T^3}$ to learn about enhanced gauge symmetry in $M$ theory.
We applied this to $G_2$-manifolds quite successfully. In this section
we will take a similar approach. The following is based upon \cite{bsaw},
see also \cite{BBmatter,Denef}.

We start by considering duality with the heterotic string.  The
heterotic string compactified on a Calabi-Yau three-fold $Z$ can
readily give chiral fermions. On the other hand, most Calabi-Yau
manifolds participate in mirror symmetry. For $Z$ to participate
in mirror symmetry means, according to Strominger, Yau and Zaslow \cite{SYZ}
that, in a suitable limit of
its moduli space, it is a ${T^3}$ fibration (with singularities and
monodromies) over a base $W$. Then, taking the ${T^3}$'s to be
small and  using on each fiber the equivalence of the heterotic
string on ${T^3}$ with $M$ theory on $K3$, it follows that the
heterotic string on $Z$ is dual to $M$ theory on a seven-manifold
$X$ that is $K3$-fibered over $W$ (again with singularities and
monodromies). $X$ depends on the gauge bundle on $Z$.  Since the
heterotic string on $Z$ is supersymmetric, $M$ theory on $X$ is
likewise supersymmetric, and hence $X$ is a manifold of $G_2$
holonomy.

The heterotic string on $Z$ will typically have a four dimensional
spectrum of chiral fermions.
Since there are many $Z$'s that could be used in this
construction (with many possible classes of gauge bundles) it
follows that there are many manifolds of $G_2$ holonomy with
suitable singularities to give nonabelian gauge symmetry with
chiral fermions.
The same conclusion can be reached using
duality with Type IIA, as many six-dimensional Type IIA
orientifolds that give chiral fermions are dual to $M$ theory on
a $G_2$ manifold \cite{ori}.

Let us try to use this construction to determine what kind of
singularity $X$ must have.  (The reasoning and the result are very
similar to that given in  \cite{kv}
for engineering matter from Type II
singularities.  In   \cite{kv}
the Dirac equation is derived directly
rather than being motivated -- as we will -- by using duality
with the heterotic string.) Suppose that the heterotic string on
$Z$ has an unbroken gauge symmetry $G$, which we will suppose to
be simply-laced (in other words, an ${\sf A}$, ${\sf D}$, or
${\sf E}$ group) and at
level one.  This means that each $K3$ fiber of $X$ will have a
singularity of type $G$. As one moves around in $X$ one will get
a family of $G$-singularities parameterized by $W$. If $W$ is
smooth and the normal space to $W$ is a smoothly varying family of
$G$-singularities, the low energy theory will be $G$ gauge theory
on $W\times\mathbb{R^{3,1}}$ without chiral multiplets.
So chiral fermions
will have to come from singularities of $W$ or points where $W$
passes through a worse-than-orbifold singularity of $X$.

We can use the duality with the heterotic string to determine
what kind of singularities are required.  The argument will
probably be easier to follow if we begin with a specific example,
so we will consider the case of the
${\sf E_8}\times {\sf E_8}$ heterotic
string with $G=SU(5)$ a subgroup of one of the $E_8$'s.  Such a
model can very easily get chiral ${\bf 5}$'s and ${\bf 10}$'s of
$SU(5)$; we want to see how this comes about, in the region of
moduli space in which $Z$ is ${T^3}$-fibered over $W$ with small
fibers, and then we will translate this description to $M$ theory
on $X$.


\newcommand{\D}{{\cal D}}
Let us consider, for example, the ${\bf 5}$'s.  The commutant of
$SU(5)$ in ${\sf E_8}$ is a second copy of $SU(5)$, which we will
denote as $SU(5)'$.  Since $SU(5)$ is unbroken, the structure
group of the gauge bundle $E$ on $Z$ reduces from
${\sf E_8}$ to
$SU(5)'$.
Massless fermions in the heterotic string transform in the adjoint
representation of ${\sf E_8}$.
The part of the adjoint representation of ${\sf E_8}$ that
transforms as ${\bf 5}$ under $SU(5)$ transforms as ${\bf 10}$
under $SU(5)'$. So to get massless chiral ${\bf 5}$'s of $SU(5)$,
we must look at the Dirac equation $\D$ on $Z$ with values in the
${\bf 10}$ of $SU(5)'$; the zero modes of that Dirac equation
will give us the massless ${\bf 5}$'s of the unbroken $SU(5)$.

We denote the generic radius of the ${T^3}$ fibers as $\alpha$,
and we suppose that $\alpha$ is much less than the characteristic
radius of $W$.  This is the regime of validity of the argument for
duality with $M$ theory on $X$ (and the analysis of mirror
symmetry \cite{SYZ}). For small $\alpha$,
 we can solve the Dirac equation on $Z$ by first solving it along the
fiber, and then along the base. In other words, we write
$\D={\D}_T +{\D}_W$, where ${\D}_T$ is the Dirac operator along the fiber
and ${\D}_W$ is the Dirac operator along the base. The eigenvalue of
${\D}_T$ gives an effective ``mass'' term in the Dirac equation on
$W$.  For generic fibers of $Z\rightarrow W$, as we explain momentarily,
the eigenvalues of ${\D}_T$ are all nonzero and  of order
$1/\alpha$. This is much too large to be canceled by the behavior
of ${\D}_W$. So zero modes of $\D$ are localized near points in $W$
above which ${\D}_T$ has a zero mode.

When restricted to a ${T^3}$ fiber, the $SU(5)'$ bundle $E$ can be
described as a flat bundle with monodromies around the three
directions in ${T^3}$. In other words, as in section \ref{heterotic},
we have three Wilson lines on each fiber.
For generic Wilson lines, every vector in
the ${\bf 10}$ of $SU(5)'$ undergoes non-trivial ``twists'' in
going around some (or all) of the three directions in ${T^3}$.
When this is the case, the minimum eigenvalue of ${\D}_T$ is of
order $1/\alpha$. This is simply because for a generic flat gauge
field on the ${T^3}$-fiber there will be no zero mode.

A zero mode of ${\D}_T$ above some point $P\in
W$  arises precisely if for some vector  in the ${\bf 10}$, the
monodromies in the fiber are all trivial.

This means that the monodromies lie in the subgroup of $SU(5)'$
that leaves fixed that vector.  If we represent the ${\bf 10}$ by
an antisymmetric $5\times 5$ matrix $B^{ij}$, $i,j=1,\dots,5$,
then the monodromy-invariant vector corresponds to an
antisymmetric matrix $B$ that has some nonzero matrix element,
say $B^{12}$; the subgroup of $SU(5)'$ that leaves $B$ invariant
is clearly then a subgroup of $SU(2)\times SU(3)$ (where in these
coordinates, $SU(2)$ acts on $i,j=1,2$ and $SU(3)$ on
$i,j=3,4,5$).  Let us consider the case (which we will soon show
to be generic)  that $B^{12}$ is the only nonzero matrix element
of $B$.  If so, the subgroup of $SU(5)'$ that leaves $B$ fixed is
precisely $SU(2)\times SU(3)$. There is actually a distinguished
basis in this problem -- the one that diagonalizes the
monodromies near $P$ -- and it is in this basis that $B$ has only
one nonzero matrix element.

The commutant of $SU(2)\times SU(3)$ in $E_8$ is $SU(6)$.  So over
the point $P$, the monodromies commute not just with $SU(5)$ but
with $SU(6)$.  Everything of interest will happen inside this
$SU(6)$. The reason for this is that the monodromies at $P$ give
large masses to all $E_8$ modes except those in the adjoint of
$SU(6)$.   So we will formulate the rest of the discussion as if
the heterotic string gauge group were just $SU(6)$, rather than
$E_8$.  Away from $P$, the monodromies break $SU(6)$ to
$SU(5)\times U(1)$ (the global structure is actually $U(5)$).
Restricting the discussion from $E_8$ to $SU(6)$ will mean
treating the vacuum gauge bundle as a $U(1)$ bundle (the $U(1)$
being the second factor in $SU(5)\times U(1)\subset SU(6)$)
rather than an $SU(5)'$ bundle.

The fact that, over $P$, the heterotic string has unbroken
$SU(6)$ means that, in the $M$ theory description, the fiber over
$P$ has an $SU(6)$ singularity.  Likewise, the fact that away from
$P$, the heterotic string has only $SU(5)\times U(1)$ unbroken
means that the generic fiber, in the $M$ theory description, must
contain an $SU(5)$ singularity only, rather than an $SU(6)$
singularity. As for the unbroken $U(1)$, in the $M$ theory
description it must be carried by the $C$-field. Indeed, over
generic points on $W$ there is a non-zero size $S^2$ which shrinks to
zero size at $P$ in order that the gauge symmetry at that point
increases. Kaluza-Klein reducing $C$ along this $S^2$ gives a $U(1)$.

If we move away from the point $P$ in the base, the vector $B$ in
the ${\bf 10}$ of $SU(5)'$ is no longer invariant under the
monodromies.  Under parallel transport around the three
directions in ${T^3}$, it is transformed by phases $e^{2\pi
i\theta_j}$, $j=1,2,3$.  Thus, the three $\theta_j$ must all
vanish to make $B$ invariant. As $W$ is three-dimensional, we
should expect generically that the point $P$ above which the
monodromies are trivial is isolated. (Now we can see why it is
natural to consider the case that, in the basis given by the
monodromies near $P$, only one matrix element of $B$ is nonzero.
Otherwise, the monodromies could act separately on the different
matrix elements, and it would be necessary to adjust more than
three parameters to make $B$ invariant.  This would be  a less
generic situation.) We will only consider the (presumably
generic) case that $P$ is disjoint from the singularities of the
fibration $Z\rightarrow W$.
Thus, the ${T^3}$ fiber over $P$ is smooth (as
we have implicitly assumed in introducing the monodromies on
${T^3}$).

In \cite{bsaw} we explicitly solved the Dirac equation in a local
model for this situation. We found that the net number of chiral
zero modes was one. We will not have time to describe the details
of the solution here.

In summary, before we translate into the $M$ theory language,
the chiral fermions in the heterotic string theory on $Z$
are localised at points on $W$ over which the Wilson lines
in the ${T^3}$-fibers are trivial. In $M$ theory this
translates into the statement that the chiral fermions are
localised at points in $W$ over which
the ${\sf ADE}$-singularity ``worsens''.
This is also consistent with what we found in the previous section.

\subsection*{M theory Description}


So we have found a local structure in the heterotic string that
gives a net chirality -- the number of massless left-handed $
{\bf 5}$'s minus right-handed ${\bf 5}$'s -- of one. Let us see in
more detail what it corresponds to in terms of $M$ theory on a
manifold of $G_2$ holonomy.

Here it may help to review the case considered in \cite{kv}
where the
goal was geometric engineering of charged matter on a Calabi-Yau
threefold in Type IIA.  What was considered there was a Calabi-Yau
three-fold $R$, fibered by $K3$'s with a base $W'$, such that over
a distinguished point $P\in W'$ there is a singularity of type
$\hat G$, and over the generic point in $W'$ this singularity is
replaced by one of type $G$ -- the rank of $\hat G$ being one
greater than that of $G$.  In our example, $\hat G=SU(6)$ and
$G=SU(5)$. In the application to Type IIA, although $R$ also has a
Kahler metric, the focus is on the complex structure.  For $\hat
G=SU(6)$, $G=SU(5)$, let  us describe the complex structure of
$R$  near the singularities. The $SU(6)$ singularity is described
by an equation $xy=z^6$ --- cf section \ref{heterotic}.
Its ``unfolding'' depends on five
complex parameters and can be written $zy=z^6 +P_4(z)$, where
$P_4(z)$ is a quartic polynomial in $z$. If -- as in the present
problem -- we want to deform the $SU(6)$ singularity while
maintaining an $SU(5)$ singularity, then we must pick $P_4$ so
that the polynomial $z^6+P_4$ has a fifth order root.  This
determines the deformation to be
\be
xy=(z+5\epsilon)(z-\epsilon)^5,
\ee
 where we interpret
$\epsilon$ as a complex parameter on the base $W'$.  Thus,
the above equation
gives the complex structure of the total space $R$.

What is described above
is the partial unfolding of the
$SU(6)$ singularity, keeping an $SU(5)$ singularity.  In our
$G_2$ problem, we need a similar construction, but we must view
the $SU(6)$ singularity as a hyper-Kahler manifold, not just a
complex manifold.  In unfolding the $SU(6)$ singularity as a
hyper-Kahler manifold,  each complex parameter in $P_4$ is
accompanied by a real parameter that controls the area of an
exceptional divisor in the resolution/deformation of the
singularity.  The parameters are thus not five complex parameters
but five triplets of real parameters. (There is an $SO(3)$
symmetry that rotates each triplet. This is the $SO(3)$ rotating
the three kahler forms in section \ref{heterotic}.)

To get a $G_2$-manifold, we must combine the complex parameter
seen in
with a corresponding real parameter.  Altogether,
this will give a three-parameter family of deformations of the
$SU(6)$ singularity (understood as a hyper-Kahler manifold) to a
hyper-Kahler manifold with an $SU(5)$ singularity.  The parameter
space of this deformation is what we have called $W$, and the
total space is a seven-manifold that is our desired singular
$G_2$-manifold $X$, with a singularity that produces the chiral
fermions that we analyzed above in the heterotic string language.

To find the hyper-Kahler unfolding of the $SU(6)$ singularity
that preserves an $SU(5)$ singularity is not difficult, using
Kronheimer's description of the general unfolding via a
hyper-Kahler quotient  \cite{kron}
At this stage, we
might as well generalize to $SU(N)$, so we consider a hyper-Kahler
unfolding of the $SU(N+1)$ singularity to give an $SU(N)$
singularity.  The unfolding of the $SU(N+1)$ singularity is
obtained by taking a system of $N+1$ hypermultiplets
$\Phi_0,\Phi_1,\dots \Phi_N$ with an action of $K=U(1)^N$.  Under
the $i^{th}$ $U(1)$ for $i=1,\dots, N$, $\Phi_i$ has charge $1$,
$\Phi_{i-1}$ has charge $-1$, and the others are neutral. This
configuration of hypermultiplets and gauge fields is known as the
quiver diagram of $SU(N+1)$ and appears in studying $D$-branes
near the $SU(N+1)$ singularity
We
let $\mathbb{H}$ denote $\mathbb{R^4}$,
so the hypermultiplets parameterize
$\mathbb{H}^{N+1}$, the product of $N+1$ copies of $\mathbb{R^4}$.
The
hyper-Kahler quotient of ${\mathbb{H}}^{N+1}$ by $K$ is obtained by
setting the $\vec D$-field (or components of the hyper-Kahler
moment map) to zero and dividing by $K$.  It is denoted
$\mathbb{H}^{N+1}//K$, and is isomorphic to the $SU(N+1)$ singularity
$\mathbb{R^4/Z_{N+1}}$. Its unfolding is described by setting the $\vec
D$-fields equal to arbitrary constants, not necessarily zero. In
all, there are $3N$ parameters in this unfolding -- three times
the dimension of $K$ -- since for each $U(1)$, $\vec D$ has three
components, rotated by an $SO(3)$ group of $R$-symmetries.

We want a partial unfolding keeping an $SU(N)$ singularity.  To
describe this, we keep $3(N-1)$ of the parameters equal to zero
and let only the remaining three vary; these three will be simply
the values of $\vec D$ for one of the $U(1)$'s.  To carry out this
procedure, we first write $K=K'\times U(1)'$ (where $U(1)'$
denotes a chosen $U(1)$ factor of $K=U(1)^N$).  Then we take the
hyper-Kahler quotient of $\mathbb{H}^{N+1}$ by $K'$ to get a hyper-Kahler
eight-manifold  $\hat X=\mathbb{H}^{N+1}//K'$, after which we take the
{\it ordinary} quotient, not the hyper-Kahler quotient, by
$U(1)'$ to get a seven-manifold $X=\hat X/U(1)'$ that should
admit a metric of $G_2$-holonomy. $X$ has a natural map to
$W=\mathbb{R^3}$ given by the value of the $\vec D$-field of $U(1)'$ --
which was not set to zero -- and this map gives the fibration of
$X$ by hyper-Kahler manifolds.

In the present example, we can easily make this explicit.  We take
$U(1)'$ to be the ``last'' $U(1)$ in $K=U(1)^N$, so $U(1)'$ only
acts on $\Phi_{N-1}$ and $\Phi_N$.  $K'$ is therefore the product
of the first $N-1$ $U(1)$'s; it acts trivially on $\Phi_N$, and
acts on $\Phi_0,\dots,\Phi_{N-1}$ according to the standard quiver
diagram of $SU(N)$.  So the hyper-Kahler quotient $\mathbb{H}^{N+1}//K'$
is just $(\mathbb{H}^N//K')\times \mathbb{H}'$, where $\mathbb{H}^N//K'$ is 
the $SU(N)$
singularity, isomorphic to $\mathbb{H}/\mathbb{Z_N}$, and $\mathbb{H}'$ is 
parameterized
by $\Phi_N$. So finally, $X$ will be
$(\mathbb{H}/\mathbb{Z_N}\times \mathbb{H}')/U(1)'$.
To make this completely explicit, we just need to identify the
group actions on $\mathbb{H}$ and $\mathbb{H}'$. If we parameterize 
$\mathbb{H}$ and
$\mathbb{H}'$ respectively by pairs of complex variables $(a,b)$ and $(a',b')$
then the $\mathbb{Z_N}$ action on $\mathbb{H}$, such that the
quotient $\mathbb{H}/\mathbb{Z_N}$ is the $SU(N)$ singularity,  is given by
\ba
\left( \begin{array}{c}a \\
b\end{array}\right)\rightarrow \left(\begin{array}{c}e^{2\pi ik/N}a \\
e^{-2\pi ik/N}b\end{array}\right)
\ea
while the $U(1)'$ action that commutes with this (and
preserves the hyper-Kahler structure)
is
\ba
\left(\begin{array}{c}a \\
b\end{array}\right)\rightarrow\left(\begin{array}{c}e^{i\psi/N}a \\
e^{-i\psi/N}b\end{array}\right)
\ea
The $U(1)'$ action on $\mathbb{H}'$ is similarly
\ba
\left(\begin{array}{c} a' \\
b' \end{array}\right)\rightarrow\left(\begin{array}{c} e^{i\psi/N}a' \\
e^{-i\psi/N}b' \end{array}\right)
\ea
In all,  if we set $\lambda=e^{i\psi/N}$,
$w_1=\overline a',$  $w_2= b'$, $w_3=a,$ $w_4=\bar b$, then the
quotient $(\mathbb{H}/\mathbb{Z_N}\times \mathbb{H}')/U(1)$ can be described 
with four
complex variables $w_1,\dots , w_4$ modulo the equivalence
\be
(w_1,w_2,w_3,w_4)\rightarrow
(\lambda^Nw_1,\lambda^Nw_2,\lambda w_3,\lambda
w_4),\;\;\;|\lambda|=1
\ee
This quotient is a cone on a weighted
projective space ${\bf \mathbb{WC}P}^3_{N,N,1,1}$.
In fact, if we impose the above equivalence relation for all nonzero complex
$\lambda$, we would get the weighted projective space itself; by
imposing this relation  only for $|\lambda|=1$, we get a cone on
the weighted projective space. Note, that the conical metric of
$G_2$-holonomy on this space does not use usual Kahler metric on
weighted projective space.

${\bf \mathbb{WC}P}^3_{N,N,1,1}$ has a family
of ${\sf A_{N-1}}$-singularities
at points  $(w_1 ,w_2 , 0,0)$. This is easily seen by
setting $\lambda$ to $e^{2\pi i/N}$. This set of points is a copy of
${\bf \mathbb{C}P}^1$ $=$ ${\bf S^2}$. Our proposed $G_2$-manifold is
a cone over weighted projective space, so it has a family of
${\sf A_{N-1}}$-singularities which are a cone over this ${\bf S^2}$.
This is of course a copy of $\mathbb{R^3}$. Away from the origin in
$\mathbb{R^3}$ the only singularities are these orbifold singularities.
At the origin however, the whole manifold develops a conical singularity.
There, the 2-sphere, which is incontractible in the bulk of the manifold,
shrinks to zero size.
This is in keeping with the anomaly inflow arguments of the previous
section. There we learned that an ${\sf ADE}$-singularity which worsens
over a point in $W$ is a good candidate for the appearance of chiral fermions.
Here, via duality with the heterotic string, we find that the conical
singularity in this example supports one chiral fermion in the ${\bf N}$
of the $SU(N)$ gauge symmetry coming from the ${\sf A_{N-1}}$-singularity.
In fact, the $U(1)$ gauge symmetry from the $C$-field in this example,
combines with the $SU(N)$ to give a gauge group which is globally
$U(N)$ and the fermion is in the fundamental representation.


Some extensions of this can be worked out in a similar fashion.
Consider the case that away from $P$, the
monodromies break $SU(N+1)$ to $SU(p)\times SU(q)\times U(1)$,
where $p+q=N+1$.  Analysis of the Dirac equation along the above lines
 shows that such a
model will give chiral fermions transforming as $({\bf
p},\bar{\bf q})$ under $SU(p)\times SU(q)$ (and charged under the
$U(1)$).  To describe a dual in  $M$ theory on a manifold of $G_2$
holonomy, we let $K=K'\times U(1)'$, where now $K'=K_1\times K_2$,
$K_1$ being the product of the first $p-1$ $U(1)$'s in $K$ and
$K_2$ the product of the last $q-1$, while $U(1)'$ is the $p^{th}$
$U(1)$. Now we must define $\hat X=\mathbb{H}^{N+1}//K'$, and the manifold
admitting a metric of $G_2$ holonomy should be  $\hat X/U(1)'$.

We can compute $\hat X$ easily, since $K_1$ acts only on
$\Phi_1,\dots,\Phi_p$ and $K_2$ only on
$\Phi_{p+1},\dots,\Phi_{N+1}$.  The hyper-Kahler quotients by
$K_1$ and $K_2$ thus simply construct the $SU(p)$ and $SU(q)$
singularities, and hence $\hat X=\mathbb{H}/\mathbb{Z_p}\times \mathbb{H}/
\mathbb{Z_q}$. $\hat X$
has planes of $\mathbb{Z_p}$ and $\mathbb{Z_q}$
singularities, which will persist
in $X=\hat X/U(1)'$, which will also have a more severe
singularity at the origin. So the model describes a theory with
$SU(p)\times SU(q)$ gauge theory and chiral fermions supported at
the origin.
 $U(1)'$ acts on $\mathbb{H}/\mathbb{Z_p}$ and $\mathbb{H}/\mathbb{Z_q}$ as 
the
familiar global symmetry that preserves the hyper-Kahler
structure of the $SU(p)$ and $SU(q)$ singularities. Representing
those singularities by pairs $(a,b)$
and $(a',b')$ modulo the usual action of
$\mathbb{Z_p}$ and $\mathbb{Z_q}$, $U(1)'$ acts by
\ba
\left(\begin{array}{c} a \\ b\end{array}\right)
\rightarrow\left(\begin{array}{c} e^{i\psi/p}a \\ e^{-i\psi/p}b\end{array}
\right)\;\;\;and\;\;\;
\left(\begin{array}{c} a' \\
b' \end{array}\right)\rightarrow\left(\begin{array}{c} e^{-i\psi/q}a' \\
e^{i\psi/q}b' \end{array}\right)
\ea

Now if $p$ and $q$ are relatively
prime, we set  $\lambda=e^{i\psi/pq}$, and we find that the
$U(1)'$ action on the complex coordinates $w_1,\dots,w_4$ (which
are defined in terms of $a,b,a',b'$ by the same formulas as
before) is
\be
(w_1,w_2,w_3,w_4)\rightarrow (\lambda^p
w_1,\lambda^p w_2,\lambda^q w_3,\lambda^q w_4).
\ee
If $p$ and $q$
are relatively prime, then the $U(1)'$ action, upon taking
$\lambda$ to be a $p^{th}$ or $q^{th}$ root of 1, generates the
$\mathbb{Z_p}{\times} \mathbb{Z_q}$ orbifolding that is part of the original
definition of $\hat X$. Hence in forming the quotient $\hat
X/U(1)'$, we need only to act on the $w$'s by the equivalence
relation.  The quotient is therefore a cone on a weighted
projective space ${\bf \mathbb{WC}P}^3_{p,p,q,q}$. If $p$ and $q$ are not
relatively prime, we let $(p,q)=r(n,m)$ where $r$ is the greatest
common divisor and $n$ and $m$ are relatively prime.  Then we let
$\lambda=\exp(i r\psi/pq)$, so the equivalence relation above
is replaced with
\be
(w_1,w_2,w_3,w_4)\rightarrow (\lambda^n w_1,\lambda^n
w_2,\lambda^m w_3,\lambda^m w_4)
\ee
To reproduce $\hat X/U(1)$ we must now also divide
by $\mathbb{Z_r}$,
acting by
\be
(w_1,w_2,w_3,w_4)\rightarrow (\zeta w_1,\zeta
w_2,w_3,w_4),
\ee
where $\zeta^r=1$.
So $X$ is a cone on
${\bf \mathbb{WC}P}^3_{n,n,m,m}/{\mathbb{Z}_r}$.

\subsection{Other Gauge Groups and Matter Representations}

We now explain how to generalise the above construction to obtain
singularities with more general gauge groups and chiral fermion
representations.
Suppose that we want to get chiral fermions in the representation
$R$ of a simply-laced group $G$.   This can be achieved for
certain representations.  We find a simply-laced group $\hat G$ of
rank one more than the rank of $G$, such that $\hat G$ contains
$G\times U(1)$ and the Lie algebra of $\hat G$ decomposes as
${\bf g}\oplus {\bf  o} \oplus {\bf r}\oplus \bar {\bf r}$, where
${\bf g}$ and ${\bf o}$ are the Lie algebras of $G$ and $U(1)$,
${\bf r}$ transforms as $R$ under $G$ and of charge 1 under
$U(1)$, and $\bar{\bf r}$ transforms as the complex conjugate.
Such a  $\hat G$ exists only for special  $R$'s, and these are
the $R$'s that we will generate from $G_2$ singularities.

Given $\hat G$, we proceed as  above on the heterotic string side.
We consider a family of ${\bf T^3}$'s, parameterized by $W$, with
monodromy that at a special point $P\in W$ leaves unbroken $\hat
G$, and at a generic point breaks $\hat G$ to $G\times U(1)$.  We
moreover assume that near $P$, the monodromies have the same sort
of generic behavior assumed above.  Then the same computation as
above will show that the heterotic string has, in this situation,
a single multiplet of fermion zero modes (the actual chirality depends
on the solving the Dirac equation)
in the representation $R$, with
$U(1)$ charge 1.

Dualizing this to an $M$ theory description, over $P$ we want a
$\hat G$ singularity, while over a generic point in $W$, we
should have a $G$ singularity.  Thus, we want to consider the
unfolding of the $\hat G$ singularity (as a hyper-Kahler
manifold) that preserves a $G$ singularity.  To do this is quite
simple.  We start with the Dynkin diagram of $\hat G$.  The
vertices are labeled with integers $n_i$, the Dynkin indices.  In
Kronheimer's construction, the $\hat G$ singularity is obtained
as the hyper-Kahler quotient of $\mathbb{H}^k$ (for some $k$) by the
action of a group $K=\prod_i U(n_i)$. Its unfolding is obtained
by allowing the $\vec D$-fields of the $U(1)$ factors (the
centers of the $U(n_i)$) to vary.

The $G$ Dynkin diagram is obtained from that of $\hat G$ by
omitting one node, corresponding to one of the $U(n_i)$ groups;
we write the center of this group as $U(1)'$.  Then we write $K$
(locally) as $K=K'\times U(1)'$, where $K'$ is defined by
replacing the relevant $U(n_i)$ by  $SU(n_i)$. We get a
hyper-Kahler eight-manifold as the hyper-Kahler quotient $\hat
X=\mathbb{H}^k//K'$, and then we get a seven-manifold $X$ by taking the
{\it ordinary} quotient $X=\hat X/U(1)'$.  This maps to $W=\mathbb{R^3}$
by taking the value of the $U(1)'$ $\vec D$-field, which was not
set to zero. The fiber over the origin is obtained by setting
this $\vec D$-field to zero after all, and gives the original
$\hat G$ singularity, while the generic fiber has a singularity
of type $G$.

One can readily work out examples of pairs $G, $
$\hat G$. We will just consider the cases most relevant for grand
unification. For $G=SU(N)$, to get chiral fields in the
antisymmetric tensor representation, $\hat G$ should be $SO(2N)$.
For $G=SO(10)$, to get chiral fields in the ${\bf 16}$, $\hat G$
should be $E_6$. For $G=SO(2k)$, to get chiral fields in the
${\bf 2k}$,  $\hat G$ should be $SO(2k+2)$. (Note in this case
that ${\bf 2k}$ is a real representation. However, the
monodromies in the above construction break $SO(2k+2)$ to
$SO(2k)\times U(1)$, and the massless ${\bf 2k}$'s obtained from
the construction are charged under the $U(1)$;  under
$SO(2k)\times U(1)$ the representation is complex.) For $2k=10$,
this example might be used in constructing $SO(10)$ GUT's.   For
$G=E_6$, to get ${\bf 27}$'s, $\hat G$ should be $E_7$.  A useful
way to describe the topology of $X$ in these examples is not
clear.

In this construction, we emphasized, on the heterotic string
side, the most generic special monodromies that give enhanced
gauge symmetry, which corresponds on the $M$ theory side to
omitting from the hyper-Kahler quotient a rather special $U(1)$
that is related to a single node of the Dynkin diagram.  We could
also consider more general heterotic string monodromies; this
would correspond in $M$ theory to omitting a more general linear
combination of the $U(1)$'s. For a more detailed discussion of some
of these examples see \cite{BBmatter}.

\newpage
\section{Topology Change in $M$ theory on Exceptional Holonomy Manifolds}
\label{transitions}

\subsection{Topology Change in $M$ theory}

In this section we discuss topology changing transitions,
by which we mean a particular behavior of the manifold $X$
(and the associated physics) in the singular limit
when one can go to a space with a different topology.
In Calabi-Yau manifolds many examples of such transitions
are known and can be understood using conformal field theory
methods, see {\it e.g.} \cite{AGM} and references therein.
Some of these transitions give rise to analogous
topology changing transitions in $G_2$ and $Spin(7)$ manifolds
obtained from finite quotients of Calabi-Yau spaces that
we discussed in section \ref{compact}.
In the context of compact manifolds with $G_2$ holonomy
this was discussed in \cite{joyce,Pioline,Lukas}.
One typically finds a transition between different
topologies of $X$, such that the following sum
of the Betti numbers remains invariant
\begin{equation}
\label{g2mirror}
b_2 + b_3 = {\rm const}
\end{equation}
which is precisely what one would expect from
the conformal field theory considerations \cite{Shatashvili}.
One can interpret the condition (\ref{g2mirror})
as a feature of the mirror phenomenon for $G_2$ manifolds
\cite{Shatashvili,BSAmirror,LeeLeung,GYZ}.

Here, we shall discuss topology change in the non-compact
asymptotically conical $G_2$ and $Spin(7)$ manifolds
of section \ref{noncompact}, which are in a sense
the most basic examples of singularities that reveal
features particular to exceptional holonomy.
Notice, that all of these manifolds have the form:
$$
X \cong B \times ({\rm contractible})
$$
where $B$ is a non-trivial cycle (a bolt),
{\it e.g.} $B = {\bf S}^3$, ${\bf S}^4$,
${\bf \mathbb{C}P}^2$, or something else.
Therefore, it is natural to ask:
\begin{center}
{\it ``What happens if ${\rm Vol}(B) \to 0$ ?''}
\end{center}

\noindent
In this limit the geometry becomes singular and,
as we discussed earlier, there are many possibilities
for $M$ theory dynamics associated with it.
One possibility is a topology change, which is
indeed what we shall find in some of the cases below.

Although we are mostly interested in exceptional holonomy manifolds,
it is instructive to start with topology changing transitions
in Calabi-Yau manifolds, where one finds two prototypical examples:

\begin{figure}
\setlength{\unitlength}{0.9em}
\begin{center}
\begin{picture}(25,10)

\put(10,2){\line(1,0){6}}
\put(16,2){\line(1,1){2}}
\put(10,2){\line(1,2){3}}
\put(16,2){\line(-1,2){3}}
\put(18,4){\line(-5,4){5}}
\put(13,0.8){${\bf S}^3$}
\put(17,2){${\bf S}^2$}
\put(11,-1){$m_{q}=0$}
\put(12,10){conifold}
\put(12,9){singularity}

\put(6,6){$\Longleftarrow$}
\put(5,7){deformation}

\put(-2,2){\line(1,0){6}}
\put(4,2){\line(1,1){2}}
\put(-2,2){\line(1,3){2}}
\put(4,2){\line(-1,3){2}}
\put(0,8){\line(1,0){2}}
\put(6,4){\line(-1,1){4}}
\put(1,0.8){${\bf S}^3$}
\put(5,2){${\bf S}^2$}
\put(-1,-1){$m_{q} \ne 0$}
\put(1,9){${\bf S}^3$}

\put(21,6){$\Longrightarrow$}
\put(19,7){resolution}

\put(22,2){\line(1,0){6}}
\put(28,2){\line(1,1){2}}
\put(22,2){\line(3,5){3}}
\put(28,2){\line(-3,5){3}}
\put(25,7){\line(1,1){2}}
\put(30,4){\line(-3,5){3}}
\put(25,0.8){${\bf S}^3$}
\put(29,2){${\bf S}^2$}
\put(23,-1){$\langle q \rangle \ne 0$}
\put(24.5,8.5){${\bf S}^2$}

\end{picture}\end{center}
\caption{Conifold transition in type IIB string theory.}
\label{figa}
\end{figure}
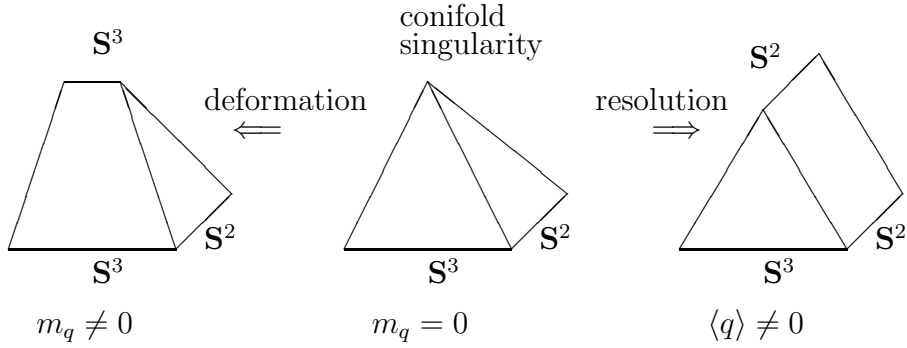

{\bf The Flop} is a transition between two geometries,
where one two-cycle shrinks to a point and
a (topologically) different two-cycle grows.
This process can be schematically described by the diagram:
$$
{\bf S}^2_{(1)} \longrightarrow \cdot \longrightarrow {\bf S}^2_{(2)}
$$
This transition is smooth in string theory \cite{greene,witten}.

{\bf The Conifold} transition is another type of topology change,
in which a three-cycle shrinks and is replaced by a two-cycle:
$$
{\bf S}^3 \longrightarrow \cdot \longrightarrow {\bf S}^2
$$
Ulike the flop, it is a real phase transition in
the low-energy dynamics which can be understood
as the condensation of massless black holes \cite{Strominger,GMS}.
Let us briefly recall the main arguments.

As the name indicates,
the conifold is a cone over a five dimensional base space. The base
has topology ${\bf S}^2 \times {\bf S}^3$ (see Figure \ref{figa}).
Two different ways to desingularize this space --- called
the deformation and the resolution --- correspond to replacing the
singularity by a finite size ${\bf S}^3$ or ${\bf S}^2$, respectively.
Thus, we have two different spaces, with topology
${\bf S}^3 \times \mathbb{R}^3$ and ${\bf S}^2 \times \mathbb{R}^4$,
which asymptotically look the same.

In type IIB string theory, the two phases of the
conifold geometry correspond to different
branches in the four-dimensional
$\mathcal{N}=2$ low-energy effective field theory.
In the deformed conifold phase, D3-branes wrapped around the
3-sphere give rise to a low-energy field $q$, with mass determined
by the size of the ${\bf S}^3$.
In the effective four-dimensional supergravity theory these
states appear as heavy, point-like, extremal black holes.
On the other hand, in the resolved conifold phase the field $q$
acquires an expectation value reflecting the condensation of
these black holes.
Of course, in order to make the transition from one phase
to the other, the field $q$ must become massless somewhere and this
happens at the conifold singularity, as illustrated
in Figure \ref{figa}.

Now, let us proceed to topology change in $G_2$ manifolds.
%
%
Here, again, one finds two kinds of topology changing transitions,
which resemble the flop and the conifold transitions in Calabi-Yau
manifolds:

\begin{figure}
\begin{center}
\epsfxsize=3in\leavevmode\epsfbox{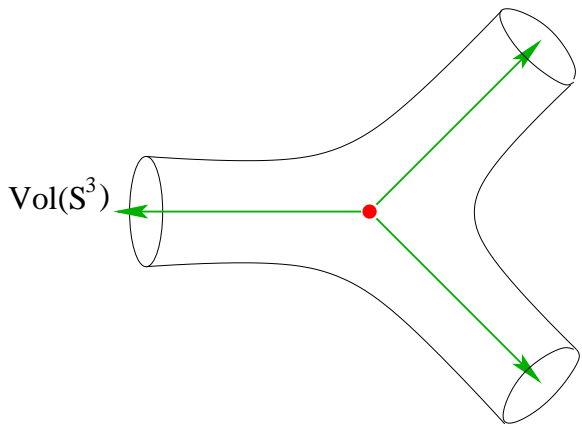}
\end{center}
\caption{Quantum moduli space of $M$ theory on $G_2$ manifold $X$
with topology ${\bf S}^3 \times \mathbb{R}^4$.
Green lines represent the `geometric moduli space' parametrized
by the volume of the ${\bf S}^3$ cycle, which is enlarged to
a smooth complex curve by taking into account $C$-field and
quantum effects. The resulting moduli space has three classical
limits, which can be connected without passing through
the point where geometry becomes singular
(represented by red dot in this picture).}
\label{gtwosthree}
\end{figure}

{\bf The $G_2$ Flop} is a transition where a 3-cycle collapses and
gets replaced by a (topologically) distinct 3-cycle:
$$
{\bf S}^3_{(1)} \longrightarrow \cdot \longrightarrow {\bf S}^3_{(2)}
$$
Note, that this is indeed very similar to the flop transition
in Calabi-Yau manifolds, where instead of a 2-cycle we have
a 3-cycle shrinking. The physics is also similar, with membranes
playing the role of string world-sheet instantons. Remember,
that the latter were crucial for the flop transition to be smooth
in string theory. For a very similar reason, the $G_2$ flop
transition is smooth in $M$ theory.
This was first prpoposed by Atiyah, Maldacena, and Vafa \cite{AMV}
(see also Acharya \cite{bsa2}),
for a 7-manifold with topology
\be
X \cong {\bf S}^3 \times \mathbb{R}^4
\label{gtwostop}
\ee
and studied further by Atiyah and Witten \cite{AW}.
In particular, they found that $M$ theory on $X$
has three classical branches, related by a triality
permutation symmetry, so that the quantum moduli
space looks as shown on Figure \ref{gtwosthree}.
Once again, the important point is that
there is no singularity in quantum theory.

If we denote the space $X$ on the $i$-th branch as $X_i$,
then each $X_i$ can be constructed as a cone over
$Y = {\bf S}^3 \times {\bf S}^3$ described in the example
of section \ref{noncompact}. Namely, as in (\ref{ghkexamp}),
we can view $Y$ as a homogeneous space $G/K$,
where $G = SU(2)^3$ and $K=SU(2)$ is its diagonal subgroup.
This description is particularly convenient because it makes all
the symmetries of $M$ theory on $X_i$ very explicit and easy to see.
Specifically, the space $Y$ can be described in terms
of three $SU(2)$ elements, $(g_1,g_2,g_3) \in G$,
with the following equivalence relation
\be
(g_1,g_2,g_3) = (g_1 h,g_2 h,g_3 h)
\label{gggrel}
\ee
Then, the seven-manifold $X_i$ is obtained by ``filling in'''
the $i$-th copy of $SU(2)$. From construction, it is clear
that the resulting manifolds $X_i$ are permuted by
the triality symmetry, and have the topology (\ref{gtwostop}).
Similar to (\ref{gggrel}), the homology of $Y$ is generated by
three 3-cycles $D_i$, subject to a linear relation
\be
D_1 + D_2 + D_3 = 0
\label{dddrel}
\ee
where the 3-cycle $D_i$ is obtained by projecting
the $i$-th copy of $SU(2)$ in $Y=SU(2)^3 / SU(2)$.
The homology of $X_i$ is obtained by imposing a further
condition, $D_i=0$, which indicates that the $i$-th
copy of $SU(2)$ is contractible in $X$.
Therefore, on the $i$-th branch we have $D_i=0$
and $D_{i+1} = - D_{i-1}$, where $i \in \mathbb{Z}$ mod 3.

Let us proceed to another kind of topology changing
transition in manifolds with $G_2$ holonomy.

\begin{figure}
\begin{center}
\epsfxsize=3in\leavevmode\epsfbox{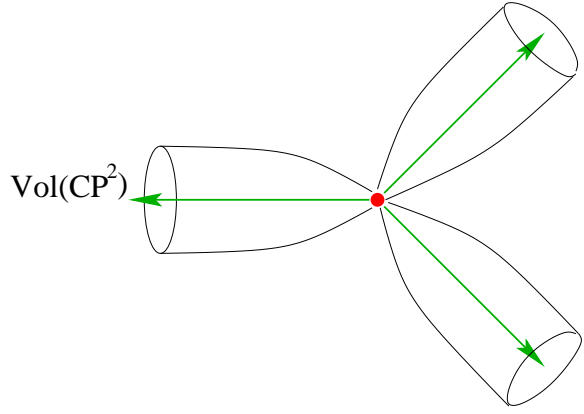}
\end{center}
\caption{Quantum moduli space of $M$ theory on $G_2$ manifold $X$
with topology $\mathbb{C}{\bf P}^2 \times \mathbb{R}^3$.
Green lines represent the `geometric moduli space' parametrized
by the volume of the $\mathbb{C}{\bf P}^2$ cycle, which is enlarged to
a {\it singular} complex curve by taking into account $C$-field and
quantum effects. The resulting moduli space has three classical
limits. In order to go from one branch to another one necessarily
has to pass through the point where geometry becomes singular
(represented by red dot in this picture).}
\label{gtwocptwo}
\end{figure}

{\bf A Phase Transition}
can be found in $M$ theory on a $G_2$ manifold with topology
$$
X \cong {\bf \mathbb{C}P}^2 \times \mathbb{R}^3
$$
A singularity develops when the ${\bf \mathbb{C}P}^2$ cycle shrinks.
As in the conifold transition, the physics of $M$ theory on this space
also becomes singular at this point, in a sense that certain
parameters jump discontinuously even in quantum theory.
Hence, this is a genuine phase transition \cite{AW}.
Note, however, that unlike the conifold transition in type IIB
string theory, this phase transition is not associated with
condensation of any particle-like states
in $M$ theory\footnote{However, such interpretation can be given
in type IIA string theory \cite{GTS}.} on $X$.
Indeed, there are no 4-branes in $M$ theory, which could result in
particle-like objects by wrapping around the collapsing
${\bf \mathbb{C}P}^2$ cycle.

Like the $G_2$ flop transition,
this phase transition has three classical branches, which
are related by triality symmetry, see Figure \ref{gtwocptwo}.
The important difference, of course, is that now one can
go from one branch to another only through the singular point.
In this transition one ${\bf \mathbb{C}P}^2$ cycle shrinks and
another (topologically different) ${\bf \mathbb{C}P}^2$ cycle grows:
$$
{\bf \mathbb{C}P}^2_{(1)} \longrightarrow \cdot \longrightarrow
{\bf \mathbb{C}P}^2_{(2)}
$$
Moreover, the global symmetry of the theory changes discontinuously
as we go through the singular point.
Specifically, the singular point is characterised by
a global $U(1) \times U(1)$ symmetry, which is broken
to different $U(1)$ subgroups on each of the three branches.
The generators of these subgroups are permuted by
the triality symmetry and add up to zero.

It was proposed in \cite{AW}, that the effective
$\mathcal{N}=1$ physics of $M$ theory on $X$ can be described
by a model with three chiral multiplets $\Phi_i$, $i=1,2,3$,
with the superpotential
\be
\mathcal{W} = \Phi_1 \Phi_2 \Phi_3
\label{wcptwo}
\ee
Extremizing the superpotential $\mathcal{W}$, it is easy to see
that the space of vacua in this theory indeed consists of three branches,
so that the $i$-th branch is parameterised by the field $\Phi_i$.
The three branches meet at the origin, $\Phi_1=\Phi_2=\Phi_3=0$,
where one finds the global $U(1) \times U(1)$ symmetry,
which acts on the chiral fields as $\Phi_i \to e^{i \theta_i} \Phi_i$
with $\theta_1 + \theta_2 + \theta_3 =0$.
On the $i$-th branch this global symmetry is spontaneously broken
to a $U(1)$ subgroup, characterized by $\theta_i =0$.

One way to see that this is indeed the right physics of
$M$ theory on $X$ is to reduce it to type IIA theory with
D6-branes in flat space-time \cite{AW}.
As a result, one finds precisely the configuration of
three intersecting D6-branes described in the example
of section \ref{typeiia}.
Then, the three branches of $M$ theory on $X$ can be easily
identified with three different deformations of the brane
configuration shown of Figure \ref{3d6s},
and the superpotential (\ref{wcptwo})
is the effect of the string world-sheet disk instantons.

Finally, we come to the last and most difficult of the
holonomy groups: $Spin(7)$.
%

\begin{figure}
\setlength{\unitlength}{0.9em}
\begin{center}
\begin{picture}(25,11)

\qbezier(10,3)(13,2)(16,3)
\put(10,3){\line(1,2){3}}
\put(16,3){\line(-1,2){3}}
\put(11,1.1){$SU(3)/U(1)$}
\put(11,-1){$m_{q}=0$}

\qbezier(0,3)(3,2)(6,3)
\put(0,3){\line(2,5){2}}
\put(6,3){\line(-2,5){2}}
\put(2,8){\line(1,0){2}}
\put(0,-1){$m_{q} \ne 0$}
\put(2.5,8.3){${\bf S}^5$}

\qbezier(20,3)(23,2)(26,3)
\put(20,3){\line(2,5){2}}
\put(26,3){\line(-2,5){2}}
\put(22,8){\line(1,0){2}}
\put(23,-1){$\langle q \rangle \ne 0$}
\put(22,8.3){${\bf \mathbb{C}P}^2$}

\put(7,6){$\Longleftarrow$}
\put(6,8){first}
\put(6,7){resolution}
\put(17,6){$\Longrightarrow$}
\put(16,8){second}
\put(16,7){resolution}

\end{picture}\end{center}
\caption{Conifold transition in $M$ theory on a manifold
with $Spin(7)$ holonomy.}
\label{figb}
\end{figure}
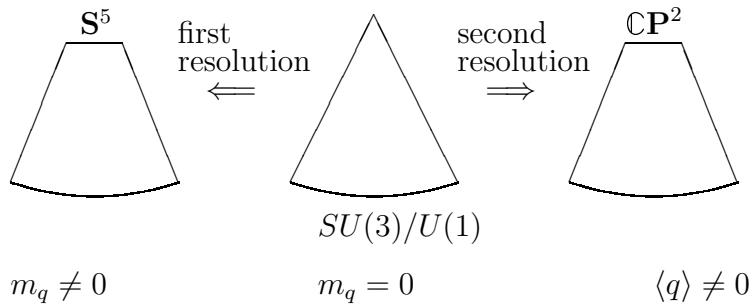

The {\bf $Spin(7)$ Conifold} is the cone on $SU(3)/U(1)$.
It was conjectured in \cite{GTS} that the effective dynamics
of $M$ theory on the $Spin(7)$ conifold is analogous to
that of type IIB string theory on the usual conifold.
Namely, the $Spin(7)$ cone on $SU(3)/U(1)$
has two different desingularizations,
obtained by replacing the conical singularity
with {\it either} a 5-sphere {\it or} with
a $\mathbb{C}{\bf P}^2$, see Figure \ref{figb}.
As a result, we obtain two different $Spin(7)$ manifolds,
with topology
$$
\mathbb{C}{\bf P}^2 \times \mathbb{R}^4
\quad {\rm and} \quad
{\bf S}^5 \times \mathbb{R}^3
$$
which are connected via the topology changing transition
$$
\mathbb{C}{\bf P}^2
\longrightarrow \cdot \longrightarrow
{\bf S}^5
$$
Like the conifold transition in string theory \cite{Strominger,GMS},
the $Spin(7)$ conifold in $M$ theory
has a nice interpretation in terms of
the condensation of branes.
Namely, in the ${\bf S}^5$ phase we have extra massive states
obtained upon quantization of the $M$5-brane wrapped
around the five-sphere. The mass of these states is related
to the volume of the ${\bf S}^5$.
At the conifold point where the five-sphere shrinks,
these $M$5-branes become massless as suggested by the classical geometry.
At this point, the theory may pass through a phase transition into
the Higgs phase, associated with the condensation of these five-brane
states, see Figure \ref{figb}.

\begin{figure}
\begin{center}
\epsfxsize=5.7in\leavevmode\epsfbox{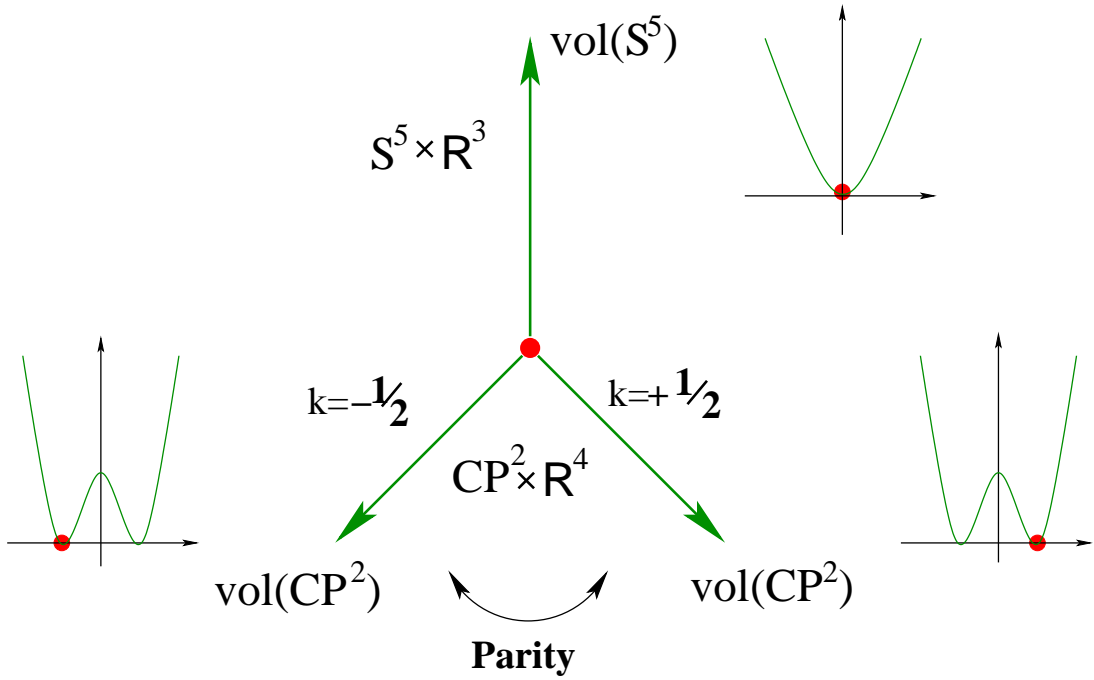}
\end{center}
\caption{The moduli space of $M$ theory on the $Spin(7)$
cone over $SU(3)/U(1)$ can be compared to the vacuum structure
of a system with spontaneous symmetry breaking.
On this picture, the G-flux is measured by
$k=\int_{\mathbb{C}{\bf P}^2} G/2\pi$.}
\label{lastfig}
\end{figure}

To continue the analogy with the Calabi-Yau conifold, recall that the
moduli space of type II string theory on the Calabi-Yau
conifold has three semi-classical regimes.
The deformed
conifold provides one of these, while there are two
large-volume limits of the resolved conifold, related
to each other by a flop transition.
In fact, the same picture emerges for the $Spin(7)$ conifold.
In this case, however, the two backgrounds
differ not in geometry, but in the $G$-flux. It was shown in
\cite{GS} that, due to the membrane anomaly of \cite{witten1},
$M$ theory on $X \cong {\bf \mathbb{C}P}^2 \times \mathbb{R}^4$
is consistent only for half-integral units of $G$-flux through
the $\mathbb{C}{\bf P}^2$ bolt.
Namely, after
the transition from $X \cong {\bf S}^5 \times \mathbb{R}^3$,
the $G$-flux may take the values $\pm 1/2$,
with the two possibilities related by
a parity transformation \cite{GTS}.
Thus,
the moduli space of $M$ theory on the $Spin(7)$ cone over $SU(3)/U(1)$
also has three semi-classical limits: one with the parity invariant
background geometry ${\bf S}^5 \times \mathbb{R}^3$,
and two with the background geometry
$X \cong {\bf \mathbb{C}P}^2 \times \mathbb{R}^4$
where parity is spontaneously broken, see Figure \ref{lastfig}.
The last two limits are mapped into each other under the
parity transformation.

This picture is reproduced in the effective low-energy theory
if we include in the spectrum light states corresponding to
$M$5-branes wrapped over the five-sphere:
\begin{center}
{\bf Effective Theory:}
${\cal N}=1$, $D=3$ Maxwell-Chern-Simons theory
with one charged complex scalar multiplet $q$
\end{center}
Here, it is the Higgs field $q$ that appears due
to quantizaion of the $M$5-branes.
In this theory, different topological phases
correspond to the Coulomb and Higgs branches:
\begin{eqnarray}
{\bf S}^5 \times \mathbb{R}^3 & \iff & {\rm ~Coulomb~ branch~} \\
\mathbb{C}{\bf P}^2 \times \mathbb{R}^4 & \iff & {\rm ~Higgs~ branch~}
\end{eqnarray}
Further agreement in favor of this identification
arises from examining the various extended
objects that exist in $M$ theory on
$\mathbb{C}{\bf P}^2 \times \mathbb{R}^4$,
obtained from wrapped $M$5 or $M$2-branes.
For example, we can consider an $M$2-brane over
$\mathbb{C}{\bf P}^1$ inside
$\mathbb{C}{\bf P}^2 \times \mathbb{R}^4$.
This non-BPS state has a semi-classical mass proportional
to the volume of $\mathbb{C}{\bf P}^1$,
and is electrically charged under the global
$U(1)_J$ symmetry of our gauge theory.
Therefore, this state can be naturally identified
with a vortex.
Note, that this state can be found only in the $\mathbb{C}{\bf P}^2$
phase ({\it i.e.} in the Higgs phase), in complete agreement
with the low-energy physics.

In view of the interesting phenomena associated to branes in the
conifold geometry, and their relationship to the conifold transition
\cite{klebwitt,KStrassler}, it would be interesting
to learn more about the $Spin(7)$ transition using membrane probes in
this background, and also to study the
corresponding holographic renormalization group flows. For work in this
area, see \cite{gt,oz,carlos}.


\subsection{Relation to Geometric Transition}

In the previous section we described the basic examples of
topology changing transitions in exceptional holonomy manifolds,
and commented on the important aspects of $M$ theory dynamics
in these transitions. As we explain in this section, some
of these transitions also have a nice interpretation in
type IIA string theory, realizing dualities between
backgrounds involving D6-branes and Ramond-Ramond fluxes
in manifolds with more restricted holonomy.
Specifically, we will consider two cases:

$\bullet~~{\bf SU(3) \to G_2:}$
We start with a relation between the conifold
transition in the presence of extra fluxes and branes
and the $G_2$ flop transition in $M$ theory \cite{bsa2,AMV}.
Note, that these two transitions are associated with
different holonomy groups\footnote{That is why we refer to
this case as $SU(3) \to G_2$.} and, in particular,
with different amount of unbroken supersymmetry.
The relation, however, appears when we introduce extra
matter fields, represented either by D-branes or by fluxes.
They break supersymmetry further, therefore, providing
a relation between two different holonomy groups.

\begin{figure}
\begin{center}
\epsfxsize=5in\leavevmode\epsfbox{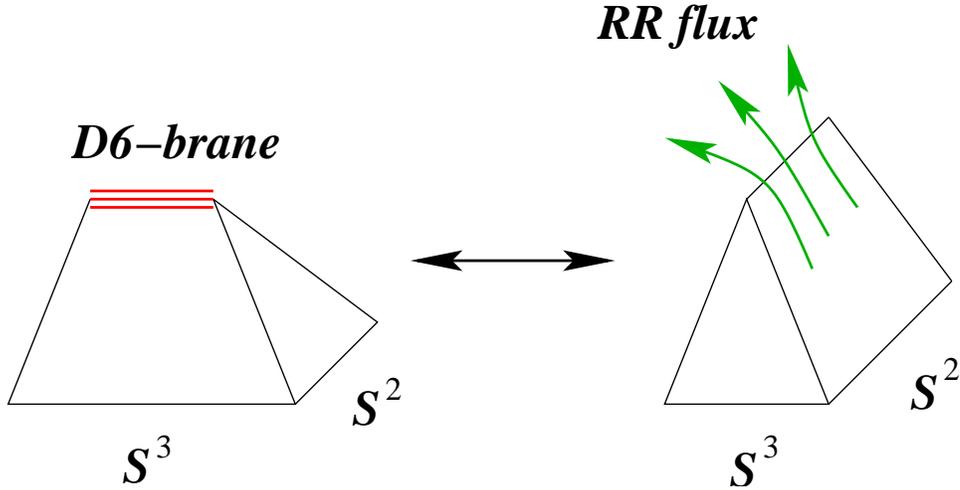}
\end{center}
\caption{Geometric transition in IIA string theory
connecting D6-branes wrapped around ${\bf S}^3$ in
the {\it deformed} conifold geometry and
{\it resolved} conifold with Ramond-Ramond 2-form flux
through the ${\bf S}^2$.}
\label{geomtr}
\end{figure}

In order to explain how this works in the case of the conifold,
let us consider type IIA theory on the deformed conifold
geometry, $T^*{\bf S}^3 \cong {\bf S}^3 \times \mathbb{R}^3$.
This already breaks supersymmetry down to $\CN=2$
in four dimensions.
One can break supersymmetry further, to $\CN=1$,
by wrapping a space-filling D6-brane around the supersymmetric
(special Lagrangian) ${\bf S}^3$ cycle in this geometry.
Then, a natural question to ask is: ``What happens if one
tries to go through the conifold transition with the
extra D6-brane?''. One possibility could be that the other
branch is no longer connected and the transition is not possible.
However, this is not what happens. Instead the physics is somewhat
more interesting.
According to \cite{vafa}, the transition proceeds,
but now the two branches are smoothly connected,
with the wrapped D6-brane replaced by Ramond-Ramond
2-form flux through the ${\bf S}^2$ cycle of the resolved
conifold, see Figure \ref{geomtr}.

As we explained in section \ref{typeiia},
both D6-branes and Ramond-Ramond 2-form tensor fields
lift to purely geometric backgrounds in $M$ theory.
Therefore, the geometric transition described above
should lift to a transition between two purely geometric
backgrounds in $M$ theory (hence, the name).
Since these geometries must preserve the same amount
of supersymmetry, namely $\CN=1$ in four dimensions,
we conclude that we deal with a $G_2$ transition.
In fact, it is the familiar flop transition
in $M$ theory on a $G_2$ manifold \cite{bsa2,AMV}:
$$
X \cong {\bf S}^3 \times \mathbb{R}^3
$$
Indeed, if we start on one of the three branches of this
manifold and choose the `$M$ theory circle'
to be the fiber of the Hopf bundle in the 3-sphere
$$
{\bf S}^1 \hookrightarrow {\bf S}^3 \to {\bf S}^2
$$
we obtain the resolved conifold as a quotient space,
$X/U(1) \cong {\bf S}^2 \times \mathbb{R}^4$.
More precisely, we obtain a resolved conifold with
Ramond-Ramond 2-form flux and no D6-branes because
the circle action has no fixed points in this case.
This gives us one side of the brane/flux duality,
namely the right-hand side on the diagram below:

\smallskip
\begin{center}
\begin{tabular}{ccc}
\framebox[5cm][c]{\parbox{4cm}{$M$ theory on $G_2$ \\
manifold ${\bf S}^3_{(1)} \times \mathbb{R}^4$}}
& ${{\rm flop} \atop \longleftarrow
\!-\!-\!-\!-\!-\!-\!-\!-\!-\!-\!-\!-\!-\! \longrightarrow}$ &
\framebox[5cm][c]{\parbox{4cm}{$M$ theory on $G_2$ \\
manifold ${\bf S}^3_{(2)} \times \mathbb{R}^4$}} \\[21pt]
$\downarrow$ && $\downarrow$ \\[7pt]
\framebox[5cm][c]{\parbox{4cm}{
IIA on ${\bf S}^3 \times \mathbb{R}^3$ with \\
D6-brane on ${\bf S}^3$}}
& ${{\rm geometric~ transition} \atop \longleftarrow
\!-\!-\!-\!-\!-\!-\!-\!-\!-\!-\!-\!-\!-\! \longrightarrow}$ &
\framebox[5cm][c]{\parbox{4cm}{
IIA on ${\bf S}^2 \times \mathbb{R}^4$ with \\
RR flux through ${\bf S}^2$}} \\
\end{tabular}
\end{center}
\smallskip

Now, let us follow the $G_2$ flop transition in $M$ theory
on the manifold $X  \cong {\bf S}^3 \times \mathbb{R}^3$.
As explained in the previous section, after the transition
we obtain a $G_2$ manifold with similar topology,
but the $M$ theory circle is now embedded in $\mathbb{R}^4$,
rather than in ${\bf S}^3$. Acting on each $\mathbb{R}^4$ fiber,
it yields $\mathbb{R}^3 = \mathbb{R}^4/U(1)$ as a quotient
space with a single fixed point at the origin of
the $\mathbb{R}^4$ (see the discussion of $M$ theory
on the Taub-NUT space in section \ref{typeiia}).
Applying this to each fiber of the $G_2$ manifold $X$,
we obtain the deformed conifold as the quotient space,
$X/U(1) \cong {\bf S}^3 \times \mathbb{R}^4,$
with the fixed point set $L={\bf S}^3$.
Since the latter is identified with the location of
the space-filling D6-brane, we recover the other side
of the brane/flux duality, illustrated in the above diagram.

Thus, we explained that the geometric transition --- which
is a highly non-trivial, non-perturbative phenomenon in
string theory --- can be understood as a $G_2$ flop transition
in $M$ theory.
Various aspects of this transition have been discussed in
\cite{bsa2, AMV, AW, BGGG, G_2conifold, Brandhuber, Cveticunif,
Dasgupta} -- \cite{Fuji}.
As we show next, there is a similar relation
between the phase transition in $G_2$ holonomy manifold
$X = \mathbb{C}{\bf P}^2 \times \mathbb{R}^3$ and the $Spin(7)$
conifold transition, discussed in the previous section.

\begin{figure}
\begin{center}
\epsfxsize=5in\leavevmode\epsfbox{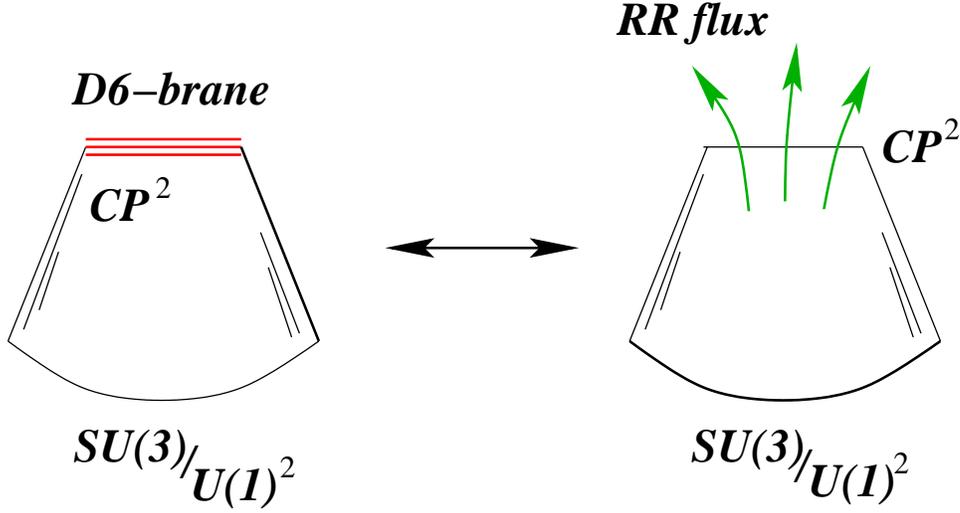}
\end{center}
\caption{Geometric transition in IIA string theory
connecting one branch of the $G_2$ manifold
$\mathbb{C}{\bf P}^2 \times \mathbb{R}^3$,
where D6-branes are warpped around
the $\mathbb{C}{\bf P}^2$ cycle and another branch,
where D6-branes are replaced by Ramond-Ramond 2-form flux
through the $\mathbb{C}{\bf P}^1 \subset \mathbb{C}{\bf P}^2$.}
\label{spin7tr}
\end{figure}

$\bullet~~{\bf G_2 \to Spin(7):}$
Consider type IIA string theory on the $G_2$ holonomy manifold
\begin{equation}
{\bf \mathbb{C}P}^2 \times \mathbb{R}^3
\label{cp2phase}
\end{equation}
which is obtained by resolving the cone over $SU(3)/U(1)^2$.
As was discussed in the previous section,
the corresponding moduli space has three classical
branches connected by a singular phase transition.
Motivated by the geometric transition in the conifold example,
one could wrap an extra D6-brane over
the $\mathbb{C}{\bf P}^2$ cycle and ask
a similar question: ``What happens if one tries to go
through a phase transition?''.

Using arguments similar to \cite{AMV},
one finds that the transition is again possible,
via $M$ theory on a $Spin(7)$ manifold \cite{GTS}.
More precisely, after the geometric transition
one finds type IIA string theory in a different
phase of the $G_2$ manifold (\ref{cp2phase}),
where the D6-brane is replaced by RR flux through
$\mathbb{C}{\bf P}^1 \subset \mathbb{C}{\bf P}^2$.
This leads to a fibration:
$$
{\bf S}^1 \hookrightarrow {\bf S}^5 \to \mathbb{C}{\bf P}^2
$$
Hence the $M$ theory lift of this configuration gives
a familiar $Spin(7)$ conifold,
$$
X \cong {\bf S}^5 \times \mathbb{R}^3.
$$
Similarly, one can identify the lift of
${\bf \mathbb{C}P}^2 \times \mathbb{R}^3$
with a D6-brane wrapped around $\mathbb{C}{\bf P}^2$
as the $Spin(7)$ manifold
${\bf \mathbb{C}P}^2 \times \mathbb{R}^4$,
which is another topological phase of the $Spin(7)$ conifold.
Summarizing, we find that the conifold transition in $M$ theory
on a $Spin(7)$ manifold is related to a geometric transition
in IIA string theory on the $G_2$ manifold (\ref{cp2phase})
with branes/fluxes, as shown in the diagram below:

\smallskip
\begin{center}
\begin{tabular}{ccc}
\framebox[5cm][c]{\parbox{4cm}{$M$ theory on $Spin(7)$ \\
manifold $\mathbb{C}{\bf P}^2 \times \mathbb{R}^4$}}
& ${{\rm conifold~ transition} \atop \longleftarrow
\!-\!-\!-\!-\!-\!-\!-\!-\!-\!-\!-\!-\!-\! \longrightarrow}$ &
\framebox[5cm][c]{\parbox{4cm}{$M$ theory on $Spin(7)$ \\
manifold ${\bf S}^5 \times \mathbb{R}^3$}} \\[21pt]
$\downarrow$ && $\downarrow$ \\[7pt]
\framebox[5cm][c]{\parbox{4cm}{
IIA on $\mathbb{C}{\bf P}^2 \times \mathbb{R}^3$ with \\
D6-brane on $\mathbb{C}{\bf P}^2$}}
& ${{\rm geometric~ transition} \atop \longleftarrow
\!-\!-\!-\!-\!-\!-\!-\!-\!-\!-\!-\!-\!-\! \longrightarrow}$ &
\framebox[5cm][c]{\parbox{4cm}{
IIA on $\mathbb{C}{\bf P}^2 \times \mathbb{R}^3$ with \\
RR flux through $\mathbb{C}{\bf P}^1$}} \\
\end{tabular}
\end{center}
\smallskip

However, unlike its prototype with larger supersymmetry,
this transition does not proceed smoothly.

\newpage
\section{Quantum Super-Yang-Mills from $G_2$ Manifolds}

Combining the results of the previous sections,
now we move on to study the relation between quantum super Yang-Mills
theory in four dimensions and properties of $M$ theory on $G_2$-manifolds.
The results of this section are based upon \cite{bsa2,AMV,AW,bsa3}.
We will be studying the physics of $M$ theory on the $G_2$-manifolds with
${\sf ADE}$-singularities whose construction we described
in sections \ref{noncompact} and \ref{heterotic}.
Specifically, we shall consider the $G_2$-manifolds that are obtained
as quotients of $\mathbb{R^4}{\times}S^3$ by $\mathbb{\Gamma_{ADE}}$.
We begin by reviewing the basic properties of super Yang-Mills theory.
We then go on to describe how these features are reflected in $M$ theory. We 
first
show how membrane instantons can be seen to generate the superpotential of the 
theory.
Then we go on to exhibit confinement and a mass gap semi-classically in $M$ 
theory.

\subsection{Super Yang-Mills Theory.}

For completeness and in order to compare easily with $M$ theory
results obtained later we briefly give a review of ${\cal N}$
$=$ $1$ pure super Yang-Mills theory. We begin with gauge group $SU(n)$.
${\cal N}$ $=$ $1$ $SU(n)$ super Yang-Mills theory in four
dimensions is an extensively studied quantum field theory.
The classical Lagrangian for the theory is
\be
{\cal L} = -{1 \over 4{g^2}}(F_{\mu\nu}^{a})^2 +
{1 \over {g^2}}{\bar \lambda}^{a}i{\not{\cal D}}{\lambda}^a +
i{\theta \over 32{\pi}^2}F_{\mu\nu}^{a}{\tilde F}^{a\mu\nu}
\ee
$F$ is the gauge field strength and $\lambda$ is the gaugino field.

It is widely believed that
this theory exhibits dynamics very similar to
those of ordinary QCD: confinement, chiral symmetry breaking,
a mass gap. There are $n$ supersymmetric vacua.
Supersymmetry constrains the dynamics of the
theory so strongly, that the values of the low energy
effective superpotential in the $n$ vacua  is known.
These are of the form
\be
W_{eff} \sim {\mu}^{3}e^{2{\pi}i{\tau}/{n}}
\ee
here $\tau$ is the complex coupling constant,
\be
\tau = {\theta \over 2\pi} + i{4\pi \over g^2}
\ee
and $\mu$ the cut-off scale. Shifting $\theta$ by $2\pi$ gives $n$ different
values for $W$.

In particular, the form of this potential suggests that it is
generated by dynamics associated with ``fractional instantons'', i.e.
instantonic objects in the theory whose quantum numbers are formally of
instanton number ${1 \over n}$. Such states are also closely
related to the spontaneously broken chiral symmetry of the theory.
Let us briefly also review some of these issues here.

Under the $U(1)$ {\sf R}-symmetry of the theory, the gauginos transform
as
\be
\lambda \rightarrow e^{i\alpha}\lambda
\ee
This is a symmetry of the classical action but not of the quantum
theory (as can easily be seen by considering the transformation
of the fermion determinant in the path integral). However, if the above
transformation is {\it combined} with a shift in the theta angle
of the form
\be
\tau \rightarrow \tau + {2n \over 2\pi}\alpha
\ee
then this cancels the change in the path integral measure.
This shift symmetry is a bona fide symmetry of the physics
if $\alpha$ $=$ ${2\pi \over 2n}$, so that even in the
quantum theory a ${\bf Z_{2n}}$ symmetry remains. Associated
with this symmetry is the presence of a non-zero value for the
following correlation function,
\be
\langle\lambda\lambda(x_1)\lambda\lambda(x_2)...
\lambda\lambda(x_n)\rangle
\ee
which is clearly invariant under the ${\bf Z_{2n}}$ symmetry.
This correlation function is generated in the 1-instanton
sector and the fact that $2n$ gauginos enter is due to the
fact that an instanton of charge 1 generates $2n$ chiral
fermion zero modes.

Cluster decomposition implies that the above correlation function
decomposes into `$n$ constituents' and therefore there exists a
non-zero value for the gaugino condensate:
\be
\langle \lambda \lambda \rangle {\neq} 0
\ee

Such a non-zero expectation value is only invariant under a
${\bf Z_2}$ subgroup of ${\bf Z_{2n}}$ implying that the
discrete chiral symmetry has been spontaneously broken.
Consequently this implies the existence of $n$ vacua in the theory.

In fact, it can be shown that
\be
\langle\lambda\lambda\rangle = 16{\pi}i{\partial \over \partial\tau}
W_{eff} \sim -{32{\pi}^2 \over n}c{\mu}^{3}e^{2{\pi}i{\tau}/n}
\ee

In view of the above facts it is certainly tempting to propose
that `fractional instantons' generate the non-zero gaugino
condensate  directly. But this is difficult to see
directly in super Yang-Mills on $\mathbb{R^{3,1}}$.
We will return to this point later.

More generally, if we replace the $SU(n)$ gauge group by
some other gauge group $H$, then the above statements are
also correct but with $n$ replaced {\it everywhere}
with ${c_2}(H)$ the
dual Coxeter number of $H$.
For {\sf A-D-E} gauge groups
${c_2}(H)$ $=$ ${\Sigma}_{i=1}^{r+1} a_i$, where $r$ is the
rank of the gauge group and the
$a_i$ are
the Dynkin indices of the affine Dynkin diagram associated
to $H$. For $A_n$, all the $a_i$ $=$ $1$; for $D_n$ groups
the four `outer' nodes have index $1$ whilst the rest have $a_i$
$=$ $2$. $E_6$ has indices $(1,1,1,2,2,2,3)$, $E_7$
has $(1,1,2,2,2,3,3,4)$ whilst $E_8$ has indices
$(1,2,2,3,3,4,4,5,6)$.

\subsection{Theta angle and Coupling Constant in M theory.}

The physics of $M$ theory supported near the singularities of
$\mathbb{C^2/{\Gamma}{\times}{R^{6,1}}}$ is described by
super Yang-Mills theory on $\mathbb{R^{6,1}}$. The gauge coupling constant
of the theory is given by
\be
{1 \over g^2_{7d}} \sim {1 \over l^3_p}
\ee
where $l_p$ is the eleven dimensional Planck length. In seven dimensions,
one analog of the theta angle in four dimensions is actually a
three-form $\Theta$. The reason for this is the seven dimensional
interaction
\be
L_I \sim \Theta {\wedge}F{\wedge}F
\ee
(with $F$ the Yang-Mills field strength). In $M$ theory $\Theta$
is given by $C$, the three-form potential for the theory.

If we now take $M$ theory on our $G_2$-manifold
$\mathbb{R^4 /{\Gamma}}{\times}W$
we have essentially compactified the seven dimensional theory on $W$ and the
four dimensional gauge coupling constant is roughly given by
\be
{1 \over g^2_{4d}} \sim {V_W  \over l^3_p}
\ee
where $V_W$ is the volume of ${W}$.
The four dimensional theta angle can be identified as
\be
\theta = \int_{W} C
\ee
The above equation is correct because under a
global gauge transformation
of $C$ which shifts the above period by $2\pi$ times
an integer --- a transformation which is a symmetry of $M$
theory --- $\theta$ changes
by $2\pi$ times an integer. Such shifts in the theta angle
are also global symmetries of the field theory.

Thus the complex gauge coupling constant of the effective four
dimensional theory may be identified as the $\tau$ parameter of $M$ theory
\be
\tau = \int_{W}  {C \over 2\pi} + i{{\Phi} \over l^3_p}
\ee

This is of course entirely natural, since $\tau$ is the only parameter in $M$ 
theory on
this space!

\subsection{Superpotential in M theory.}

There is a very elegant way to calculate the superpotential of super 
Yang-Mills theory on
$\mathbb{R^{3,1}}$ by first compactifying it on a circle to three dimensions
\cite{three}.  The
three dimensional theory has a perturbative expansion since the Wilson lines
on the circle behave as Higgs fields whose vev's break the gauge symmetry to 
the
maximal torus. The theory has a perturbative expansion in the Higss vevs,
which can be used to
compute  the superpotential of the compactified theory.
One then takes the four dimensional limit. In order to compute the field 
theory superpotential
we will mimick this idea in $M$ theory \cite{bsa1}.
Compactifying the theory on a small circle is equivalent
to studying pertubative Type IIA string theory on our $G_2$-manifold.

\subsection*{Type IIA theory on $X=\mathbb{R^4}/$$\mathbb{\Gamma_{ADE}} 
{\times} S^3$  }

Consisder Type IIA string theory compactified to three dimensions on a
seven manifold $X$ with holonomy $G_2$. If $X$ is smooth we can determine the
massless spectrum of the effective supergravity theory in three dimensions
as follows. Compactification on $X$ preserves four of the 32 supersymmetries
in ten dimensions, so the supergravity theory has three dimensional
${\cal N}$ = 2
local supersymmetry.  The relevant bosonic fields of the ten dimensional
supergravity theory are the metric, $B$-field, dilaton plus the Ramond-Ramond
one- and three-forms. These we will denote by $g,B,{\phi},{A_1},{A_3}$
respectively.
Upon Kaluza-Klein reduction the metric gives rise to a three-metric
and ${b_3}(X)$ massless scalars. The latter parametrise the moduli space of
$G_2$-holonomy metrics on $X$. $B$ gives rise to
${b_2}(X)$ periodic scalars ${\varphi}_i$.
$\phi$ gives a three dimensional dilaton. $A_1$ reduces to
a massless vector, while $A_3$ gives ${b_2}(X)$ vectors and
${b_3}(X)$ massless
scalars.
In three dimensions a vector is dual to a periodic scalar, so
at a point in moduli space where the vectors are free we can dualise them.
The dual of the vector field originating from $A_1$ is the period of the
RR 7-form on $X$, whereas the duals of the vector fields coming from $A_3$
are given by the periods of the RR 5-form $A_5$ over a basis of 5-cycles
which span the fifth homology group of $X$. Denote these by scalars
by ${\sigma}_i$.
All in all, in the dualised theory
we have in addition to the supergravity multiplet, ${b_2}(X) +{b_3}(X)$
scalar
multiplets. Notice that ${b_2}(X)$
of the scalar multiplets contain two real
scalar fields, both of which are periodic.

Now we come to studying the
Type IIA theory on $X=\mathbb{R^4 /{\Gamma}_{ADE}} {\times} S^3$.
Recall that $X=\mathbb{R^4 /{\Gamma}_{ADE}} {\times} S^3$ is defined as
an orbifold of the standard spin bundle of $S^3$.
To determine the massless spectrum of IIA string theory on
$X$ we can use standard orbifold techniques. However, the answer can be 
phrased
in a simple way. $X$ is topologically
$\mathbb{R^4}/\mathbb{\Gamma_{ADE}} {\times} S^3$. This
manifold can be desingularised to give a smooth seven manifold
$M^{\mathbb{\Gamma_{ADE}}}$
which is topologically ${X^{\mathbb{\Gamma_{ADE}}}}{\times}S^3$,
where $X^{\mathbb{\Gamma_{ADE}}}$ is
homeomorphic to an $ALE$ space.
The string theoretic cohomology groups of $X$ are isomorphic to the usual
cohomology groups of $M^{\mathbb{\Gamma_{ADE}}}$.
The reason for this is simple: $X$ is a global
orbifold of $S({S^3})$. The string theoretic cohomology groups count
massless string states in the orbifold CFT. The massless string states in
the twisted sectors are localised near the fixed points of the action of
$\mathbb{\Gamma_{ADE}}$ on the spin bundle. Near the fixed points we can work 
on
the tangent space of $S(S^3)$ near a fixed point and the action of 
$\mathbb{\Gamma_{ADE}}$
there is just its natural action on $\mathbb{R^4}{\times}\mathbb{R^3}$.

Note that blowing up $X$ to give $M^{\mathbb{\Gamma_{ADE}}}$
cannot give a metric with $G_2$-holonomy which is continuosly connected
to the singular $G_2$-holonomy metric on $X$, since this would require
that the addition to homology in passing from $X$ to
$M^{\mathbb{\Gamma_{ADE}}}$ receives contributions
from four-cycles. This is necessary since these are dual to elements of
$H^3 (M)$ which generate metric deformations preserving the $G_2$-structure.
This argument does not rule out the possibility that
$M^{\mathbb{\Gamma_{ADE}}}$ admits `disconnected'
$G_2$-holonomy metrics, but is consistent with the fact that pure
super Yang-Mills theory in four dimensions does not have a Coulomb branch.

The important points to note are that the twisted sectors contain massless
states consisting of $r$ scalars and $r$ vectors where $r$ is the rank of
the corresponding {\sf {ADE}} group associated to
$\mathbb{\Gamma}$.
The $r$ scalars can
intuitively thought of as the periods of the $B$ field through $r$ two cycles.
In fact, for a generic point in the moduli space of the orbifold conformal
field theory the spectrum contains massive particles charged under the
$r$ twisted vectors.
These can be interpreted as wrapped D2-branes whose quantum
numbers are precisely those of $W$-bosons associated with the breaking of
an {\sf {ADE}} gauge group to $U(1)^r$. This confirms our interpretation of 
the
origin of this model from $M$ theory:
the values of the $r$ $B$-field scalars can be interpreted
as the expectation values of Wilson lines around the eleventh dimension
associated with this symmetry breaking. At weak string coupling and large
$S^3$ volume these states are very massive and the extreme low energy
effective dynamics of the twisted sector states is described by ${\cal N}$=2
$U(1)^r$ super Yang-Mills in three dimensions.
Clearly however, the underlying
conformal field theory is not valid when the $W$-bosons become massless. The
appropriate description is then the pure super Yang-Mills theory on
$\mathbb{R^{2,1}}{\times}{S^1}$ which corresponds to a sector of $M$ theory on
$X{\times}S^1$.
In this section however, our strategy will be to
work at a generic point in the CFT moduli space which corresponds to being
far out along the Coulomb branch of the gauge theory.
We will attempt to calculate the
superpotential there and then continue the result to four dimensions.
This exactly mimics the strategy of \cite{three} in field theory.
Note that we are implicitly ignoring gravity here. More precisely, we are
assuming that in the absence of gravitational interactions with the twisted
sector, the low energy physics of the twisted sectors of the CFT
is described by the Coulomb branch of the gauge theory. This is natural since
the twisted sector states are localised at the singularities of
$J{\times}\mathbb{R^{2,1}}$ whereas the gravity propagates in bulk.

In this approximation,
we can dualise the photons to obtain a theory of $r$ chiral
multiplets, each of whose bosonic components ({\mbox{\boldmath ${\varphi}$}}
and {\mbox{\boldmath ${\sigma}$}})
is periodic.
But remembering
that this theory arose from a non-Abelian one we learn that the moduli space
of classical vacua is
\be
{\cal M}_{cl} = {{\mathbb C}^r \over {{\Lambda}_{W}^{\mathbb{C}}\rtimes W_g}}
\ee

where ${\Lambda}^{\mathbb{C}}_W$ is the
complexified weight lattice of the {\sf ADE} group and $W_g$ is
the Weyl group.

We can now ask about quantum effects. In particular is there a
non-trivial superpotential for these chiral multiplets? In a theory with four
supercharges BPS instantons with only two chiral fermion zero modes can 
generate
a superpotential. Are there instantons in Type IIA theory on J ?
BPS instantons come from branes wrapping supersymmetric cycles and Type IIA
theory on a $G_2$-holonomy space can have instantons corresponding to 
D6-branes
wrapping the space itself or D2-brane instantons which wrap supersymmetric
3-cycles. For smooth $G_2$-holonomy manifolds these were studied in 
\cite{greg}.
In the case at hand the D6-branes would generate a superpotential
for the dual of the graviphoton multiplet which lives in the gravity multiplet
but since we wish to ignore gravitational physics for the moment, we will 
ignore
these. In any case, since $X$ is non-compact, these configurations have
infinite action. The D2-branes on the other hand are much more interesting.
They can wrap the supersymmetric $S^3$ over which the singularities of $X$
are fibered. We can describe the dynamics of a wrapped D2-brane as follows.
At large volume, where the sphere becomes flatter and flatter the world-volume
action is just the so called `quiver gauge theory'
described in \cite{quiver}. Here
we should describe this theory not just on $S^3$ but on a
supersymmetric  $S^3$ embedded in
a space with a non-trivial $G_2$-holonomy structure. The upshot is that the
world-volume theory is in fact a cohomological field theory \cite{bsv}
so we can trust it
for any volume as long as the ambient space has $G_2$-holonomy. This is 
because
the supersymmetries on the world-volume are actually
scalars on $S^3$ and so must square to zero.

Note
that, since we are ignoring gravity, we are implicitly ignoring higher
derivative corrections which could potentially also affect this claim.
Another crucial
point is that the $S^3$ which sits at the origin in
$\mathbb{R^4}$ in the covering
space of $X$ is the supersymmetric cycle, and the spheres away from the origin
are not supersymmetric,
so that the BPS wrapped
D2-brane is constrained to live on the singularities of $X$. In the quiver
gauge theory, the origin is precisely the locus in moduli space at which the
single D2-brane can fractionate (according to the quiver diagram) and this
occurs by giving expectation values to the scalar fields which parametrise
the Coulomb branch which corresponds to the position of our D2-brane in
the dimensions normal to $X$.

What contribution to the superpotential do the fractional D2-branes make?
To answer this we need to identify the configurations which possess only
two fermionic zero modes. We will not give a precise string theory argument
for this, but using the correspondence
between this string theory and field theory will identify exactly which
D-brane instantons we think are responsible for generating the superpotential.
This may sound like a strong assumption, but as we hope will become clear,
the fact that the fractional D2-branes are wrapped D4-branes is actually
anticipated by the field theory! This makes this assumption, in our opinion,
somewhat weaker and adds credence to the overall picture being presented here.

In  \cite{frac}, it was shown that the fractionally
charged D2-branes are actually D4-branes which wrap the `vanishing' 2-cycles
at the origin in $\mathbb{{R^4}/{\mathbb{\Gamma}}}$. More precisely, each 
individual
fractional D2-brane, which originates from a single D2-brane
possesses D4-brane charge, but the total configuration, since it began life
as a single D2-brane has zero D4-brane charge. The possible contributions to
the superpotential are constrained by supersymmetry and must be
given by a holomorphic function of the $r$ chiral superfields and
also of the holomorphic gauge coupling constant $\tau$ which corresponds
to the complexified volume of the $S^3$ in eleven dimensional $M$ theory.
We have identified
above the bosonic components of the chiral
superfields above. $\tau$ is given by
\be
\tau = \int \Phi + i C
\ee
where $\Phi$ is the $G_2$-structure defining 3-form on $X$. The period of
the $M$ theory 3-form through $S^3$ plays the role of the theta angle.

The world-volume action of a D4-brane contains the couplings
\be
L = B \wedge A_3 + A_5
\ee

Holomorphy dictates that there is also a term
\be
B \wedge \Phi
\ee

so that the combined terms are written as
\be
B \wedge \tau + A_5
\ee

Since the fourbranes wrap the `vanishing cycles' and the $S^3$ we see that
the contribution of the D4-brane corresponding to the $k$-th
fractional D2 takes the form
\be
S = - \mbox{\boldmath ${\beta}_{k}.z$}
\ee
where we have defined
\be
{\mbox{\boldmath $z$}} = \tau {\mbox{\boldmath ${\varphi}$}} +
{\mbox{\boldmath ${\sigma}$}}
\ee
and the \mbox{\boldmath ${\beta}_k$} are charge vectors. The
$r$ complex fields {\mbox{\boldmath $z$}} are the natural holomorphic
functions upon which the superpotential will depend.

The wrapped D4-branes are the magnetic duals of the massive D2-branes which
we identified above as massive $W$-bosons. As such they are magnetic monopoles
for the original ${\sf ADE}$ gauge symmetry.
Their charges are therefore given
by an element of the co-root lattice of the Lie algebra and thus each of
the $r$ + 1 \mbox{\boldmath ${\beta}$}'s is a rank $r$ vector in this space.
Choosing a basis
for this space corresponds to choosing a basis for the massless states in
the twisted sector Hilbert space which intuitively we can think of as a basis
for the cohomology groups Poincare dual to the `vanishing' 2-cycles. A natural
basis is provided by the simple co-roots of the Lie algebra of ${\sf ADE}$, 
which
we denote by \mbox{\boldmath ${\alpha}^*_k$}
for $k = 1,...,r$.
This
choice is natural, since these, from the field theory point of view are
the fundamental monopole charges.

At this point it is useful to mention that the $r$
wrapped D4-branes whose magnetic charges are given by the simple co-roots of
the Lie algebra correspond in field theory to monopoles with charges
\mbox{\boldmath ${\alpha}^*_k$} and each of these is known to possess 
precisely
the right number of zero modes to contribute to the superpotential. Since we
have argued that in a limit of the Type IIA theory on $X$, the dynamics
at low energies is governed
by the field theory studied in \cite{three} it is natural to expect that these
wrapped fourbranes also contribute to the superpotential. Another striking
feature of the field theory is that these monopoles also possess a fractional
instanton number - the second Chern number of the gauge field on
$\mathbb{R^{2,1}}{\times}S^1$. These are precisely in correspondence with the
fractional D2-brane charges. Thus, in this sense, the field theory anticipates
that fractional branes are wrapped branes.

In the field theory on $\mathbb{R^{2,1}}{\times}S^1$
it is also important
to realise that there is precisely one additional BPS state which
contributes to the superpotential. The key point is that this state, unlike
the previously discussed monopoles have dependence on the periodic direction
in spacetime.
This state is associated with the affine
node of the Dynkin diagram. Its monopole charge is given by
\be
 - {\Sigma}_{k=1}^r  \mbox{\boldmath ${\alpha}^*_{k}$}
\ee
and it also carries one unit of instanton number.

The action for this state is
\be
S =  {\Sigma}_{k=1}^r \mbox{\boldmath ${\alpha}^*_{k}.z$} -2\pi i \tau
\ee

Together, these $r$ + 1 BPS states can be regarded as fundamental in the sense
that all the other finite action BPS configurations can be thought of as
bound states of them.

Thus, in the correspondence with string theory it is
also natural in the same sense as alluded to above that a state with these
corresponding quantum numbers also contributes to the superpotential. It
may be regarded as a bound state of anti-D4-branes with a charge one D2-brane.
In the case of $SU(n)$ this is extremely natural, since
the total D4/D2-brane charge of the $r$ +1 states is zero/one,
and this is precisely the charge of the D2-brane configuration on $S^3$
whose world-volume action is the quiver gauge theory for the affine
Dynkin diagram for $SU(n)$. In other words, the entire superpotential is
generated by a single D2-brane which has fractionated.

In summary, we have seen that the correspondence between the Type IIA string
theory on $X$ and the super Yang-Mills theory on
$\mathbb{R^{2,1}}{\times}S^1$ is quite
striking. Within the context of this correspondence we considered a smooth
point in the moduli space of the perturbative Type IIA CFT, where the spectrum
matches that of the Yang-Mills theory along its Coulomb branch. On the
string theory side we concluded that the possible instanton contributions
to the superpotential are from wrapped D2-branes. Their world volume theory
is essentially topological, from which we concluded that they can fractionate.
As is well known, the fractional D2-branes are really wrapped fourbranes.
In the correspondence with field theory, the wrapped fourbranes are magnetic
monopoles, whereas the D2-branes are instantons. Thus if, these branes
generate a superpotential they correspond, in field theory to
monopole-instantons.
This is exactly how the field theory potential is known to be generated.
We thus expect
that the same occurs in the string theory on $X$.

Finally, the superpotential generated by these instantons is of affine-Toda
type and is known to possess $c_2 ({\sf ADE})$ minima corresponding to the 
vacua of the
${\sf ADE}$ super
Yang-Mills theory on $\mathbb{R^{3,1}}$. The value of the superpotential
in each of these vacua is of the form ${e}^{2{\pi} i\tau  \over c_2}$.
As such it formally looks as though it was generated by fractional instantons,
and in this context fractional $M$2-brane instantons. This result holds
in the four dimensional $M$ theory limit because of holomorphy.

Let us demonstrate the vacuum structure in the simple case when the gauge 
group is
$SU(2)$. Then there is only one scalar field, $z$.
There are two fractional D2-brane instantons whose actions are
\be
S_1 = -z \;\;\;and\;\;\; S_2 = z - 2\pi i\tau
\ee
Both of these contribute to the superpotential as
\be
W = e^{-S_1} + e^{-S_2}
\ee
Defining $z=lnY$ we have
\be
W = Y + {e^{2\pi i\tau} \over Y}
\ee
The critical points of $W$ are
\be
Y= \pm e^{2\pi i \tau \over 2}
\ee

This result about the superpotential suggests strongly  that there is a limit 
of $M$ theory
near an ${\sf ADE}$ singularity in a $G_2$-manifold which is precisely super 
Yang-Mills
theory. We will now go on to explore other limits of this $M$ theory 
background.

\subsection{M theory Physics on ADE-singular $G_2$-manifolds.}

We saw previously in section \ref{transitions},
that before taking the quotient by $\mathbb{\Gamma}$,
the $M$ theory physics on $\mathbb{R^4}{\times}S^3$,
with its $G_2$-metric, was smoothly varying as a function of $\tau$.
In fact the same is true in the case with ${\sf ADE}$-singularities.
One hint for this was that we explicitly saw just now that
the superpotential is non-zero in the various vacua
and this implies that the $C$-field is non-zero.
This suggestion was concretely proven in \cite{AW},
see also \cite{Curiowww}.

Before orbifolding by $\mathbb{\Gamma}$ we saw
there were three semiclassical limits
of $M$ theory in the space parameterised by $\tau$.
These were described by $M$ theory
on three large and smooth $G_2$-manifolds $X_i$,
all three of which were of the form
$\mathbb{R^4}{\times}{S^3}$.
There are also three semiclassical i.e.
large volume $G_2$-manifolds when we orbifold by $\mathbb{\Gamma}$.
These are simply the quotients by $\mathbb{\Gamma}$ of the $X_i$.
One of these is
the $G_2$-manifold $\mathbb{R^4 /\Gamma_{ADE}}{\times} S^3$.
The other two are
both of the form $S^3 /\mathbb{\Gamma_{ADE}} {\times}\mathbb{R^4}$.
To see this, note
that the three $S^3$'s in the three $G_2$-manifolds $X_i$
of the form $\mathbb{R^4}{\times}{S^3}$
correspond to the three ${S^3}$ factors in
$G=SU(2)^3 = {S}^3 {\times}{S}^3{\times}{S}^3$,
{\it c.f.} section \ref{noncompact}.
$\mathbb{\Gamma_{ADE}}$ is a subgroup of one of these $S^3$'s.
If $\mathbb{\Gamma_{ADE}}$ acts on the $\mathbb{R^4}$ factor
of $X_1$ in the standard way, then it must act on $S^3$
in $X_2$ - since $X_2$ can be thought of
as the same manifold but with the two $S^3$'s at infinity exchanged.
Then, because of the permutation symmetry it also acts on the $S^3$
in $X_3$. Thus, $X_2 /\mathbb{\Gamma_{ADE}}$
is isomorphic to  $S^3 /\mathbb{\Gamma_{ADE}} {\times}\mathbb{R^4}$,
as is $X_3$. Thus, varying $\tau$ can take us from $M$ theory
on the singular $G_2$-manifold $X_1 /\mathbb{\Gamma_{ADE}}$ to
the smooth $G_2$-manifolds $X_{2,3} /\mathbb{\Gamma_{ADE}}$.

On $X_1 /\mathbb{\Gamma_{ADE}}$ in the large volume limit,
we have a semi-classical description of the four dimensional physics
in terms of perturbative super Yang-Mills theory.
But, at extremely low energies, this theory becomes strongly coupled,
and is believed to confine and get a mass gap.
So, apart from calculating the superpotential in each vacuum,
we can't actually calculate the spectrum here.

What about the physics in the other two semiclassical limits, namely large
$X_{2,3}/\mathbb{\Gamma_{ADE}}$? These $G_2$-manifolds are completely smooth.
So supergravity is a good approximation to the $M$ theory physics.
What do we learn about the $M$ theory physics in this approximation?

\subsection{Confinement from $G_2$-manifolds.}

If it is true that the qualitative physics of $M$ theory on  $X_2 
/\mathbb{\Gamma_{ADE}}$
and $X_3 /\mathbb{\Gamma_{ADE}}$ is the same as that of $M$ theory
on $X_1 /\mathbb{\Gamma_{ADE}}$, then some of the properties of super 
Yang-Mills
theory at low energies ought to be visible. The gauge theory is believed
to confine ${\sf ADE}$-charge at low energies. If a gauge theory confines in 
four dimensions,
electrically charged confining flux tubes (confining strings) should be 
present. If the classical
fields of the gauge theory contain only fields in the adjoint representation 
of the
gauge group $G$, then these strings are charged with respect to the center of 
$G$, $Z(G)$.
Can we see these strings in $M$ theory on $X_2 /\mathbb{\Gamma_{ADE}}$?
As shown in \cite{bsa3}, the answer is yes.

The natural candidates for such strings are $M2$-branes which wrap around 
1-cycles in
$X_2 /\mathbb{\Gamma_{ADE}}$ or $M5$-branes which wrap 4-cycles in
$X_2 /\mathbb{\Gamma_{ADE}}$. Since $X_2 /\mathbb{\Gamma_{ADE}}$ is 
homeomorphic
to ${S^3}/{\mathbb{\Gamma_{ADE}}}{\times}{\mathbb{R^4}}$ which is contractible 
to
${S^3}/{\mathbb{\Gamma_{ADE}}}$, the homology groups of $X_2 
/\mathbb{\Gamma_{ADE}}$
are the same as those of the three-manifold  ${S^3}/{\mathbb{\Gamma_{ADE}}}$. 
Thus, our
space has no four cycles to speak of, so the confining strings can only come 
from
$M2$-branes wrapping one-cycles in ${S^3}/{\mathbb{\Gamma_{ADE}}}$.
The string charges are classified by the first homology group
$H_1 ({S^3}/{\mathbb{\Gamma_{ADE}}}, \mathbb{Z})$.    For any manifold, the 
first homology
group is isomorphic to the abelianisation of its fundamental group, $\Pi_1$. 
The abelianisation
is obtained by setting all commutators in $\Pi_1$ to be trivial i.e.
\be
H_1 (M, \mathbb{Z}) \cong {\pi_1 (M) \over [{\pi_1 (M)},{\pi_1 (M)}]}
\ee

The fundamental group of ${S^3}/\mathbb{\Gamma_{ADE}}$ is 
$\mathbb{\Gamma_{ADE}}$.
Hence, in order to calculate the charges of
our candidate confining strings we
to compute the abelianisations of all of the finite subgroups of $SU(2)$.

$\mathbb{\Gamma_{A_{n-1}}}$ $\cong$ $\mathbb{Z_n}$. The gauge group is locally 
$SU(n)$.
Since $\mathbb{Z_n}$ is abelian, its commutator subgroup is trivial and
hence the charges of our strings take values in $\mathbb{Z_n}$. Since this is 
isomorphic to the
center of $SU(n)$ this is the expected answer for a confining $SU(n)$ theory.

For  $\mathbb{\Gamma}$ $\cong$ $\mathbb{D_{k-2}}$, the binary dihedral group 
of order
$4k-8$, the local gauge group of the Yang-Mills theory is $SO(2k)$.
We remind that the binary dihedral group is generated by
two elements $\alpha$ and $\beta$ which obey the relations
\be
{\alpha}^2 = {\beta}^{k-2}
\ee
\be
\alpha \beta = {\beta}^{-1} \alpha
\ee
\be
{\alpha}^4 = {\beta}^{2k-4} = 1
\ee

To compute the abelianisation of  $\mathbb{D_{k-2}}$, we simply take these 
relations and
impose that the commutators are trivial. From the second relation this implies 
that
\be
\beta = {\beta}^{-1}
\ee
which in turn implies that
\be
{\alpha}^2 = 1 \;\;\;\; for\;\;\;\;\; k = 2p
\ee
and
\be
{\alpha}^2 = \beta \;\;\;\;\; for\;\;\;\;\; k = 2p+1
\ee
Thus, for $k=2p$ we learn that the abelianisation of $\mathbb{D_{k-2}}$ is 
isomorphic
to $\mathbb{{Z_2}{\times}Z_2}$, whereas for $k=2p+1$ it is isomorphic to 
$\mathbb{Z_4}$.
These groups are respectively the centers of $Spin(4p)$ and $Spin(4p+2)$. This 
is the
expected answer for the confining strings in $SO(2k)$ super Yang-Mills which 
can be coupled
to spinorial charges.

To compute the abelianisations of the binary tetrahedral (denoted 
${\mathbb{T}}$),
octahedral (${\mathbb{O}}$) and icosahedral (${\mathbb{I}}$)
groups which correspond respectively to $E_6$, $E_7$ and $E_8$ super 
Yang-Mills theory,
we utilise the fact that the order of $F/[F,F]$ - with $F$ a finite group -
is the number of
inequivalent one dimensional representations of $G$. The representation theory 
of the
finite subgroups of $SU(2)$ is described through the Mckay correspondence by 
the
representation theory of the corresponding Lie algebras. In particular the 
dimensions
of the irreducible representations of ${\mathbb{T,O}}$ and ${\mathbb{I}}$ are 
given by the
coroot integers (or dual Kac labels) of the affine Lie algebrae associated to 
$E_6$,
$E_7$ or $E_8$ respectively. From this we learn that the respective orders
of ${\mathbb{T}}/[{\mathbb{T}},{\mathbb{T}}]$,  ${\mathbb{O}}/[{\mathbb{O}},
{\mathbb{O}}]$ and
${\mathbb{I}}/[{\mathbb{I}},{\mathbb{I}}]$ are three, two and one. Moreover, 
one can easily
check that ${\mathbb{T}}/[{\mathbb{T}},{\mathbb{T}}]$ and 
${\mathbb{O}}/[{\mathbb{O}},{\mathbb{O}}]$ are
$\mathbb{Z_3}$ and $\mathbb{Z_2}$ respectively, by examining their group 
relations.
Thus we learn that   ${\mathbb{T}}/[{\mathbb{T}},{\mathbb{T}}]$,  
${\mathbb{O}}/[{\mathbb{O}},{\mathbb{O}}]$ and
${\mathbb{I}}/[{\mathbb{I}},{\mathbb{I}}]$ are, respectively isomorphic to the 
centers
$Z({E_6})$, $Z({E_7})$ and $Z({E_8})$ in perfect agreement with the 
expectation that
the super Yang-Mills theory confines. Note that the ${\sf E_8}$-theory does 
not confine,
since the strings are uncharged.

This result is also natural from the following point of view. In the singular
$X_1 /\mathbb{\Gamma_{ADE}}$ (where the actual gauge theory dynamics is) the 
gauge
bosons correspond to $M2$-branes wrapped around zero-size cycles. When we vary
$\tau$ away from the actual gauge theory limit until we reach $M$ theory on a 
large and
smooth $X_2 /\mathbb{\Gamma_{ADE}}$ the confining strings are also wrapped 
$M2$-branes.
In the gauge theory we expect the confining strings to be ``built'' from the 
excitations of the
gauge fields themselves. In $M$ theory, the central role played by the gauge 
fields is actually
played by the $M2$-brane.

\subsection{Mass Gap from $G_2$-manifolds.}

We can also see the mass gap expected of the gauge theory,
by studying the spectrum of $M$ theory on the smooth $G_2$-manifolds
$X_2 /\mathbb{\Gamma_{ADE}}$ and $X_3 /\mathbb{\Gamma_{ADE}}$.
One can easliy show that there are no $L^2$-normalisable fluctuations
of the $G_2$-holonomy metric on $X_i$.
In order to see this, one has to look at the $L^2$ norm of
the metric fluctuations,
\be
|\delta g|^2 =
\int_X d^7 x \sqrt{g} g^{ii'} g^{jj'} \delta g_{ij} \delta g_{i'j'}
\label{dgnorm}
\ee
{}From this expression, it follows that a deformation
of an asymptotically conical metric like (\ref{conicalmet})
is $L^2$ normalizable if and only if $\delta g /g$ goes to
zero faster than $r^{-7/2}$, where $r$ is the radial coordinate.
Using the explicit form (\ref{bsgppmet}) of the $G_2$ metric on $X_i$,
we find $\delta g / g \sim r^{-3}$, where $\delta g$ is the variation
of the metric (\ref{bsgppmet}) with respect to a change in $r_0$.
Therefore, this deformation is not $L^2$-normalisable,
\be
|\delta g|^2 \to \infty
\ee
This implies that the spectrum of $M$ theory on the $X_i$ is massive.
Since $X_{2,3} /\mathbb{\Gamma_{ADE}}$ is a {\it smooth} quotient
of $X_{2,3}$, the $M$ theory spectrum also has a mass gap.
Since $M$ theory on $X_{2,3} /\mathbb{\Gamma_{ADE}}$ is smoothly
connected to $M$ theory on $X_1 /\mathbb{\Gamma_{ADE}}$ by varying $\tau$,
this demonstrates that pure super Yang-Mills theory has a mass gap.

\bigskip
\centerline{\bf Acknowledgments}
\noindent

Various topics covered in this review are based on the work done
together with A.~Brandhuber, J.~Gomis, S.~Gubser X.~de la Ossa,
D.~Tong, J.~Sparks, C.~Vafa, E.~Witten, S.-T.~Yau, and E.~Zaslow,
whom we wish to thank for many useful discussions and collaboration.
This report was largely completed during the period
B.S.A. served as a research associate
at New High Energy Theory Center, Rutgers University,
and S.G. served as a Clay Mathematics Institute Long-Term Prize Fellow.
We thank both institutions for their support.
The work of S.G. is also supported in part by RFBR grant 04-02-16880
and RFBR grant for Young Scientists 02-01-06322.


\end{document}